\documentclass[reprint,twocolumn,showpacs,preprintnumbers,amsmath,amssymb,floatfix,aps,pra]{revtex4-2}

\usepackage{graphicx}
\usepackage{dcolumn}
\usepackage{bm}
\usepackage{color}
\usepackage{ulem}

\newcommand{\bra}[1]{\langle #1|}
\newcommand{\ket}[1]{|#1\rangle}

\newcommand{\vek}[1]{\mathbf{#1}}

\usepackage{caption}
\usepackage{subcaption}
\definecolor{darkgreen}{rgb}{0.0, 0.5, 0.0}

\newcommand{\affA}{Zentrum f\"ur Optische Quantentechnologien, Universit\"at Hamburg, Luruper Chaussee 149, 22761 Hamburg, Germany}

\newcommand{\affC}{Van der Waals Zeeman Institute, Institute of Physics, University of Amsterdam, Science Park 904, 1098 XH Amsterdam, The Netherlands}

\begin{document}


\title{Dynamics of a trapped ion in a quantum gas: effects of particle statistics}


\author{Lorenzo Oghittu, Melf Johannsen}
\affiliation{\affA}
\author{Rene Gerritsma}
\affiliation{\affC}
\author{Antonio Negretti}
\affiliation{\affA}

\date{\today}


\begin{abstract}
We study the quantum dynamics of an ion confined in a radiofrequency trap in interaction with either a Bose or spin-polarized Fermi gas. To this end, we derive quantum optical master equations in the limit of weak coupling and the Lamb-Dicke approximations. For the bosonic bath, we also include the so-called ``Lamb-shift" correction to the ion trap due to the coupling to the quantum gas as well as the extended Fr\"ohlich interaction within the Bogolyubov approximation that have been not considered in previous studies. We calculate the ion kinetic energy for various atom-ion scattering lengths as well as gas temperatures by considering the intrinsic micromotion and we analyse the damping of the ion motion in the gas as a function of the gas temperature. We find that the ion's dynamics depends on the quantum statistics of the gas and that a fermionic bath enables to attain lower ionic energies. 
\end{abstract}

\maketitle


\section{Introduction}
\label{sec:intro} 

The interest in studying both experimentally and theoretically quantum mixtures of ions and ultracold gases is increasing rapidly. Indeed, such a compound system offers various fascinating perspectives both on fundamental quantum few- and many-body physics research and on technological applications that the two systems separately cannot afford. For instance, the exploration of novel polaronic states~\cite{CasteelsJLTP11,Gregory2021,BruunPRL2020} and quantum simulation of the electron-phonon coupling~\cite{BissbortPRL13,Michelsen2019,JachymskiPRR2020}. For an extensive overview on atom-ion physics research, we refer to the review~\cite{TomzaRMP,CoteAAMOP16,HarterCP14}. 
Experimentally, a considerable effort has been undertaken in the last few years in order to cool the atom-ion compound system down to the quantum regime, namely when only s-wave atom-ion collisions take place. To this end, three experimental approaches have been pursed so far: ionization of a highly excited Rydberg atom in a Bose-Einstein condensate~\cite{Kleinbach2018,Engel2018}; an ion confined in a radiofrequency (rf) trap interacting with an optically trapped atomic gas~\cite{Feldker2019,HirzlerPRA2020}; sympathetic cooling of ions and atoms in optical dipole traps~\cite{SchmidtPRL2020,weckesser2021observation}. 
In the former approach, the ion is not trapped after ionisation and a controlled momentum kick via external electric fields is imparted in order to investigate charge transport in a bosonic medium~\cite{DieterlePRL2021,DieterlePRA2020}. The second relies on the well-established laser cooling and manipulation techniques of trapped ions, which are confined by means of a combination of time-dependent and time-independent electric fields. The exquisite control of the ion motion enables to prepare various non-classical states~\cite{Leibfried2003,SchneiderRPP12} and, in principle, to infer on environment properties by reading out the ion quantum state. The time-dependent fields, however, can seriously jeopardise the attainment of the ultracold atom-ion collisional regime, a notorious issue named micromotion. 
Finally, the third approach is somehow a hybrid of the former two, as it reduces drastically the impact of ion micromotion in Paul traps and, at least in principle, it enables to control the ion motion optically. At the same time, however, since the optical trap is not as deep as the rf-trap, unavoidable stray fields can reduce the ion lifetime in the trap~\cite{SchneiderPRA12,Lambrecht2016}. Moreover, an intense laser light can induce undesired chemical reactions between the ion and the particles of the gas.

Here, we focus our attention on the second approach and investigate the cooling dynamics of a trapped ion immersed in either a bosonic or fermionic environment. Albeit the s-wave regime in hybrid traps has been attained with only fermionic atoms so far~\cite{Feldker2019,hirzler2021observation}, several ongoing experiments involve bosonic ensembles~\cite{RaviNatCommun12,EberleCPC16,KrukowPRL16,MeirPRL16,DuttaPRL17,Wessels2017,SchmidtPRL2020}. Up until now, however, a few theoretical studies have been undertaken in order to assess the impact of ion micromotion on the atom-ion quantum dynamics: a quantum mechanical calculation in one dimension (1D)~\cite{NguyenPRA12,Joger2014}, a semiclassical investigation of confinement-induced resonances in quasi-1D~\cite{MelezhikPRA2019,MelezhikPRA2021}, and a 3D master equation analysis~\cite{KrychPRA15}. Apart from the latter, all others concerned a single trapped atom and ion, and therefore only the emerging two-body physics has been investigated. In Ref.~\cite{KrychPRA15}, however, the Fr\"ohlich model was assumed, while the Lamb-shift and, specifically for the bosonic case, the contribution of the non-condensed fraction were not considered. Here, by Lamb-shift we refer to the renormalisation of the ion trap parameters due to the coupling to the quantum gas. In condensed-matter and for a free impurity such a shift is named polaron shift, whereas in this work we use the quantum optics terminology. The aforementioned studies confirmed that the impact of the ion micromotion can be mitigated by choosing a small atom-ion mass ratio. With the present study, we first aim at developing a formalism for mobile quantum impurities based on an open quantum system approach, which does not rely on the (linear) Fr\"ohlich impurity-bath interaction model and rotating-wave approximation. Moreover, we want to understand the role of the gas quantum statistics on the ion cooling dynamics and whether atom-ion pairs different than Li/Yb$^+$ can reach the s-wave limit. As far as the first objective is concerned, we find that the corrections causing the Lamb-shift yield additional dissipative contributions in the master equation, as a consequence of the non-applicability of the rotating-wave approximation. Furthermore, the quantum statistics of the ultracold gas affects significantly the ion dynamics. While for temperatures larger than the Fermi temperature $\mathcal{T}_F$ and the critical temperature of condensation $\mathcal{T}^0_{\mathrm{c}}$ the ion dynamics reproduces essentially the one corresponding to the interaction with a buffer gas, at low temperature the fermionic and bosonic nature of the gas is observable in a distinct gas temperature dependence of the ion energy. Both for the fermionic and bosonic gas we observe a minimum in the ion energy for a temperature $\mathcal{T}_{\min}$ close to $\mathcal{T}^0_{\mathrm{c}}$. Below $\mathcal{T}_{\min}$ the ion energy increases again, but only marginally for fermions. On the other hand, for a bosonic gas the ion energy dependence on the gas temperature $\mathcal{T}$ exhibits a more rich structure. Indeed, after the enhancement of the ion energy for $\mathcal{T}^0_{\mathrm{c}} < \mathcal{T} < \mathcal{T}_{\min}$, below $\mathcal{T}^0_{\mathrm{c}}$ it presents a maximum due to the interplay between the condensate contribution and that of the normal part of the gas. Moreover, the damping rate of the ion motion exhibits the same dependence on the gas temperature as the condensate fraction for $ \mathcal{T} < \mathcal{T}^0_{\mathrm{c}}$. In addition, we elucidate the role of the long-range character of the atom-ion polarisation potential on the ion quantum dynamics by comparing our findings with those of the pseudopotential. Contrary to neutral impurities, the non-equilibrium dynamics of the ion in the quantum gas is non-universal, that is, it cannot be uniquely characterised by the impurity-gas scattering length like for neutral impurities. Since there is no clear separation of length scales in the many-body problem at typical gas densities, the tail of the atom-ion interaction plays a crucial role in the ion dynamics and the effective range corrections cannot be neglected.

The paper is organised as follows: In Sec.~\ref{sec:ai-interaction} we introduce the atom-ion potential, while in Sec.~\ref{sec:sPb} the system plus bath description is outlined, which is kept on purpose quite general. A master equation including the contribution of the non-condensed fraction is obtained in a way that is valid for both neutral and charged impurities. In Sec.~\ref{sec:LDA-PL_approx} we focus on the trapped ion case and make use of the well-known Lamb-Dicke approximation, which enables us to further simplify the description of the ion dynamics. We continue with Sec.~\ref{sec:2ndMom} by providing the equations of motion of observables of interest, while in Sec.~\ref{sec:results} we present our results. In Sec.~\ref{sec:disc} we draw our conclusions and provide an outlook for future work. 


\section{Atom-ion interaction potential}
\label{sec:ai-interaction}

\begin{figure}
\centering
\includegraphics[width=0.4\textwidth]{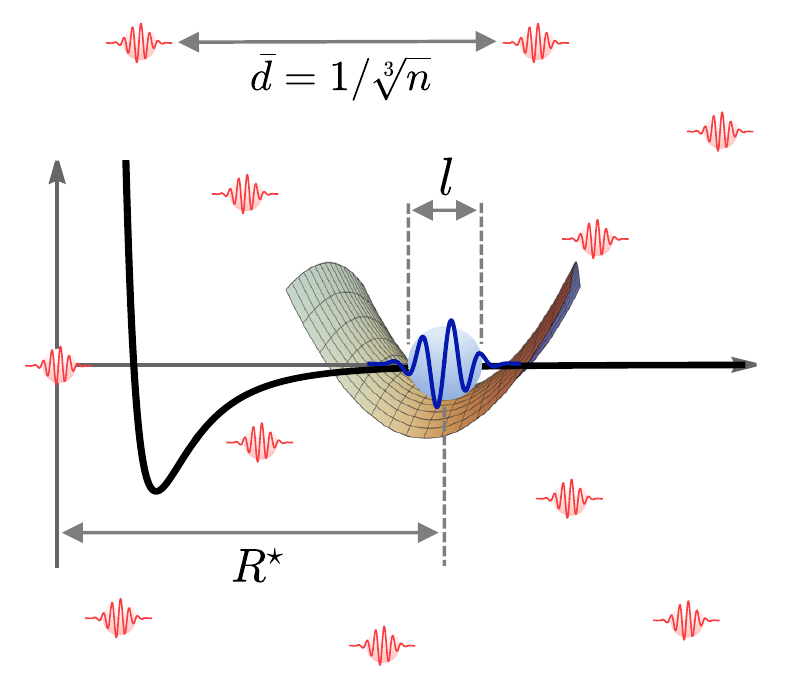}
\caption{\label{fig:general_scheme} (Color online). Schematic view of the open quantum system with the main length scales. The width $l$ of the Paul trap potential (represented by the saddle) corresponds to the size of the ion in the ground state (large blue wave) and is (much) smaller than the two other lengths: the average distance $\bar{d}$ among the gas particles (small red waves), which is defined by atomic density $n$; the characteristic length $R^\star$ of the atom-ion potential (black thick line).}
\end{figure}

In this section we discuss briefly the interaction between an atom and an ion and how we model it for the master equation calculation. 

{\it Polarisation potential.} The interaction between an atom and an ion in free space is described asymptothically by ($r\equiv\vert\mathbf{r}\vert$)
\begin{align}
\label{eq:Vai}
V_{\mathrm{ai}}(\mathbf{r}) = -\frac{C_4}{r^4}
\end{align}
with $C_{4} = \frac{\alpha e^2}{2}\frac{1}{4\pi\epsilon_0}$ (in SI units)~\footnote{We note that for lithium additive and nonadditive interaction coefficients for the three-body Li-Li-Li$^+$ system have been recently computed~\cite{Babb2020}.}, where $\alpha$ is the static polarisability of the atom, $e$ is the elementary electronic charge, and $\epsilon_0$ the vacuum permittivity. Here, $r$ denotes the separation between the atom and the ion. The potential is characterised by the length $R^\star = (2\mu C_4/\hbar^2)^{1/2}$ and energy $E^\star= \hbar^2 / [2\mu (R^\star)^2]$ scales, with $\mu = m M /(m+M)$ the reduced mass, $m$ the atom mass, and $M$ the ion mass. 

{\it Length scales.} In our setting there are several relevant length scales (see Fig.~\ref{fig:general_scheme}). Firstly, the aforementioned $R^\star$, which is typically in the range of hundreds of nanometers and gives, as a rule of thumb, the order of magnitude of the 3D zero-energy s-wave atom-ion scattering length (see also Refs.~\cite{IdziaszekPRA07,IdziaszekNJP11}). For instance, for the atom-ion pair $^7$Li/$^{174}$Yb$^+$ we have $R^\star \simeq 75.15\,$nm, for $^{23}$Na/$^{174}$Yb$^+$ $R^\star\simeq 129.85\,$nm, and for $^{87}$Rb/$^{174}$Yb$^+$ $R^\star\simeq 307.23\,$nm. Secondly, the atom-atom (background) scattering length $a_{\mathrm{aa}}^s$ which is typically on the order of a few nanometers. Thirdly, for a condensate, we have the healing length, which is defined as $\xi = (8\pi n a_{\mathrm{aa}}^s)^{-1/2}$ with $n$ being the gas density. For a typical gas density $n = 10^{14}/\mathrm{cm}^3$ we have, e.g., $\xi \simeq 273.61\,$nm for $^{87}$Rb, and $\xi \simeq 380.38\,$nm for $^{23}$Na. Instead, for a Fermi gas, the inverse of the Fermi wave vector $\lambda_F = 2 \pi / k_F = (3 \pi^2 n)^{-1/3}$ gives another length scale. For $n = 10^{14}/\mathrm{cm}^3$, we have $\lambda_F \simeq 437.56$ nm. Fourthly, the mean path length $\bar{d}$ that at the aforementioned typical gas density is about $215.44\,$nm. Hence, all lengths are comparable and therefore  there is no separation of length scales as in the neutral counterpart. As a consequence, non-universal behaviour in the ion statical and dynamical properties is expected (see, for instance, Refs.~\cite{GooldPRA10,SchurerPRA14,SchurerNJP15} for a static ion analysis). Thus, either very low atomic densities are considered, where a universal behaviour is expected, or else the long-range tail of the atom-ion interaction strongly matters.\\
Finally, the last relevant length for a trapped ion system is the ion trap length $l$, which corresponds to the ion ground state size. This length is about a few tens of nms for $^{174}$Yb$^+$, as we discuss in Sec.~\ref{sec:LDA-PL_approx}. While $l$ is rather small compared to $R^\star$ for heavy atoms, for lithium it is roughly half of the corresponding $R^\star$. This means that scattering of an atom and an ion should not be analysed in free space, as we do in the next paragraph, as the confinement affects the atom-ion collision as for neutrals in waveguides. Here, however, we neglect effects like confinement-induced resonances as a dedicated study of such a phenomenology is required.

{\it Quantum regime condition.} It is important to note that the energy $E^\star$ sets the onset of s-wave atom-ion collisions, namely for energies smaller than $E^\star$ the quantum regime can be attained~\footnote{We note that this is a necessary condition, but not yet a sufficient one, as trap-shaped induced resonances can still occur.}. Indeed, $E^\star$ corresponds to the height of the centrifugal barrier for the $\ell=1$ partial wave from threshold (see, e.g., Fig.~1 of Ref.~\cite{HarterCP14}). Assuming that the kinetic energy of the atom is negligible, since it is ultracold compared to the ion, the collisional energy in the relative atom-ion coordinate frame is given by~\cite{SaitoPRA17,Furst2018}
\begin{align}
\label{eq:Ecoll}
E_{\mathrm{coll}} = k_{\mathrm{B}} \mathcal{T}_{\mathrm{coll}} \simeq \frac{\mu}{M}E_{\mathrm{kin}}
\end{align}
with $E_{\mathrm{kin}} = k_{\mathrm{B}}\mathcal{T}_{\mathrm{kin}}$ the ion's average kinetic energy in the laboratory frame of reference and $k_{\mathrm{B}}$ being the Boltzmann constant~\footnote{The relation~(\ref{eq:Ecoll}) is obtained easily as follows. Since what matters is the relative atom-ion kinetic energy, the collision energy is thus given by $E_{\mathrm{coll}} = \mu \upsilon_{\mathrm{rel}}^2 /2 = \mu (\boldsymbol{\upsilon}_{\mathrm{a}} - \boldsymbol{\upsilon}_{\mathrm{i}})^2 /2 = \mu/2 (\upsilon_{\mathrm{i}}^2+\upsilon_{\mathrm{i}}^2 - 2 \boldsymbol{\upsilon}_{\mathrm{a}}\cdot \boldsymbol{\upsilon}_{\mathrm{i}})$. Here, $\upsilon_{\mathrm{i}}\equiv |\boldsymbol{\upsilon}_{\mathrm{i}}|$ ($\upsilon_{\mathrm{a}}\equiv |\boldsymbol{\upsilon}_{\mathrm{a}}|$) denotes the magnitude of the ion (atom) velocity. Now, if $\upsilon_{\mathrm{i}}\gg\upsilon_{\mathrm{a}}$, that is, the atoms are very slow (i.e. ultracold) compared to the ion micromotion, then $E_{\mathrm{coll}} \simeq \mu \upsilon_{\mathrm{i}}^2 /2 = \mu E_{\mathrm{kin}}/M$ with $E_{\mathrm{kin}} = M \upsilon_{\mathrm{i}}^2 /2$.}. 
Hence, in order to enter the quantum regime of s-wave collisions, the ion's kinetic energy must be smaller than $E^\star$, that is,
\begin{align}
\label{eq:Es}
E_{\mathrm{kin}} \ll E_{\mathrm{s}} = \frac{M}{\mu} E^\star = \left(1+\frac{M}{m}\right) E^\star.
\end{align}
In case of a light atom and a heavy ion we have $\mu\simeq m$ so that $M/\mu\gg 1$, and thus a significant gain in the upper limit for s-wave collisions is obtained. For example, for $^7$Li/$^{174}$Yb$^+$ we have $E_{\mathrm{s}}/k_{\mathrm{B}}\simeq 164.26\,\mu$K and for $^{23}$Na/$^{174}$Yb$^+$ we find $E_{\mathrm{s}}/k_{\mathrm{B}}\simeq 6.07\,\mu$K. 
This shows that there is a rather broad range of temperatures and that these are at least an order of magnitude smaller than those of ultracold neutral collisions (
on the mK scale). 



{\it Regularized potential.} Given the fact that the potential~(\ref{eq:Vai}) is singular and that later in the master equation we need to compute the Fourier transform of the atom-ion potential, we introduce the regularization~\cite{KrychPRA15}
\begin{align}
\label{eq:Vaireg}
V_{\mathrm{ai}}^{r}(\mathbf{r}) = - C_4 \frac{r^2 - c^2}{r^2 + c^2} \frac{1}{(b^2 + r^2)^2}.
\end{align}
Here, $b$ and $c$ are tuneable parameters that have units of a length and control the energy spectrum of the potential as well as the atom-ion scattering length. The Fourier transform of~(\ref{eq:Vaireg}) is linked to the scattering amplitude in the first-order Born approximation, which is defined as
\begin{align}
\label{eq:scattamp}
f(q)=-\frac{\mu}{2\pi\hbar^2}\int_{\mathbb{R}^3}\mathrm{d}\mathbf{r}\,e^{i\mathbf{q}\cdot\mathbf{r}}V_{\mathrm{ai}}^{r}(\mathbf{r}).
\end{align}
By using spherical coordinates and by integrating out the angular part, we obtain 
\begin{align}
\label{eq:fq}
f(q)&=-\frac{2\mu}{q \hbar^2}\int_{\mathbb{R}^+}\mathrm{d}r\,r \sin(q r)V_{\mathrm{ai}}^{r}(r) \nonumber\\
\phantom{=} & = \frac{c^2 \pi (R^\star)^2}{(b^2 - c^2)^2 q}\left\{
e^{-b q}\left[ 1 + \frac{(b^4 - c^4)q}{4 b c^2}\right] - e^{-c q}
\right\},
\end{align}
where we used the fact that $(R^\star)^2 = 2\mu C_4 / \hbar^2$. The determination of $b$ and $c$ is discussed in appendix~\ref{sec:bc_reg}. 


\section{Impurity master equation}
\label{sec:sPb}

In this section we provide an open system description of an impurity in a quantum gas of either bosons or fermions by following the approach of Ref.~\cite{CarmichaelQOBook}, where the impurity is described in first quantisation, whereas the quantum bath in the second one. We focus mainly on the bosonic case, for which we apply Bogolyubov theory, but we consider also the quadratic terms of the bosonic field operators, which result in an extended Fr\"ohlich interaction Hamiltonian. The inclusion of this interaction has been proven to be crucial in the description of the many-body response function of Rydberg~\cite{SchmidtPRA18} and Bose polarons in vicinity of a Feshbach resonance~\cite{ShchadilovaPRL16}. The fermionic case is considered only for a normal gas, i.e. not superfluid BCS theory, and it is obtained as a special case of the master equation for a bosonic bath for gas temperatures above the critical temperature of condensation. We note that in the literature a master equation treatment of an impurity in a degenerate Bose gas has already been  undertaken~\cite{DaleyPRA04,LenaPRA20,KrychPRA15,MitchisonPRA16}, but (i) only the (linear) Fr\"ohlich interaction has been considered and (ii) the Lamb-shift have been not taken into account. Moreover and specifically for the ionic impurity, the fermionic bath has been not investigated in Ref.~\cite{KrychPRA15}.


\subsection{System plus bath Hamiltonian}
\label{sec:splusr}

The total Hamiltonian of the system, the atomic impurity in interaction with a bosonic bath, is given by $\hat H = \hat H_S + \hat H_B + \hat H_{BS}$, where $\hat H_S$ is the impurity time-dependent Hamiltonian~\footnote{For the specific case of the trapped ion, $\hat H_S\equiv \hat H_I^{trap}$ with $\hat H_I^{trap}$ given in Appendix~\ref{sec:ionHquant}},
\begin{align}
\label{eq:Hr}
\hat H_B = \int_{\mathbb{R}^3}\mathrm{d}\mathbf{r}_b\,\hat\Psi_b^\dag(\vek{r}_b)\left[
\frac{\hat{\vek{p}}_b^2}{2 m} + \frac{g}{2} \hat\Psi_b^\dag(\vek{r}_b) \hat\Psi_b(\vek{r}_b)
\right]\hat\Psi_b(\vek{r}_b),
\end{align}
and 
\begin{align}
\label{eq:Hrs}
\hat H_{BS} = \int_{\mathbb{R}^3}\mathrm{d}\mathbf{r}_b\,\hat\Psi_b^\dag(\vek{r}_b)V_{\mathrm{ib}}(\vek{r}_b - \hat{\vek{r}}) \hat\Psi_b(\vek{r}_b).
\end{align}
Here, $V_{\mathrm{ib}}$ denotes the interaction between the impurity and a particle of the bath. Besides, we assume that the bath is confined in a box of length $L$~\footnote{$L$ is assumed to be much larger than other length scale involved in the system description.} and that the interaction between the bosons is given by a contact potential with coupling constant $g=4\pi\hbar^2 a_{\mathrm{bb}}^s / m$ and $a_{\mathrm{bb}}^s$ the 3D s-wave atom-atom scattering length, while for the fermions we assume a spin-polarised gas with no intra-particle interaction. Let us note that at this stage the models describing the bosonic and fermionic baths are different, i.e. the former is interacting while the latter is not. Nonetheless, and specifically for the ion in a Paul trap, it turns out that the interaction among bosons is much smaller than other energies involved, so that we can safely neglect it (see Sec.~\ref{sec:LDA-PL_approx} for details). Hence, the differences we observe in the ionic dynamics in the two baths are owed to their quantum statistics and not to the interaction. For the sake of completeness, however, we keep here the derivation of the master equation as general as possible. 

The bosonic quantum field is expanded as 
\begin{align}
\hat\Psi_b(\vek{r}_b) =\sqrt{n_0} + \delta\hat\Psi_b(\vek{r}_b)
\end{align}
where $n_0=N_0/L^3$ is the density of the condensate, i.e. the zero momentum component, while $N_0$ is the atom number. Fluctuations around the condensate mode are described in terms of Bogolyubov modes
\begin{align}
\delta\hat\Psi_b(\vek{r}_b) = L^{-3/2}\sum_{\vek{q}} u_{\vek{q}} \hat b_{\vek{q}} e^{i\vek{q}\cdot\vek{r}_b} + v_{\vek{q}} \hat b_{\vek{q}}^\dag e^{-i\vek{q}\cdot\vek{r}_b}
\end{align}
where $[\hat b_{\vek{q}},\hat b_{\vek{q}^\prime}^\dag] = \delta_{\vek{q},\vek{q}^\prime}$. Using this expression in Eq.~(\ref{eq:Hr}), we arrive at
\begin{align}
\hat H_B = E_0 + \sum_{\vek{q}} \hbar\omega_{\vek{q}} \hat b_{\vek{q}}^\dag \hat b_{\vek{q}}.
\end{align}
Here, $E_0 = g N_0^2 / (2 L^3)$ is the ground state energy of the condensate and $\mu_{\mathrm{G}} = \partial_{N_0} E_0 = g n_0$ is the chemical potential at zero temperature. The corresponding dispersion relation is given by~\cite{LifshitzVol10}
\begin{align}
\label{eq:Bogo-disp}
\varepsilon(\vek{q}) = \hbar\omega_{\vek{q}} = \sqrt{
\left(\frac{\hbar^2 q^2}{2 m}\right)^2 + (\hbar c_s q)^2
}
\end{align}
with $c_s = (gn_0/m)^{1/2}$ being the speed of sound, and the Bogolyubov amplitudes $u_{\vek{q}}$ and $v_{\vek{q}} $ are given in Ref.~\cite{LifshitzVol10}.
Given this, the atomic density operator is 
\begin{align}
\hat\Psi_b^\dag(\vek{r}_b) \hat\Psi_b(\vek{r}_b) = n_0 + \Delta\hat n(\vek{r}_b).
\end{align}
The first term provides a constant energy term in Eq.~(\ref{eq:Hrs}) for a homogeneous gas, as we consider here, and it can be neglected, since it shifts merely the energy minimum. The second term is given by
\begin{align}
\Delta\hat n(\vek{r}_b) & = \hat\Psi_b^\dag(\vek{r}_b) \hat\Psi_b(\vek{r}_b) - n_0 \nonumber\\
\phantom{=} & = \sqrt{n_0} [\delta\hat\Psi_b(\vek{r}_b) + \delta\hat\Psi_b^\dag(\vek{r}_b)] + \delta\hat\Psi_b^\dag(\vek{r}_b) \delta\hat\Psi_b(\vek{r}_b) \nonumber\\
\phantom{=} & = \delta\hat n (\vek{r}_b) + \delta^2\hat n (\vek{r}_b)
\end{align}
with $\delta\hat n (\vek{r}_b) = \sqrt{n_0} [\delta\hat\Psi_b(\vek{r}_b) + \delta\hat\Psi_b^\dag(\vek{r}_b)]$. Hence, we have 
\begin{align}
\delta\hat n (\vek{r}_b) = \sqrt{\frac{n_0}{L^3}} \sum_{\vek{q}} (u_{\vek{q}} + v^*_{\vek{q}}) \hat b_{\vek{q}} e^{i\vek{q}\cdot\vek{r}_b} + (u^*_{\vek{q}} + v_{\vek{q}}) \hat b_{\vek{q}}^\dag e^{-i\vek{q}\cdot\vek{r}_b},
\end{align}
and
\begin{align}
&\delta^2\hat n (\vek{r}_b) = L^{-3}  \sum_{\vek{q},\vek{q}^\prime}
\left[
u^*_{\vek{q}} u_{\vek{q}^\prime} e^{-i(\vek{q} - \vek{q}^\prime)\cdot\vek{r}_b} \hat b_{\vek{q}}^\dag \hat b_{\vek{q}^\prime}\right.\nonumber\\
&\left.
+ u^*_{\vek{q}} v_{\vek{q}^\prime} e^{-i(\vek{q} + \vek{q}^\prime)\cdot\vek{r}_b} \hat b_{\vek{q}}^\dag \hat b_{\vek{q}^\prime}^\dag
+ v^*_{\vek{q}} u_{\vek{q}^\prime} e^{i(\vek{q} + \vek{q}^\prime)\cdot\vek{r}_b} \hat b_{\vek{q}} \hat b_{\vek{q}^\prime}\right.\nonumber\\
&\left.
+ v^*_{\vek{q}} v_{\vek{q}^\prime} e^{i(\vek{q} - \vek{q}^\prime)\cdot\vek{r}_b} \hat b_{\vek{q}} \hat b_{\vek{q}^\prime}^\dag
\right].
\end{align}
Given this, the system-bath Hamiltonian becomes 
\begin{align}
\label{eq:HRS-Bogo}
& \hat H_{BS} = \int_{\mathbb{R}^3}\mathrm{d}\mathbf{r}_b\,V_{\mathrm{ib}}(\vek{r}_b - \hat{\vek{r}}) \Delta\hat n(\vek{r}_b) 
\nonumber\\
& = \hbar \sum_{\vek{q}} \left(\hat S_{\vek{q}} \hat\Gamma_{\vek{q}} + \hat S^{\dag}_{\vek{q}} \hat\Gamma_{\vek{q}}^\dag\right)
+ \hbar \sum_{\vek{q},\vek{q}^\prime} \left(\hat S_{\vek{q},\vek{q}^\prime}^{(u,u^\prime)} \hat\Gamma_{\vek{q}}^\dag \hat\Gamma_{\vek{q}^\prime}
\right. 
\nonumber\\
& \left. + \hat S_{\vek{q},\vek{q}^\prime}^{(u,v^\prime)} \hat\Gamma_{\vek{q}}^\dag \hat\Gamma_{\vek{q}^\prime}^\dag 
+ \hat S_{\vek{q},\vek{q}^\prime}^{(v,u^\prime)} \hat\Gamma_{\vek{q}} \hat\Gamma_{\vek{q}^\prime} 
+ \hat S_{\vek{q},\vek{q}^\prime}^{(v,v^\prime)} \hat\Gamma_{\vek{q}} \hat\Gamma_{\vek{q}^\prime}^\dag\right)
\nonumber\\
&= \hat H_{BS}^{(1)} + \hat H_{BS}^{(2)},
\end{align}
where we used the notation of Ref.~\cite{CarmichaelQOBook}: $\hat\Gamma_{\vek{q}} \equiv \hat b_{\vek{q}}$, and 
\begin{align}
\hat S_{\vek{q}} & = \frac{\sqrt{n L^3}}{\hbar} (u_{\vek{q}} + v^*_{\vek{q}}) e^{i\vek{q}\cdot\hat{\vek{r}}} c_{\vek{q}},\nonumber\\
\hat S_{\vek{q},\vek{q}^\prime}^{(u,u^\prime)} & = \frac{u^*_{\vek{q}} u_{\vek{q}^\prime}}{\hbar} e^{i(\vek{q}^\prime-\vek{q})\cdot\hat{\vek{r}}}c_{\vek{q}^\prime-\vek{q}},\nonumber\\
\hat S_{\vek{q},\vek{q}^\prime}^{(u,v^\prime)} & = \frac{u^*_{\vek{q}} v_{\vek{q}^\prime}}{\hbar} e^{-i(\vek{q}^\prime+\vek{q})\cdot\hat{\vek{r}}}c^*_{\vek{q}^\prime+\vek{q}},\nonumber\\
\hat S_{\vek{q},\vek{q}^\prime}^{(v,u^\prime)} & = \frac{v^*_{\vek{q}} u_{\vek{q}^\prime}}{\hbar} e^{i(\vek{q}^\prime+\vek{q})\cdot\hat{\vek{r}}}c_{\vek{q}^\prime+\vek{q}},\nonumber\\
\hat S_{\vek{q},\vek{q}^\prime}^{(v,v^\prime)} & = \frac{v^*_{\vek{q}} v_{\vek{q}^\prime}}{\hbar} e^{-i(\vek{q}^\prime-\vek{q})\cdot\hat{\vek{r}}}c^*_{\vek{q}^\prime-\vek{q}},\nonumber\\
c_{\vek{q}} & = \frac{1}{L^3} \int_{\mathbb{R}^3}\mathrm{d}\vek{y}\, e^{i\vek{q}\cdot\vek{y}} V_{\mathrm{ib}}(\vek{y}).
\label{eq:S}
\end{align}
In addition, $\hat H_{BS}^{(1)}$ is the first sum over $\vek{q}$ in Eq.~(\ref{eq:HRS-Bogo}) and it denotes the so-called Fr\"ohlich model Hamiltonian in the context of the electron-phonon coupling in solid-state~\cite{Mahan2000}, while $\hat H_{BS}^{(2)}$ refers to the double sum over $\vek{q}$, $\vek{q}^\prime$, which describes the so-called extended Fr\"ohlich model~\cite{RathPRA13,SchmidtPRA18}. Specifically for the ionic impurity, the coefficient $c_{\vek{q}}$ is linked to the scattering amplitude~(\ref{eq:scattamp}) as
\begin{align}
\label{eq:cq}
c_{\vek{q}} = - \frac{2\pi\hbar^2}{\mu L^3} f(q).
\end{align}
Apart from the Bogolyubov approximation, the expression~(\ref{eq:HRS-Bogo}) is exact for a bosonic bath. For a normal Fermi gas, the interaction Hamiltonian reduces to
\begin{align}
\label{eq:HRS-Fermi}
\hat H_{BS} = \hbar \sum_{\vek{q},\vek{q}^\prime} \hat S_{\vek{q},\vek{q}^\prime}\hat\Gamma_{\vek{q}}^\dag \hat\Gamma_{\vek{q}^\prime}
\end{align}
with $\hat S_{\vek{q},\vek{q}^\prime} = e^{i(\vek{q}^\prime-\vek{q})\cdot\hat{\vek{r}}}c_{\vek{q}^\prime-\vek{q}}/\hbar$ and $\hat\Gamma_{\vek{q}}^\dag\equiv\hat c_{\vek{q}}^\dag$ ($\hat\Gamma_{\vek{q}}\equiv\hat c_{\vek{q}}$) being the creation (annihilation) operator of a free fermion of momentum ${\vek{q}}$ with $\{\hat c_{\vek{q}},\hat c_{\vek{q}^\prime}^\dag\} = \delta_{\vek{q},\vek{q}^\prime}$.
 
If the impurity-bath interaction is described by the pseudopotential, as for neutral impurities, we have~\footnote{Here we neglect the differential operator $\partial_r (r\cdot)_{\vert_{r=0}}$, since its application to a plane wave has unity effect.}
\begin{align}
\label{eq:Vib-pseudo}
V_{\mathrm{ib}}(\vek{r}) = g_{\mathrm{ib}} \delta(\vek{r}),
\end{align}
where $g_{\mathrm{ib}} = 2\pi\hbar^2 a_{\mathrm{ib}}^s / \mu$ with $a_{\mathrm{ib}}^s$ being the 3D $s$-wave impurity-bath scattering length. Thus, Eq.~(\ref{eq:cq}) becomes
\begin{align}
c_{\vek{q}} = \frac{2\pi\hbar^2}{\mu L^3} a_{\mathrm{ib}}^s. 
\end{align}
Specifically for the atom-ion case, we see that $f(q) \rightarrow - a_{\mathrm{ib}}^s$. Thus, if we would replace the atom-ion interaction~(\ref{eq:Vaireg}) by Eq.~(\ref{eq:Vib-pseudo}), in all subsequent equations one has simply to replace the scattering amplitude in the first-Born approximation with the scattering length $a_{\mathrm{ib}}^s$. 


\subsection{Validity requirements and chemical potential}
\label{sec:val}

Let us comment on the validity of the Bogolyubov approximation, which implies that both the quantum and thermal depletion must be small~\cite{Castin2001}. 
As we shall discuss in Sec.~\ref{sec:LDA-PL_approx}, we are mostly interested in the high temperature regime, that is, $k_{\mathrm{B}} \mathcal{T}\gg g n_{\mathrm{t}}$, which means that the intra-particle interactions are essentially negligible. Here, $n_{\mathrm{t}}$ is the total gas density that in the zero temperature limit corresponds to the condensate density $n_0$. In this regime, the Bogolyubov approximation is valid only if $n_{\mathrm{t}} \lambda_{\mathrm{dB}}(\mathcal{T})^{3}\gg \zeta(3/2)$. Here, $\lambda_{\mathrm{dB}}(\mathcal{T}) = [2\pi\hbar^2/(m k_{\mathrm{B}} \mathcal{T})]^{1/2}$ denotes the thermal de Broglie wavelength, $\zeta(x)$ is the zeta-Riemann function. 
The last inequality can be also rewritten as $\mathcal{T} \ll \mathcal{T}_{\mathrm{c}}^0 = \frac{2\pi\hbar^2 n_{\mathrm{t}}^{2/3}}{m k_{\mathrm{B}} [\zeta(3/2)]^{2/3}}$, where $\mathcal{T}_{\mathrm{c}}^0$ is the critical temperature of condensation of a non-interacting and untrapped Bose gas. 
Thus, in order to perform the Bogolyubov approximation 
all conditions have to be fulfilled simultaneously, namely the non-condensed fraction $n_{\mathrm{t}} (\mathcal{T}/\mathcal{T}_{\mathrm{c}}^0)^{3/2}\ll 1$. Here, $n_{\mathrm{t}} = n_0 + n_{\mathrm{n}}$ is the total density of the gas with $n_{\mathrm{n}}$ being the normal (i.e. non-condensed) component, from which we retrieve the condensate density as
\begin{align}
\label{eq:n0}
n_0 = n_{\mathrm{t}}  - n_{\mathrm{n}} = n_{\mathrm{t}} \left[1 - 
(\mathcal{T}/\mathcal{T}_{\mathrm{c}}^0)^{3/2}
\right].
\end{align}
In table~\ref{tb:Tc} we provide some values of the critical temperature of condensation at typical quantum gas densities. 

\begin{table}
\hrule
\vspace{0.15cm}
\begin{tabular}{cccccccccccccccccccccccccccccccccccccccccccccccccccccccccccccccc}
\vspace{0.1cm}
Boson 
	&  & & & & & $n_{\mathrm{t}} =10^{12}\,\mathrm{cm}^{-3}$
	& &  & & & & $10^{13}\,\mathrm{cm}^{-3}$
	& &  & & & & $10^{14}\,\mathrm{cm}^{-3}$
\\[1ex]
$^{7}$Li &  & & & & & 229 & &  & & & & 1063 & &  & & & & 4934  \\ 
$^{23}$Na &  & & & & & 70 & &  & & & & 324 & &  & & & & 1506  \\ 
$^{87}$Rb &  & & & & & 18 & &  & & & & 86 & &  & & & & 398 \\ 
\end{tabular}
\vspace{0.1cm}
\hrule
\vspace{0.15cm}
\begin{tabular}{cccccccccccccccccccccccccccccccccccccccccccccccccccccccccccccccc}
\vspace{0.1cm}
Fermion 
	&  & & & & & $n_{\mathrm{t}} =10^{12}\,\mathrm{cm}^{-3}$
	& &  & & & & $10^{13}\,\mathrm{cm}^{-3}$
	& &  & & & & $10^{14}\,\mathrm{cm}^{-3}$
\\[1ex]
$^{6}$Li &  & & & & & 613 & &  & & & & 2844 & &  & & & & 13198  \\ 
$^{40}$K &  & & & & & 92 & &  & & & & 428 & &  & & & & 1987  \\ 
\end{tabular}
\vspace{0.1cm}
\hrule
\caption{\label{tb:Tc} 
Top: Critical temperature for condensation $\mathcal{T}_{\mathrm{c}}^0$ of a non-interacting gas for three bosonic species and densities. 
Bottom: Fermi temperature $\mathcal{T}_{F}$ for two fermionic species and three densities. Temperature values are given in units of nK. 
}
\end{table}

Afterwards, it will be important to compute the chemical potential for a non-interacting and homogenous Bose and Fermi gas at temperature $\mathcal{T}$. To this end, let us remind that the chemical potential for the bosons reads~\cite{Huang1987}
\begin{align}
\label{eq:muB}
\mu_{\mathrm{G}} = \left\{
\begin{array}{l}
\text{0, if $\mathcal{T}\le \mathcal{T}_{\mathrm{c}}^0$},\\
\\
\text{root of $n_{\mathrm{t}}\lambda_{\mathrm{dB}}^3=g_{3/2}(z)$, if $\mathcal{T} > \mathcal{T}_{\mathrm{c}}^0$}. 
\end{array}
\right.
\end{align}
Here, $z=e^{\frac{\mu_{\mathrm{G}}}{k_{\mathrm{B}} \mathcal{T}}}$ is the so-called fugacity and $g_{3/2}(z) = \sum_{l=1}^\infty z^l l^{-3/2}$. We note that the chemical potential for $\mathcal{T} > \mathcal{T}_{\mathrm{c}}^0$ is negative or else $g_{3/2}(z)$ does not converge. 

For the fermions, the chemical potential is obtained by solving numerically the equation of state~\cite{Huang1987}
\begin{align}
\label{eq:muF}
n_{\mathrm{t}}\lambda_{\mathrm{dB}}^3=f_{3/2}(z),
\end{align}
where $f_{3/2}(z) = \sum_{l=1}^\infty (-1)^{l+1} z^l l^{-3/2}$. At $\mathcal{T} = 0$, the chemical potential corresponds to the Fermi energy $E_F$:
\begin{align}
\mu_{\mathrm{G}} \equiv E_F = \frac{\hbar^2}{2 m} \left(
6\pi^2 n_{\mathrm{t}}
\right)^{2/3}.
\end{align}
We note that for $\mathcal{T} > \mathcal{T}_{F}$, $\mu_{\mathrm{G}}$ is negative, similarly to the bosonic case, where $\mathcal{T}_{F} = E_F/k_{\mathrm{B}}$ is the Fermi temperature. 
In the high temperature limit $\mathcal{T}\gg \mathcal{T}_{\mathrm{c}}^0,\,\mathcal{T}_{F}$, the chemical potential of both the bosons and the fermions are well described by that of the Boltzmann gas
\begin{align}
\mu_{\mathrm{G}} = k_{\mathrm{B}} \mathcal{T} \ln( n_{\mathrm{t}}\lambda_{\mathrm{dB}}^3 ).
\end{align}


\subsection{Markovian master equation}
\label{sec:vNe}

In this section we describe the relevant steps of the derivation of the master equation for the bosons, while for the fermionic bath we simply provide the final result, since the derivation is analogous. 

We start from the full system-bath density matrix $\hat\chi(t)$, which obeys the von Neumann equation
\begin{align}
\frac{d}{d t}\hat\chi(t) = -\frac{i}{\hbar}[\hat H , \hat\chi].
\end{align}
Writing the density operator in the interaction picture as
\begin{align}
\label{eq:intpict}
\tilde{\chi}(t) = \hat U^\dag(0,t) e^{i\hat H_B t/\hbar} \hat\chi(t) e^{-i\hat H_B t/\hbar} \hat U(0,t),
\end{align}
where 
\begin{align}
\hat U(t_1,t_2) = \mathsf{T} \exp\left[
-\frac{i}{\hbar}\int_{t_1}^{t_2}\mathrm{d}t\,\hat H_S(t)
\right]
\end{align}
with $\mathsf{T}$ the time-ordered evolution operator, we have
\begin{align}
\label{eq:dtchi}
\frac{d}{d t}\tilde\chi(t) = -\frac{i}{\hbar}[\tilde H_{BS}(t) , \tilde\chi(t)].
\end{align}
Here, $\tilde H_{BS}$ is the interaction Hamiltonian in the interaction picture, which is defined similarly to Eq.~(\ref{eq:intpict}). 
The formal solution of Eq.~(\ref{eq:dtchi}) is 
\begin{align}
\tilde\chi(t) = \tilde\chi(0) -\frac{i}{\hbar}\int_0^t\mathrm{d}t^\prime\,[\tilde H_{BS}(t^\prime) , \tilde\chi(t^\prime)]
\end{align}
and substituting it into the commutator~(\ref{eq:dtchi}) we obtain
\begin{align}
\label{eq:exactCHI}
\frac{d}{d t}\tilde\chi(t) & = -\frac{i}{\hbar}[\tilde H_{BS}(t) , \tilde\chi(0)] \nonumber\\
\phantom{=}& -\frac{1}{\hbar^2} \int_0^t\mathrm{d}t^\prime\,[\tilde H_{BS}(t) , [\tilde H_{BS}(t^\prime) , \tilde\chi(t^\prime)]].
\end{align}
Thus, we assume that initially, at $t=0$, the system and the bath are uncorrelated, namely $\hat\chi(0) = \tilde\chi(0) = \hat \rho(0)\otimes \hat B_0$, where $\hat B_0$ is the initial bath density matrix. This is a reasonable assumption if the impurity and the bath are initially well-separated such that no interaction occurs. 
By tracing over the bath degrees of freedom in Eq.~(\ref{eq:exactCHI}) we arrive to the equation~\footnote{Here, $\mathrm{Tr}_R[\tilde\chi(t)] = \tilde\rho(t)$ and we eliminated the term Tr$_B\{[\tilde H_{BS}(t) , \tilde\chi(0)]\}$ by assuming Tr$_B\{\tilde H_{BS}(t) \hat B_0\} =0$ (see Ref.~\cite{CarmichaelQOBook} for details).}
\begin{align}
\label{eq:ME1}
\frac{d}{d t}\tilde\rho(t) & = -\frac{1}{\hbar^2} \int_0^t\mathrm{d}t^\prime\,\mathrm{Tr}_B\{[\tilde H_{BS}(t) , [\tilde H_{BS}(t^\prime) , \tilde\chi(t^\prime)]]\}.
\end{align}
The next step consists in performing the so-called Born approximation, namely we assume that the impurity-bath coupling is weak and that the bath is so large that $\tilde\chi(t^\prime) \simeq \tilde \rho(t^\prime)\otimes \hat B_0$ $\forall t^\prime$ holds. Thus, Eq.~(\ref{eq:ME1}) becomes 
\begin{align}
\label{eq:ME2}
\frac{d}{d t}\tilde\rho(t) & = -\frac{1}{\hbar^2} \int_0^t\mathrm{d}t^\prime\,\mathrm{Tr}_B\{[\tilde H_{BS}(t) , [\tilde H_{BS}(t^\prime) , \tilde \rho(t^\prime)\otimes \hat B_0]]\}.
\end{align}
In order to further simplify this equation, we make the Markov approximation, namely we replace $\tilde \rho(t^\prime)$ by $\tilde \rho(t)$ in order to obtain a time-local master equation 
\begin{align}
\label{eq:ME3}
\frac{d}{d t}\tilde\rho(t) & = -\frac{1}{\hbar^2} \int_0^t\mathrm{d}t^\prime\,\mathrm{Tr}_B\{[\tilde H_{BS}(t) , [\tilde H_{BS}(t^\prime) , \tilde \rho(t)\otimes \hat B_0]]\}.
\end{align}
This equation is known in the literature as the Redfield equation~\cite{Breuer2002}. The Hamiltonian $\tilde H_{BS}(t)$ keeps the original structure of the Schr\"odinger picture, but with time-dependent system and bath operators:
\begin{align}
\hat S_{\vek{q}}(t) &= \hat U^\dag(0,t) \hat S_{\vek{q}} \hat U(0,t)\nonumber\\
\hat\Gamma_{\vek{q}}(t) & = e^{i\hat H_B t/\hbar} \hat\Gamma_{\vek{q}} e^{-i\hat H_B t/\hbar} = e^{-\frac{i}{\hbar}\varepsilon({\vek{q}}) t} \hat\Gamma_{\vek{q}}\nonumber\\
\hat S_{\vek{q},\vek{q}^\prime}^{(u,u^\prime)}(t) & = \hat U^\dag(0,t)\hat S_{\vek{q},\vek{q}^\prime}^{(u,u^\prime)}\hat U(0,t).
\end{align}
Next, we need to perform the partial trace over the bath degrees of freedom, namely we need to assess
\begin{align}
\label{eq:ME3bis}
&\mathrm{Tr}_B\{[\tilde H_{BS}(t) , [\tilde H_{BS}(t^\prime) , \tilde \rho(t)\otimes \hat B_0]]\} = 
\nonumber\\
&\mathrm{Tr}_B\{[\tilde H_{BS}^{(1)}(t) , [\tilde H_{BS}^{(1)}(t^\prime) , \tilde \rho(t)\otimes \hat B_0]]\} 
\nonumber\\ 
&+\mathrm{Tr}_B\{[\tilde H_{BS}^{(1)}(t) , [\tilde H_{BS}^{(2)}(t^\prime) , \tilde \rho(t)\otimes \hat B_0]]\} 
\nonumber\\ 
&+\mathrm{Tr}_B\{[\tilde H_{BS}^{(2)}(t) , [\tilde H_{BS}^{(1)}(t^\prime) , \tilde \rho(t)\otimes \hat B_0]]\} 
\nonumber\\ 
&+\mathrm{Tr}_B\{[\tilde H_{BS}^{(2)}(t) , [\tilde H_{BS}^{(2)}(t^\prime) , \tilde \rho(t)\otimes \hat B_0]]\}.
\end{align}
Thus, we consider the bath's thermal density matrix 
\begin{align}
\!\!\!\hat B_0 = \frac{e^{-\beta (\hat H_B -\mu_{\mathrm{G}}\hat N)}}{\mathcal{Z}},
\qquad
\mathcal{Z} = \mathrm{Tr}_B\{e^{-\beta(\hat H_B-\mu_{\mathrm{G}}\hat N)}\},
\end{align}
where $\hat N$ is the bath number operator. The mixed terms in the third and fourth line of Eq.~(\ref{eq:ME3bis}) are zero, since they contain an odd number of bath operators. Thus, only the terms of the second (i.e. with $\tilde H_{BS}^{(1)}(t)$ only) and the last line (i.e. with $\tilde H_{BS}^{(2)}(t)$ only) of Eq.~(\ref{eq:ME3bis}) remain. 

First, we consider the thermal average of the double commutator involving $\tilde H_{BS}^{(1)}(t)$, which includes the averages of two bath operators only, for example, $\langle \tilde\Gamma_{\vek{q}}(t)\tilde\Gamma_{\vek{q}^\prime}(t^\prime)\rangle_{B_0}$= Tr$_B\{\hat B_0\tilde\Gamma_{\vek{q}}(t)\tilde\Gamma_{\vek{q}^\prime}(t^\prime)\}$. On the other hand, $\langle \tilde\Gamma_{\vek{q}}(t)\tilde\Gamma_{\vek{q}^\prime}(t^\prime)\rangle_{B_0} = \langle \tilde\Gamma^\dag_{\vek{q}}(t)\tilde\Gamma^\dag_{\vek{q}^\prime}(t^\prime)\rangle_{B_0} = 0$, while
\begin{align}
\langle \tilde\Gamma_{\vek{q}}(t)\tilde\Gamma^\dag_{\vek{q}^\prime}(t^\prime)\rangle_{B_0} = e^{-\frac{i}{\hbar}\varepsilon({\vek{q}}) (t - t^\prime) } (n_{\vek{q}} + 1) \delta_{\vek{q},\vek{q}^\prime}.
\end{align}
Here, $n_{\vek{q}} = \langle \hat b^\dag_{\vek{q}} \hat b_{\vek{q}}\rangle = [e^{\beta_{\mathcal{T}}(\varepsilon({\vek{q}})-\mu_{\mathrm{G}})}-1]^{-1}$ is the Bose-Einstein occupation number and $\beta_{\mathcal{T}} = 1/(k_{\mathrm{B}} \mathcal{T})$. 

The average of the double commutator with only $\tilde H_{RS}^{(2)}(t)$ has terms that vanish when the number of raising and lowering bath operators is not the same, while the non-zero contributions are given in appendix~\ref{sec:appendix:thermalave}. Putting all together into Eq.~(\ref{eq:ME3}) and performing the change of variable $\tau = t- t^\prime$, and finally transforming back to the Schr\"odinger picture, we arrive at the following final master equation
\begin{widetext}
\begin{align}
\label{eq:ME5}
&\frac{d}{d t}\hat\rho(t) = -\frac{i}{\hbar} [\hat H_S, \hat\rho] - \sum_{\vek{q}}\int_0^t\mathrm{d}\tau\,\Omega^2_{\vek{q}}\left\{
(n_{\vek{q}} + 1) [\hat Z_{\vek{q}},\hat W_{\vek{q}}(t,\tau)\hat\rho(t)] + n_{\vek{q}} [\hat\rho(t) \hat W_{\vek{q}}(t,\tau), \hat Z_{\vek{q}}] + \text{H. c.}
\right\}
\nonumber\\
&- \sum_{\vek{q},\vek{q}^\prime}\int_0^t\mathrm{d}\tau\,
\left\{
n_{\vek{q}} (n_{\vek{q}^\prime} + 1) [\hat Z_{\vek{q}^\prime-\vek{q}},\hat W_{\vek{q}^\prime-\vek{q}}(t,\tau)
\hat\rho(t)]\Omega_{u,u^\prime}^{u,u^\prime}(\vek{q}^\prime-\vek{q})
+2 n_{\vek{q}} (n_{\vek{q}^\prime} + 1) [\hat Z_{\vek{q}^\prime-\vek{q}},\hat W_{\vek{q}^\prime-\vek{q}}(t,\tau)
\hat\rho(t)]\Omega_{u,u^\prime}^{v,v^\prime}(\vek{q}^\prime-\vek{q})\right.
\nonumber\\
&
\left(
(1 + n_{\vek{q}} + n_{\vek{q}^\prime} + n_{\vek{q}} n_{\vek{q}^\prime})
[\hat Z_{\vek{q}^\prime+\vek{q}},\hat W_{\vek{q}^\prime+\vek{q}}(t,\tau)\hat\rho(t)]
+ n_{\vek{q}} n_{\vek{q}^\prime}
[\hat Z_{\vek{q}^\prime+\vek{q}},\hat\rho(t)\hat W_{\vek{q}^\prime+\vek{q}}(t,\tau)]
\right)
\left(\Omega_{v,u^\prime}^{u,v^\prime}(\vek{q}^\prime+\vek{q}) + 
\tilde{\Omega}_{v,u^\prime}^{u,v^\prime}(\vek{q}^\prime+\vek{q})
\right)
\nonumber\\
&\left.
n_{\vek{q}^\prime} (n_{\vek{q}} + 1) [\hat Z_{\vek{q}-\vek{q}^\prime},\hat W_{\vek{q}-\vek{q}^\prime}(t,\tau)
\hat\rho(t)]\Omega_{v,v^\prime}^{v,v^\prime}(\vek{q}-\vek{q}^\prime) + \text{H. c.}
\right\}.
\end{align}
\end{widetext}
Here, we have defined the operators
\begin{align}
\label{eq:ZW}
&\hat Z_{\vek{q}} = e^{i\vek{q}\cdot\hat{\vek{r}}} \qquad \hat W_{\vek{q}}(t,\tau) = 
e^{-\frac{i}{\hbar}\varepsilon(\vek{q})\tau} e^{-i\vek{q}\cdot\hat{\vek{r}}(t,\tau)},
\nonumber\\
&\hat W_{\vek{q}^\prime - \vek{q}}(t,\tau) = 
e^{-\frac{i}{\hbar}(\varepsilon(\vek{q}^\prime)-\varepsilon(\vek{q}))\tau} e^{-i(\vek{q}^\prime-\vek{q})\cdot\hat{\vek{r}}(t,\tau)},
\nonumber\\
&\hat W_{\vek{q}^\prime + \vek{q}}(t,\tau) = 
e^{-\frac{i}{\hbar}(\varepsilon(\vek{q}^\prime)+\varepsilon(\vek{q}))\tau} e^{-i(\vek{q}^\prime+\vek{q})\cdot\hat{\vek{r}}(t,\tau)},
\end{align}
and the coefficients 
\begin{align*}
\Omega^2_{\vek{q}} & = \frac{\vert u_{\vek{q}} + v^*_{\vek{q}}\vert^2}{\hbar^2} \vert c_{\vek{q}} \vert^2  n_0 L^3,
\nonumber\\
\Omega_{u,u^\prime}^{u,u^\prime}(\vek{q}^\prime-\vek{q}) &=
\frac{\vert u_{\vek{q}} \vert^2 \vert u_{\vek{q}^\prime}\vert^2}{\hbar^2} \vert c_{\vek{q}^\prime-\vek{q}} \vert^2,
\nonumber\\
\Omega_{v,v^\prime}^{v,v^\prime}(\vek{q}^\prime-\vek{q}) &=
\frac{\vert v_{\vek{q}} \vert^2 \vert v_{\vek{q}^\prime}\vert^2}{\hbar^2} \vert c_{\vek{q}^\prime-\vek{q}} \vert^2,
\end{align*}
\begin{align}
\label{eq:Omega2}
\Omega_{u,u^\prime}^{v,v^\prime}(\vek{q}^\prime-\vek{q}) &=
\frac{u_{\vek{q}}^*v_{\vek{q}}^*u_{\vek{q}^\prime}v_{\vek{q}^\prime}}{\hbar^2} \vert c_{\vek{q}^\prime-\vek{q}} \vert^2,
\nonumber\\
\Omega_{v,u^\prime}^{u,v^\prime}(\vek{q}^\prime+\vek{q}) &=
\frac{u_{\vek{q}}^*v_{\vek{q}}^*u_{\vek{q}^\prime}v_{\vek{q}^\prime}}{\hbar^2} \vert c_{\vek{q}^\prime+\vek{q}} \vert^2,
\nonumber\\
\tilde{\Omega}_{v,u^\prime}^{u,v^\prime}(\vek{q}^\prime+\vek{q}) &=
\frac{\vert v_{\vek{q}} \vert^2 \vert u_{\vek{q}^\prime}\vert^2}{\hbar^2} \vert c_{\vek{q}^\prime+\vek{q}} \vert^2,
\end{align}
whereas  
\begin{align}
\label{eq:rttau}
\hat{\vek{r}}(t,\tau) = \hat U(0,t) \hat U^\dag(0,t-\tau)\, \hat{\vek{r}}\, \hat U(0,t-\tau)\hat U^\dag(0,t).
\end{align}
This relation describes the impurity position evolution in absence of the gas. 
Equation~(\ref{eq:ME5}) is not yet a Markovian master equation, even though the time development of the system density matrix relies only on the density matrix at time $t$. Indeed, the impurity density matrix in Eq.~(\ref{eq:ME5}) still depends on the specific choice for the system preparation at $t=0$ via the impurity's trajectory in $W_{\vek{q}-\vek{q}^\prime}(t,\tau)$ or, in a more precise mathematical language, it is not yet a dynamical semigroup~\cite{Breuer2002}. Hence, to render~(\ref{eq:ME5}) a Markovian master equation, we let the upper limit of the integral go to infinity, which is permissible if the integrand disappears sufficiently fast for $\tau\gg \tau_R=\hbar/(k_{\mathrm{B}} \mathcal{T})$. This is justifiable if the time scale of the system, $\tau_S$, over which the system density matrix $\hat\rho(t)$ varies appreciably, is much larger than the time scale of the bath, $\tau_R$. In other words, we require the bath correlation functions to decay much faster than $\tau_S$~\cite{Breuer2002}. Hence, the Markov approximation is justified when the bath correlation functions, e.g., $\langle \tilde\Gamma_{\vek{q}}(t)\tilde\Gamma_{\vek{k}}(t^\prime)\rangle_{B_0}$=Tr$_B\{\hat B_0\tilde\Gamma_{\vek{q}}(t)\tilde\Gamma_{\vek{k}}(t^\prime)\}$, are proportional to $\delta(t - t^\prime)$. 
Instead, the Born approximation is fulfilled if the dissipative damping rate is smaller than the relevant system's transition frequencies. We shall come back to this point later in the paper. 
Note that with the upper limit of the integral going to infinity, the first line of Eq.~\eqref{eq:ME5} is equal to Eq.~(23) of Ref.~\cite{KrychPRA15}. However, while in the latter the time integral in the definition of $W_{\mathbf{k},\mathbf{k}'}(t)$ is included, we prefer here to write it explicitly.

Finally, we underline that up until now Eq.~(\ref{eq:ME5}) is valid for any impurity in a condensate (not only for an ion), provided that the Fourier transform~(\ref{eq:cq}) can be computed. Indeed, the solution to Eq.~(\ref{eq:rttau}) depends on the impurity dynamics only and the Hamiltonian $\hat H_S(t)$ can also represent the free evolution of a not trapped ion in a BEC or an impurity atom in an optical lattice. For a normal Fermi gas, however, the master equation~(\ref{eq:ME5}) reduces to the double sum only, namely the sum over $\vek{q}$ in the first line disappears. Moreover, only one term of the double sum contributes, as we have single-particle energy states and not Bogolyubov modes. In practice, one sets in Eq.~(\ref{eq:ME5}) the Bogolyubov amplitudes $u\equiv 1$ and $v\equiv 0$ and replaces $n_{\vek{q}^\prime} + 1$ by $1 - n_{\vek{q}^\prime}$, because of the anticommutation relations of the fermionic field operators. Here, $n_{\vek{q}} = [e^{\beta_{\mathcal{T}}(\varepsilon({\vek{q}})-\mu_{\mathrm{G}})}+1]^{-1}$ is the Fermi-Dirac occupation number with $\mu_{\mathrm{G}}$ being the chemical potential obtained from solving Eq.~(\ref{eq:muF}).


\section{Trapped ion master equation}
\label{sec:LDA-PL_approx}

The Markovian master equation~(\ref{eq:ME5}) can be further simplified for an ion in a radiofrequency trap, because of the separation of energy and length scales between the atomic ensemble and the trapped ion system. Indeed, we are going to make two further approximations:
\begin{itemize}
\item[(a)] The particle-like approximation;
\item[(b)] The Lamb-Dicke approximation.
\end{itemize} 
The former concerns the bosonic energy dispersion~(\ref{eq:Bogo-disp}). Because of the large energy difference between the bosonic bath and the ion system, only particle-like excitations couple to the ion motion. This implies that the Bogolyubov dispersion relation~(\ref{eq:Bogo-disp}) is essentially quadratic in the wave vector $q$. For a linear Paul trap, for instance, we have (see Appendix~\ref{sec:nuj} for details): $a_{x,y} = -0.001$, $a_{z} = 0.002$, $q_x = 0.2$, $q_y = -0.2$, $q_z=0$, and $\Omega_{rf} = 2\pi\times$2 MHz, as obtained from the trap design of Ref.~\cite{Joger2017}. These parameters yield the following reference trap frequencies for an ytterbium ion (see appendix \ref{sec:ionHquant}): $\nu_x \simeq 2 \pi\times$169 kHz, $\nu_y \simeq 2 \pi\times$112 kHz, $\nu_z \simeq 2 \pi\times$45 kHz (the $z$ axis is the longitudinal direction, where a system with two or more ions would form a linear crystal). For the sake of convenience, we rescale the dispersion relation~(\ref{eq:Bogo-disp}) as
\begin{align}
\frac{\omega_{\vek{q}}}{\nu_\xi} = \sqrt{
\left(\frac{\bar{q}^2}{2}\right)^2 + (\bar{c}_s \bar{q})^2
}
\qquad\xi=x,y,z
\end{align}
where $\bar{q} = \ell_\xi q$ with $\ell_\xi=\sqrt{\hbar/(m\nu_\xi)}$, and $\bar{c}_s = 4\pi (a_{\mathrm{aa}}^s/\ell_\xi) n \ell_\xi^3$. For sodium atoms with a density $n_0=10^{14}$cm$^{-3}$ we have: $\bar{c}_s \simeq 0.009$ ($\xi \equiv x$), $\bar{c}_s \simeq 0.014$ ($\xi \equiv y$), and $\bar{c}_s \simeq 0.034$ ($\xi \equiv z$). 
The bosons speed of sound ($c_s = \bar{c}_s\ell_\xi\nu_\xi$) is therefore quite small compared to the ion motion in the secular trap such that only phonons of quite low $\bar{q}\sim\bar{c}_s$ (i.e., large wavelength) yield an appreciable difference in the dispersion relation compared to the free particle energy $\bar{q}^2/2$. On the other hand, only phonons in the condensate of comparable energy as $\hbar\nu_\xi$ will couple to the ion motion, so that we can safely assume a particle-like dispersion relation, $\varepsilon(\vek{q}) = \hbar^2 q^2/(2 m)$, and set $u_{\vek{q}}\simeq 1$, $v_{\vek{q}}\simeq 0$, namely the bosonic bath can be treated as a non-interacting Bose gas. This corresponds to an atom velocity of $\sqrt{2\hbar\nu_\xi/m}$ in the $\xi$-th direction. For example, $v_x \simeq 0.077\,$m/s or, in rescaled units, $\bar{v}_x \simeq 1.414$, which is much larger than $\bar{c}_s$, and therefore the atom is moving at supersonic velocities. To such an atomic velocity it is associated the wavelength $\lambda_x\simeq 226.63\,$nm. As a consequence of the energy separation, several terms of the quadratic corrections of the atom-ion interaction in Eq.~(\ref{eq:ME5}) can be safely discarded. 

\begin{figure}
    \centering
    \includegraphics[width=.35\textwidth]{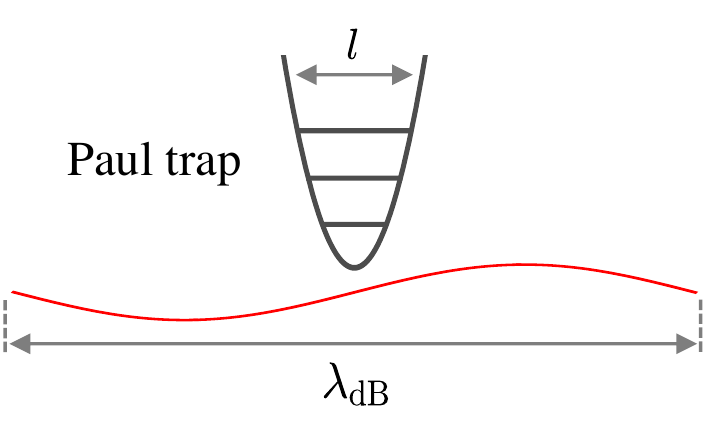}
    \caption{(Color online). Schematic representation of the Lamb-Dicke approximation: the size $l$ of the ion trap (corresponding to the size of the ion in its ground state) is much smaller than the de Broglie wavelength $\lambda_\mathrm{dB}$ of the atoms in the gas.}
    \label{fig:schemeLD}
\end{figure}

On the other hand, the approximation~(b) implies that the typical wavelength of the bosons or of the fermions, i.e. the thermal de Broglie wavelength $\lambda_{\mathrm{dB}}(\mathcal{T})$, is much larger than the $\xi$-th width of the ion ground state, $l_\xi = \sqrt{\hbar/(M\nu_\xi)}$ (see Fig.~\ref{fig:schemeLD}). Let us still consider the example of bosonic sodium atoms at a temperature of $\mathcal{T} = 200$ nK. Thus, we have $\lambda_{\mathrm{dB}}(\mathcal{T})\simeq 814.18\,$nm, while for a trapped ytterbium ion the width of the ground state in the secular trap is $l_z \simeq 36.05\,$nm. This yields a ratio $l_z / \lambda_{\mathrm{dB}}(\mathcal{T})\simeq 0.044$ (similarly for the other directions). Even if we consider the previously estimated supersonic atom velocity, we get $l_\xi/\lambda_\xi\simeq 0.045$ for lithium atoms, $l_\xi/\lambda_\xi\simeq 0.082$ for sodium atoms, and $l_\xi/\lambda_\xi\simeq 0.159$ for rubidium atoms. Hence, the ion spreading is quite localised compared to that of the bath's particle, and therefore the Lamb-Dicke approximation holds very well in the regime we are interested in. 


\subsection{Simplified master equation}
\label{sec:LDAbec}

Under the applicability conditions of the Lamb-Dicke approximation, we can legitimately expand the exponential functions appearing in Eq.~(\ref{eq:ME5}) up to second-order in $\mathbf{q}\cdot\hat{\mathbf{r}}$. For instance, the commutator reduces to
\begin{align}
&[\hat Z_{\vek{q}},\hat W_{\vek{q}}(t,\tau)\hat\rho(t)] \simeq e^{-\frac{i}{\hbar}\varepsilon(\vek{q})\tau}\left\{
i [\vek{q}\cdot\hat{\vek{r}},\hat\rho(t)] \right.\nonumber\\
\phantom{=} & \left. + [\vek{q}\cdot\hat{\vek{r}},\vek{q}\cdot\hat{\vek{r}}(t,\tau)\hat\rho(t)] 
 - \frac{1}{2} [(\vek{q}\cdot\hat{\vek{r}})^2,\hat\rho(t)]
\right\}.
\end{align}
Here, the operators $\hat W_{\vek{q}}(t,\tau) \equiv \hat W_{\vek{q}}(t,t-\tau)$ and $\hat{\vek{r}}(t,\tau) \equiv \hat{\vek{r}}(t-\tau)$ [see also Eq.~(\ref{eq:rttau})]. Given this, the three directions are decoupled from each other, because odd powers of the wave vectors $\mathbf{q}$ vanish, as a consequence of the symmetric summation in the master equation, that is, since the bath is homogeneously confined. Therefore, Eq.~(\ref{eq:ME5}) can be rewritten as follows
\begin{widetext}
\begin{align}
\label{eq:MElda}
& \frac{d}{d t}\hat\rho(t) =-\frac{i}{\hbar} [\hat H_S, \hat\rho] - \sum_{\vek{q},\xi}\int_0^\infty\mathrm{d}\tau\,\Omega^2_{\vek{q}}q_\xi^2\left\{
i\sin\left(\frac{\varepsilon(\vek{q})\tau}{\hbar}\right) [\hat r_\xi^2,\hat\rho(t)]  + 
e^{-i\frac{\varepsilon(\vek{q})\tau}{\hbar}}[\hat r_\xi,\hat r_\xi(t,\tau)\hat\rho(t)] 
- e^{i\frac{\varepsilon(\vek{q})\tau}{\hbar}}[\hat r_\xi,\hat\rho(t)\hat r_\xi(t,\tau)]
\right.\nonumber\\
\phantom{=} & 
\left.
+ 2 \,n_{\vek{q}} 
\cos\left(\frac{\varepsilon(\vek{q})\tau}{\hbar}\right)\left(
[\hat r_\xi,\hat r_\xi(t,\tau)\hat\rho(t)]
-[\hat r_\xi,\hat\rho(t)\hat r_\xi(t,\tau)]
\right)
\right\}
-\sum_{\vek{q},\vek{q}^\prime}\sum_{\xi}\int_0^\infty\mathrm{d}\tau\,
n_{\vek{q}} (n_{\vek{q}^\prime} + 1) (q_\xi^\prime - q_\xi)^2 \Omega_{u,u^\prime}^{u,u^\prime}(\vek{q}^\prime-\vek{q})
\nonumber\\
\phantom{=} & 
\times
\left\{
i\sin\left(\frac{\varepsilon(\vek{q}^\prime) - \varepsilon(\vek{q})}{\hbar}\tau\right) [\hat r_\xi^2,\hat\rho(t)] 
+e^{-i\frac{[\varepsilon(\vek{q}^\prime) - \varepsilon(\vek{q})]\tau}{\hbar}}[\hat r_\xi,\hat r_\xi(t,\tau)\hat\rho(t)] 
- e^{i\frac{[\varepsilon(\vek{q}^\prime) - \varepsilon(\vek{q})]\tau}{\hbar}}[\hat r_\xi,\hat\rho(t)\hat r_\xi(t,\tau)]
\right\}
\end{align}
\end{widetext}
with $\xi=x,y,z$. The first sum over $\vek{q}$ in the first two lines of Eq.~(\ref{eq:MElda}) refers to the Fr\"olich model, while the double sum in the last two lines refers to the extended Fr\"olich model~(\ref{eq:HRS-Bogo}), that is, $\hat H_{RS}^{(2)}$. Because of the particle-like approximation, however, only the term $\Omega_{u,u^\prime}^{u,u^\prime}$ remains, while for the fermions no Fr\"olich interaction appears. 

In order to make further progress, we need to explicitly use the solution of the ion dynamics in the Paul trap in the absence of the gas. The full solution $\hat r_{\xi}(t)$ is provided with various details in Ref.~\cite{KrychPRA15}, which is given by
\begin{align}
\label{eq:IonPosEvolution}
\hat r_{\xi}(t,\tau) & = \sum_{s,s^\prime}C_s^{\xi} C_{s^\prime}^{\xi}\left[
\hat r_\xi\left(\frac{\beta_\xi}{2} + s^\prime\right)\frac{\Omega_{rf}}{\nu_\xi}\cos(\mathcal{I}^{\xi}_{s,s^\prime}(t,\tau)) \right.\nonumber\\
\phantom{=}& \left.- \frac{\hat p_\xi}{\nu_\xi M} \sin(\mathcal{I}^{\xi}_{s,s^\prime}(t,\tau))
\right]
\end{align}
for $\xi=x,y,z$, $[\hat{\vek{r}},\hat{\vek{p}}]=i\hbar$ ($\hat{\vek{p}}$ is the ion momentum operator), and  
\begin{align}
\label{eq:Inm}
\mathcal{I}^{\xi}_{s,s^\prime}(t,\tau) = \Omega_{rf}\left[
\left(\frac{\beta_\xi}{2}+s\right)\tau-(s-s^\prime)t
\right].
\end{align}
The coefficients $C_s^\xi$, the trap parameters $\beta_\xi$, and the frequencies $\nu_\xi$ are introduced in the appendix~\ref{sec:nuj}. 

Since we consider a gas confined in a box of volume $L^3$, the wave vector $\vek{q}$ assumes quantised values: $2\pi s_{\xi}/L$ with $s_\xi\in\mathbb{Z}$ (periodic boundary conditions) and $\xi=x,y,z$. Hence, in the limit $L\rightarrow +\infty$ the allowed values of $\vek{q}$ in momentum space become closely spaced, and since their density is d$\vek{s} = L^3/(2\pi)^3$d$\vek{q}$, we make the replacement
\begin{align}
\label{eq:replacement}
\sum_{\vek{q}} \rightarrow \frac{L^3}{(2\pi)^3}\int_{\mathbb{R}^3}\mathrm{d}\vek{q}.
\end{align} 
Such a continuum limit approximation is reasonable for a large bath. Furthermore, we use the identity~\cite{CarmichaelQOBook}
\begin{align}
\label{eq:delta-CPV}
\int_0^\infty\mathrm{d}\tau\, e^{-i(\omega-\omega_0)\tau} = \pi\delta(\omega-\omega_0)-i\mathcal{P}\left(\frac{1}{\omega - \omega_0}\right),
\end{align}
where $\mathcal{P}$ denotes the Cauchy principal value (CPV), whose action on a test function $\varphi(\omega)$ is:
\begin{align}
\label{eq:CPV}
\mathcal{P}\left(\frac{1}{\omega}\right)(\varphi) = \lim_{\epsilon\rightarrow 0^+}\int_{\mathbb{R} \slash [-\epsilon,\epsilon]}\mathrm{d}\omega\,\frac{\varphi(\omega)}{\omega}.
\end{align}

Now, we apply these results and we focus first our attention on the Fr\"olich contribution to the master equation, namely the incoherent term in Eq.~(\ref{eq:MElda}) involving the summation over $\vek{q}$. Besides, at the moment, we neglect the contribution due to the CPV and look at the $\delta$-contributions only. Thus, $\omega \equiv \varepsilon(\vek{q})/\hbar$, while $\omega_0 \equiv 0$ or $\omega_0 \equiv \Omega_{rf}(\beta_\xi/2+s)$ in Eq.~(\ref{eq:delta-CPV}). When $\omega_0 \equiv 0$, however, the contribution of the term coming from the sine function in the first line of Eq.~(\ref{eq:MElda}) vanishes, as it can be verified by performing the integration~(\ref{eq:replacement}) in spherical coordinates. On the other hand, when $\omega_0 \equiv \Omega_{rf}(\beta_\xi/2+s)$ and after having moved to spherical coordinates, we first write the Dirac's delta as
\begin{align}
\delta(\omega - \omega_0)=\frac{m}{\hbar q}\delta(q - q_{s,\xi})
\end{align}
with $q = \vert \vek{q} \vert$, and~\footnote{We note that the definition of $q_{s,\xi}$ is a mere definition originated by the fact that $q_{s,\xi}^2 = 2 m \Omega_{rf}(\beta_\xi/2+s) /\hbar$ has been introduced when manipulating the exponential and trigonometric functions in the master equation~(\ref{eq:MElda}). Importantly, $q_{s,\xi}^2$ can assume negative or positive values, upon the sign of $\beta_\xi/2+s$.}
\begin{align}
\label{eq:qsj}
q_{s,\xi} = \sqrt{\frac{2 m \Omega_{rf}}{\hbar}\left\vert
\frac{\beta_\xi}{2} + s
\right\vert}.
\end{align}
Hence, the integration in momentum space yields 
\begin{align}
\int_{\mathbb{R}^3}\mathrm{d}\vek{q}\,\frac{f^2(q) q^2_\xi}{q}\delta(q - q_{s,\xi}) = \frac{4}{3}\pi q_{s,\xi}^3 f^2(q_{s,\xi})
\end{align}
with $f$ being the scattering amplitude~(\ref{eq:fq}) evaluated in $q=q_{s,\xi}$.

As far as the contribution of the extended Fr\"ohlich model is concerned, namely the terms due to the double summation over $\vek{q}$ and $\vek{q}^\prime$ in Eq.~(\ref{eq:MElda}), we proceed in a very similar manner with the exception that we now have to assess a double integration in momentum space. In the appendix~\ref{sec:doubleintegral} we provide details of this calculation. On the other hand, in order to assess the contribution due the Lamb-shift we have to compute the integral~(\ref{eq:CPV}), whose details are outlined in the appendix~\ref{sec:cpv}. We note, however, that we performed such a Lamb-shift calculation only for the linear terms of Eq.~(\ref{eq:MElda}), since the contribution of the extended Fr\"ohlich model is much smaller, and therefore it can be neglected. 

Under the above outlined approximations, we arrive at
\begin{align}
\label{eq:ME-LDA-SIunits}
&\frac{d}{d t}\hat\rho(t) = - \frac{i}{\hbar}
[\hat H_S + \delta\hat H_S, \hat\rho]  
- \Gamma\sum_{\xi=x,y,z}\left\{
\Lambda_\xi [\hat r_\xi,\hat\rho(t) \hat p_\xi] \right.\nonumber\\
&- \Lambda_\xi^* [\hat r_\xi,\hat p_\xi \hat\rho(t)] -\Phi_\xi  [\hat r_\xi,\hat\rho(t) \hat r_\xi] 
\left.+ \Phi_\xi^*  [\hat r_\xi, \hat r_\xi \hat\rho(t)] \right\}.
\end{align}
Here, $\Gamma = \frac{2}{3}\frac{m n_0\pi\hbar}{\mu^2}$ and $\delta\hat H_S$ is the correction to the free ion Hamiltonian due to the interaction to the quantum gas (i.e., the Lamb-shift), which is given by 
\begin{align}
\delta\hat H_S = (1-\phi) \frac{M}{2} \sum_{\xi=,x,y,z}\delta W_\xi(t) \hat r_\xi^2
\end{align}
with $\phi = 0$ for the bosons and $\phi = 1$ for the fermions, 
\begin{align}
\label{eq:dW}
& \delta W_\xi(t) = \frac{\Omega_{rf}^2}{4}\left[
\delta a_\xi - 2 \delta q_\xi\cos(\Omega_{rf} t) - 2 \delta q_\xi^\prime \,g_\xi(t)
\right],
\end{align}
and
\begin{align}
\label{eq:delta_aq}
& g_\xi(t) = \!\sum_{s,s^\prime\notin S_{\mathrm{i}}} \!F_{s,s^\prime}^\xi \cos[(s-s^\prime)\Omega_{rf} t]
\left[{\mathcal{J}}_+^\prime(q_{s,\xi}) - {\mathcal{J}}_-^\prime(q_{s,\xi})\right],\nonumber\\
&\delta a_\xi = - Q{\mathcal{J}}_-^\prime(0) + Q\sum_s\left\{
F^\xi_{s,s}\left[
{\mathcal{J}}_-^\prime(q_{s,\xi}) - {\mathcal{J}}_+^\prime(q_{s,\xi})
\right]\right\},\nonumber\\
& \delta q_\xi = \frac{Q}{2}\sum_{\vert s - s^\prime\vert = 1}\left\{
F^\xi_{s,s^\prime}\left[
{\mathcal{J}}_+^\prime(q_{s,\xi}) - {\mathcal{J}}_-^\prime(q_{s,\xi})
\right]
\right\},
\end{align}
where ${\mathcal{J}}_\pm^\prime(q_{s,\xi})$ are defined in Eq.~(\ref{eq:J-CPV}) and Eq.~(\ref{eq:J-CPVplus}), $\delta q_\xi^\prime = Q/2$, and  
\begin{align}
S_{\mathrm{i}} &= \{(s,s^\prime): \vert s - s^\prime \vert = 0\,\, \text{or} \,\, 1\}, \nonumber\\
F_{s,s^\prime}^\xi & = C_s^\xi C_{s^\prime}^\xi \left(
\frac{\beta_\xi}{2} + s^\prime
\right)\frac{\Omega_{rf}}{2\nu_\xi},
\qquad
S_{s,s^\prime}^\xi = i \frac{C_s^\xi C_{s^\prime}^\xi}{M\nu_\xi 2}, \nonumber\\
Q &= \frac{32}{3}\frac{m}{M}\frac{\hbar^2 n_0}{\mu^2\Omega_{rf}^2}.
\end{align}
We see that the coupling to the quantum gas renormalises the 
geometric Paul trap parameters as $a_\xi\mapsto a_\xi + \delta a_\xi$  and $q_\xi\mapsto q_\xi + \delta q_\xi$ 
[see also Eq.~(\ref{eq:PTfreq})] and it yields additional time-dependent driving terms [i.e., $g_\xi(t)$]. 
Moreover, we introduced the functions $\Phi_\xi(t) = \Phi_\xi^\delta(t) + (1-\phi) \Phi_\xi^\mathcal{P}(t)$, $\Lambda_\xi(t) = \Lambda_\xi^\delta(t) + (1-\phi) \Lambda_\xi^\mathcal{P}(t)$, where 

\begin{equation}
    \begin{split}
    \Phi_\xi^\delta(t)=&\sum_{s,s'}F_{s,s'}^\xi\Bigg\{|q_{s,\xi}|^3f(q_{s,\xi})^2(1-\phi)\\[1ex]
    \times&\bigg[\cos\big[(s-s')\Omega_{rf}t\big](1+2n_{q_{s,\xi}})\\[1ex]
    &+i\sin\big[(s-s')\Omega_{rf}t\big]\mathrm{sgn}\bigg(\frac{\beta_\xi}{2}+s\bigg)\bigg]\\[1ex]
    &+\eta^-_{s,\xi}e^{-i(s-s')\Omega_{rf}t}+\eta^+_{s,\xi}e^{i(s-s')\Omega_{rf}t}\Bigg\},
\end{split}
\label{eq:Phi}
\end{equation}
\begin{equation}
    \begin{split}
       \Lambda_\xi^\delta(t)=&\sum_{s,s'}S_{s,s'}^\xi\Bigg\{|q_{s,\xi}|^3f(q_{s,\xi})^2(1-\phi)\\[1ex]
       \times&\bigg[i\sin\big[(s-s')\Omega_{rf}t\big](1+2n_{q_{s,\xi}})\\[1ex]
       &+\cos\big[(s-s')\Omega_{rf}t\big]\mathrm{sgn}\bigg(\frac{\beta_\xi}{2}+s\bigg)\bigg]\\[1ex]
       &-\eta^+_{s,\xi}e^{-i(s-s')\Omega_{rf}t}+\eta^-_{s,\xi}e^{i(s-s')\Omega_{rf}t}\Bigg\}
    \end{split}
    \label{eq:Lambda}
\end{equation}
with
\begin{equation}
\label{eq:eta_sxi}
    \eta^\pm_{s,\xi}=\frac{m}{16\pi^2\hbar^2\beta_\mathcal{T}n_0}\Big(\mathcal{F}_{s,\xi}^{(1),\pm}+(-1)^\phi\mathcal{F}_{s,\xi}^{(2)}\Big).
\end{equation}
Here, $\mathcal{F}_{s,\xi}^{(1),\pm}$ and $\mathcal{F}_{s,\xi}^{(2)}$ are defined in Eq.~(\ref{eq:Fnj1}) and Eq.~(\ref{eq:Fnj2}), respectively, and

\begin{align}
\label{eq:PhiLambda-lamb}
&\Phi_\xi^\mathcal{P}(t) = \frac{2}{\pi}\sum_{s,s^\prime} F_{s,s^\prime}^\xi 
\sin[(s-s^\prime)\Omega_{rf} t] \left[{\mathcal{J}}_+^\prime(q_{s,\xi}) + {\mathcal{J}}_-^\prime(q_{s,\xi})\right.
\nonumber\\
&+ \left. 
2\left({\mathcal{J}}_+^{n_q}(q_{s,\xi}) + {\mathcal{J}}_-^{n_q}(q_{s,\xi})\right)
\right]
\left[
2\Theta(\mathrm{sgn}(\beta_\xi/2 + s)) - 1
\right] ,\nonumber\\[2ex]
&\Lambda_\xi^\mathcal{P}(t) = i\frac{2}{\pi}\sum_{s,s^\prime} S_{s,s^\prime}^\xi \left\{
\left[
1 - 2\Theta(\mathrm{sgn}(\beta_\xi/2 + s)) 
\right]
\left[{\mathcal{J}}_+^\prime(q_{s,\xi})\right.
\right.
\nonumber\\
&\left. + {\mathcal{J}}_-^\prime(q_{s,\xi})
+ 2\left({\mathcal{J}}_+^{n_q}(q_{s,\xi}) + {\mathcal{J}}_-^{n_q}(q_{s,\xi})\right)
\right] 
\cos[(s-s^\prime)\Omega_{rf} t] 
 \nonumber\\
&\left.-i \sin\left[(s-s^\prime)\Omega_{rf} t\right]
\left[{\mathcal{J}}_-^\prime(q_{s,\xi}) - {\mathcal{J}}_+^\prime(q_{s,\xi})\right]\right\}.
\end{align}
The labels $\delta$ and $\mathcal{P}$ indicate the origin of the contribution, namely with $\delta$ from the Dirac's delta in Eq.~(\ref{eq:delta-CPV}), whereas with $\mathcal{P}$ from the Cauchy principal value, i.e. the Lamb-shift. On the other hand, the terms involving the functions $\mathcal{F}_{s,\xi}^{(1,2)}$ stem from the extended Fr\"ohlich model. Besides, the integrals ${\mathcal{J}}_\pm^{n_q}(q_{s,\xi})$ are defined in Eq.~(\ref{eq:J-CPVnq}) and they stem from the first term in the second line of Eq.~(\ref{eq:MElda}), i.e. the one involving the cosine function. Interestingly, this term does not contribute to the renormalisation of the Paul trap parameters~(\ref{eq:delta_aq}), as the sum of each single exponential function, when the cosine is written as a linear combination of exponentials, vanishes for the part that concerns the unitary dynamics. 
Furthermore, because of the Lamb-Dicke approximation we made, we note that the three directions are decoupled. 
Thus, Eq.~(\ref{eq:ME-LDA-SIunits}) can be split into three components, each representing a direction of the ion motion. In other words, we have a master equation per each direction. 

A few remarks, however, are in order. First, we note that even the contribution stemming from the CPV yields incoherent terms, as indicated by the functions $\Phi_\xi^\mathcal{P}(t)$ and $\Lambda_\xi^\mathcal{P}(t)$. Second, the contribution stemming from the Dirac's delta yields coherent (i.e., unitary) dynamics as well, since the functions $\Phi_\xi(t)$ and $\Lambda_\xi(t)$ are complex. This is because we did not perform the rotating-wave approximation (RWA), that is, we did not neglect the non-secular terms with $\omega \ne \Omega_{rf}(\beta_\xi/2+s)$, which is typically applied in quantum optics. Given the fact that the most relevant Paul trap coefficients $C_s^\xi$ for a linear geometry are those with $s=0,\pm 1$, the contributions in $\Phi_\xi^\mathcal{P}(t)$ and $\Lambda_\xi^\mathcal{P}(t)$ with $\sin\left[(s-s^\prime)\Omega_{rf}t\right]$ are in general small, and therefore one could in principle apply the RWA also in this context. Nonetheless, we find that the term in $\Lambda_\xi^\mathcal{P}(t)$ involving $\cos\left[(s-s^\prime)\Omega_{rf}t\right]$ has a non-negligible contribution. This means that an application of the RWA would underestimate the overall dissipative dynamics. A similar reasoning applies for the coherent dynamics stemming from the imaginary contributions of the $\Phi_\xi^\delta(t)$ and $\Lambda_\xi^\delta(t)$ functions. This is also the reason why we cannot transform Eq.~(\ref{eq:ME-LDA-SIunits}) in Lindbland form (see also appendix~\ref{sec:lindbland}), as an essential requisite is the RWA~\cite{Breuer2002}. We note that up until now, these effects have been not taken into account in investigations in the context of impurity physics for settings similar to ours~\cite{DaleyPRA04,KrychPRA15,NielsenNJP19,LenaPRA20}.


\section{Ion energy and first order moments} 
\label{sec:2ndMom}

The energy of the ion at time $t$ is given by the expectation value of the ion Hamiltonian (\ref{eq:Hion}):
\begin{align}
\label{eq:meanHion}
\langle\hat H_I^{trap}(t) \rangle = \langle\hat{H}_I^{kin}(t)\rangle 
+\frac{M}{2}{W}_{\xi}^\prime(t)\langle\hat{r}_\xi^2\rangle, 
\end{align}
where ${W}_{\xi}^\prime(t) = W_{\xi}(t) + \delta W_{\xi}(t)$, with $W_{\xi}(t)$ given by 
\begin{align}
W_{\xi}(t) = \frac{\Omega_{rf}^2}{4}\left[a_{\xi} - 2 q_{\xi}\cos(\Omega_{rf}t)\right], 
\end{align}
$\delta W_{\xi}(t)$ defined in Eq.~(\ref{eq:dW}) and

\begin{equation}
    \langle\hat{H}_I^{\mathrm{kin}}(t)\rangle=\sum_{\xi=x,y,z}\frac{\langle\hat p_{\xi}^2\rangle}{2M} 
    \label{eq:meanKin}
\end{equation}

In order to calculate it, we have to determine the expectation values $\langle\hat{p}_\xi^2(t)\rangle = $ Tr$\{\hat{p}_\xi^2\hat\rho_\xi(t)\}$ and $\langle\hat{r}_\xi^2(t)\rangle = $ Tr$\{\hat{r}_\xi^2\hat\rho_\xi(t)\}$ $\forall\,\xi = x,y,z$ with $\hat\rho_\xi(t)$ being the ion density matrix corresponding to the $\xi$-th direction, whose equation of motion is obtained by considering only the pertinent direction in the sum appearing in Eq.~(\ref{eq:ME-LDA-SIunits}). Instead of solving the full master equation, however, it is computationally less expensive to solve the corresponding differential equations for the expectation values of the square of the position and momentum, which are coupled to the covariance $\langle \hat c_\xi \rangle = \langle(\hat{r}_\xi\hat p_{\xi} + \hat p_{\xi} \hat{r}_\xi)(t)\rangle$. Therefore, by using the definition of the expectation value of an observable, $\langle\hat{\mathcal{O}}\rangle = $Tr$\{\hat{\mathcal{O}}\hat\rho(t)\}$, and the master equation~(\ref{eq:ME-LDA-SIunits}), one arrives at the set of coupled differential equations:
\begin{align}
\label{eq:EOM2ndMom}
&\frac{d}{dt} \langle \hat r_\xi^2\rangle = \frac{\langle \hat c_{\xi}\rangle}{M}, \nonumber\\
&\frac{d}{dt} \langle \hat p_{\xi}^2\rangle\! = \!\left\{2\hbar \Gamma \Im[\Phi_\xi(t)] - M{W}_\xi^\prime(t)\right\} \langle \hat c _{\xi} \rangle 
- 4\hbar\Gamma \Im[\Lambda_\xi(t)] \langle \hat p_{\xi}^2 \rangle \nonumber\\
&+ 2 \hbar^2 \Gamma \Re[\Phi_\xi(t)],  \nonumber\\
&\frac{d}{dt} \langle \hat c_{\xi} \rangle = 2 \left\{2\hbar \Gamma \Im[\Phi_\xi(t)] - M{W}_\xi^\prime(t)\right\} \langle \hat r_{\xi}^2 \rangle + 2 \frac{\langle \hat p_{\xi}^2 \rangle}{M}
\nonumber\\
& -2 \hbar \Gamma \Im[\Lambda_\xi(t)] \langle \hat c_{\xi} \rangle + 2 \hbar^2 \Gamma \Re[\Lambda_\xi(t)].
\end{align} 
In the limit for which the Lamb-shift and the extended Fr\"ohlich model are not considered, the equations of motion (58) of Ref.~\cite{KrychPRA15} are retrieved. We note that the set of equations~(\ref{eq:EOM2ndMom}) holds for both the bosonic and fermionic bath, but with different $\Phi_\xi$ and $\Lambda_\xi$ functions. 

The radiofrequency fields set the smallest time scale in the open quantum system. It is therefore useful to evaluate the time averaged energy over a rf-period, namely 
\begin{align}
\label{eq:meanH-Trf}
\langle\langle\hat H_I^{\mathrm{kin}} (t) \rangle\rangle_{T_{rf}} = 
\frac{1}{T_{rf}} \int_t^{t+T_{rf}} \mathrm{d}t^\prime \langle\hat H_I^{\mathrm{kin}}(t^\prime) \rangle
\end{align}
where $T_{rf} = 2\pi/\Omega_{rf}$. Here, the notation $\langle\langle\dots \rangle\rangle_{T_{rf}}$ denotes the time average over $T_{rf}$. 
In this way we average out the fast oscillations due to the rf-field and the final ion energy at thermal equilibrium with the atomic gas can be assessed. 

Finally, we provide equations of motion of the first order moments of the ion position and momentum operators. Exactly with the same procedure that we outlined previously for Eq.~(\ref{eq:EOM2ndMom}), the coupled differential equations for the first order moments read:
\begin{align}
\label{eq:EOM1stMom}
&\frac{d}{dt} \langle \hat r_\xi\rangle = \frac{\langle \hat p_{\xi}\rangle}{M}, \\
&\frac{d}{dt} \langle \hat p_{\xi}\rangle\! = \!\left\{2\hbar \Gamma \Im[\Phi_\xi(t)] - M{W}_\xi^\prime(t)\right\} \!\langle \hat r _{\xi} \rangle \!
- \!2\hbar\Gamma \Im[\Lambda_\xi(t)] \langle \hat p_{\xi} \rangle.\nonumber
\end{align}
The first equation of motion of $\langle \hat r_\xi\rangle$ is simply the definition of the ion momentum in the $\xi$-direction, while the second one provides the average force acting on the ion. The latter is, on the one side, due to the ion trap, i.e. the term proportional to ${W}_\xi^\prime(t)$, and, on the other side, to the atom-ion interaction, namely the term proportional to $\Im[\Phi_\xi(t)]$. Moreover, the term proportional to $\Im[\Lambda_\xi(t)]$ relies on the ion momentum, that is, it corresponds to a velocity-dependent force, which results in a damped ion motion because of the presence of the gas, unless the initial conditions for $\langle \hat{r}_\xi \rangle$ and $\langle \hat{p}_\xi \rangle$ vanish. It is interesting to note that the form of the equations of motion~(\ref{eq:EOM1stMom}) resembles that of $\langle \hat r_\xi^2\rangle$ and $\langle \hat p_\xi^2\rangle$, where $\langle \hat c_\xi\rangle$ is replaced by $\langle \hat p_\xi\rangle$ or $\langle \hat r_\xi\rangle$ and $\langle \hat p_\xi^2\rangle$ by $\langle \hat p_\xi\rangle$, but Eq.~(\ref{eq:EOM1stMom}) does not have source terms.\\


\section{Numerical Results}
\label{sec:results}

Before we present and discuss our numerical findings, we note that hereafter they are based on the parameters $b$ and $c$ computed as explained in appendix~\ref{sec:bc_reg}. Moreover, the bare Paul trap parameters are chosen as: $a_{x,y} = -0.001$, $a_{z} = 0.002$, $q_x = -q_y = 0.2$, $q_z=0$, and 
$\Omega_{rf} = 2\pi\times$2 MHz. Such parameters correspond to a linear trap geometry, whose frequencies are: $\nu_x \simeq 2 \pi\times$169 kHz, $\nu_y \simeq 2 \pi\times$112 kHz, $\nu_z \simeq 2 \pi\times$45 kHz (a $^{174}$Yb ion is always assumed). Thus, we have $\beta_{x,y} \simeq 0.1389$ and $\beta_z \simeq 0.0447$. 
Details on the choice of the initial density matrix are provided in the appendix~\ref{sec:ionHquant}. 


\subsection{Renormalised Paul trap parameters}
\label{sec:dadq}

To begin with, we analyse the impact of the coupling to the quantum gas on the ion dynamics by considering the renormalised trap $a$- and $q$-parameter. 
The absolute amount of change of those parameters from their bare values, i.e. without the gas, provides us a rule of thumb to assess how strong can be the coupling, that is, how large can be the atomic density such that the master equation can yield a faithful description of the ion dynamics. 
The modification of those parameters relies on the particular atom-ion species via the mass ratio as well as on the condensate density.

\begin{figure}
\centering
\includegraphics[width=.4\textwidth]{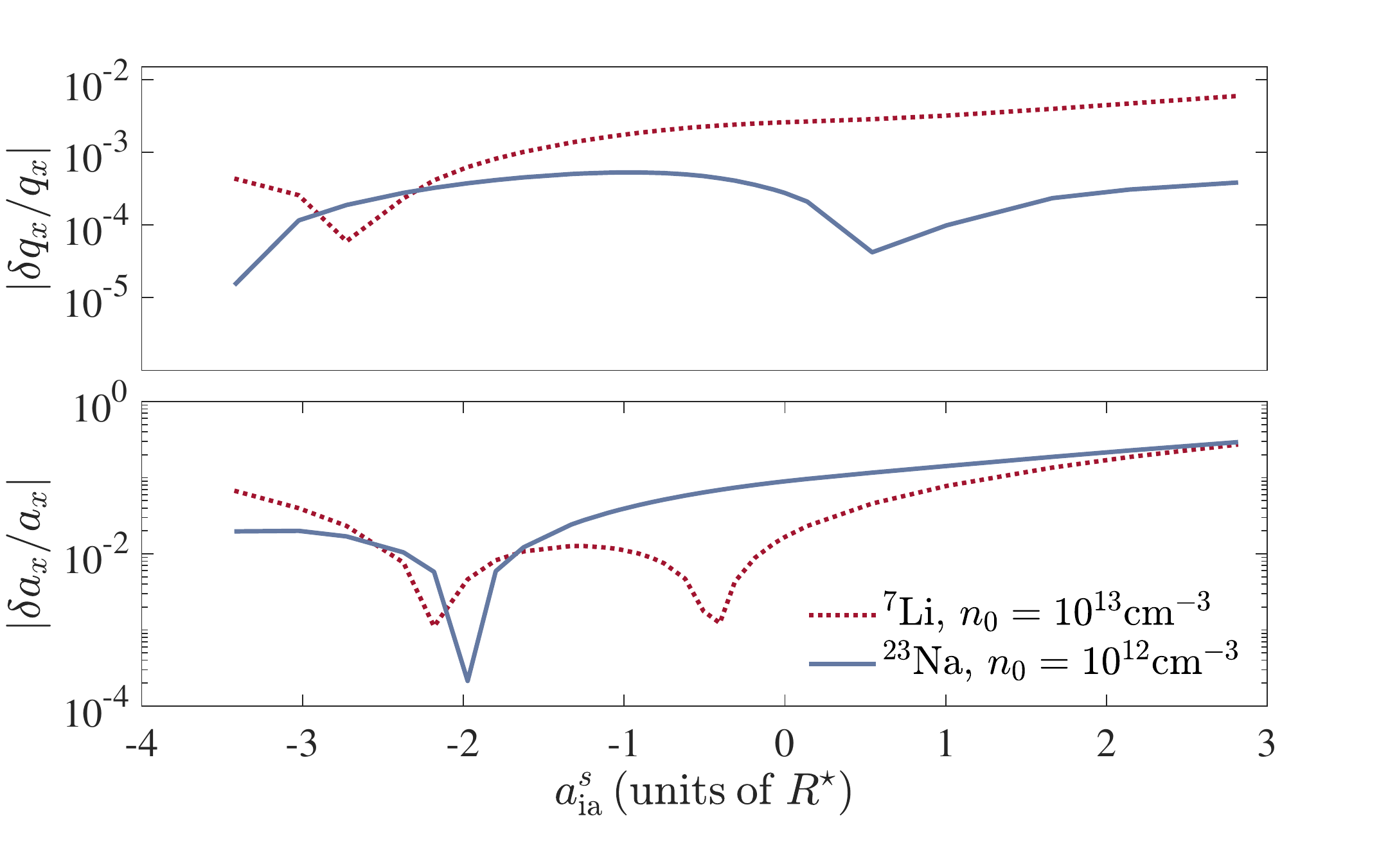}
\caption{\label{fig:delta_aq} (Color Online). Corrections to the Paul trap parameters along the $x$ direction for sodium and lithium atoms for two densities as a function of the atom-ion scattering length. Top panel: relative correction to the $q$-parameter. Bottom panel: relative correction to the $a$-parameter.}
\end{figure}

In Fig.~\ref{fig:delta_aq} we show the corrections to the Paul trap parameters $\delta a_\xi$ and $\delta q_\xi$ relatively to their bare values as a function of the atom-ion scattering length for $n_0 = 10^{12}/$cm$^3$ and $n_0 = 10^{13}/$cm$^3$ for sodium and lithium, respectively, along the transverse direction $\xi \equiv x$. Note that for lower densities the values of $\delta a_\xi$ and $\delta q_\xi$ are reduced by $n_0/(10^{12}/$cm$^3$) for sodium and by $n_0/(10^{13}/$cm$^3$) for lithium due to their definition.   
As it can be seen, the $q$-parameter, namely that of the driving rf-field, is very weakly affected by the coupling to the gas (top panel). The $a$-parameter, instead, assumes larger values, especially for positive scattering lengths (bottom panel). Furthermore, we see that the heavier the atom, the larger is the impact on the trap (for equal densities), as expected (in the figure the result for lithium has to be divided by ten to be compared with that of sodium). These results show that, while for lithium densities up to $n_0 = 10^{13}/$cm$^3$ can be considered (ideally the ratio should be smaller than unity), it is better not to exceed $n_0 = 10^{12}/$cm$^3$ for sodium atoms because of the strong modification to the $a$-parameter. 


\subsection{Ion in a lithium gas}
\label{sec:Li}


\begin{figure*}
\centering
\includegraphics[scale=0.4]{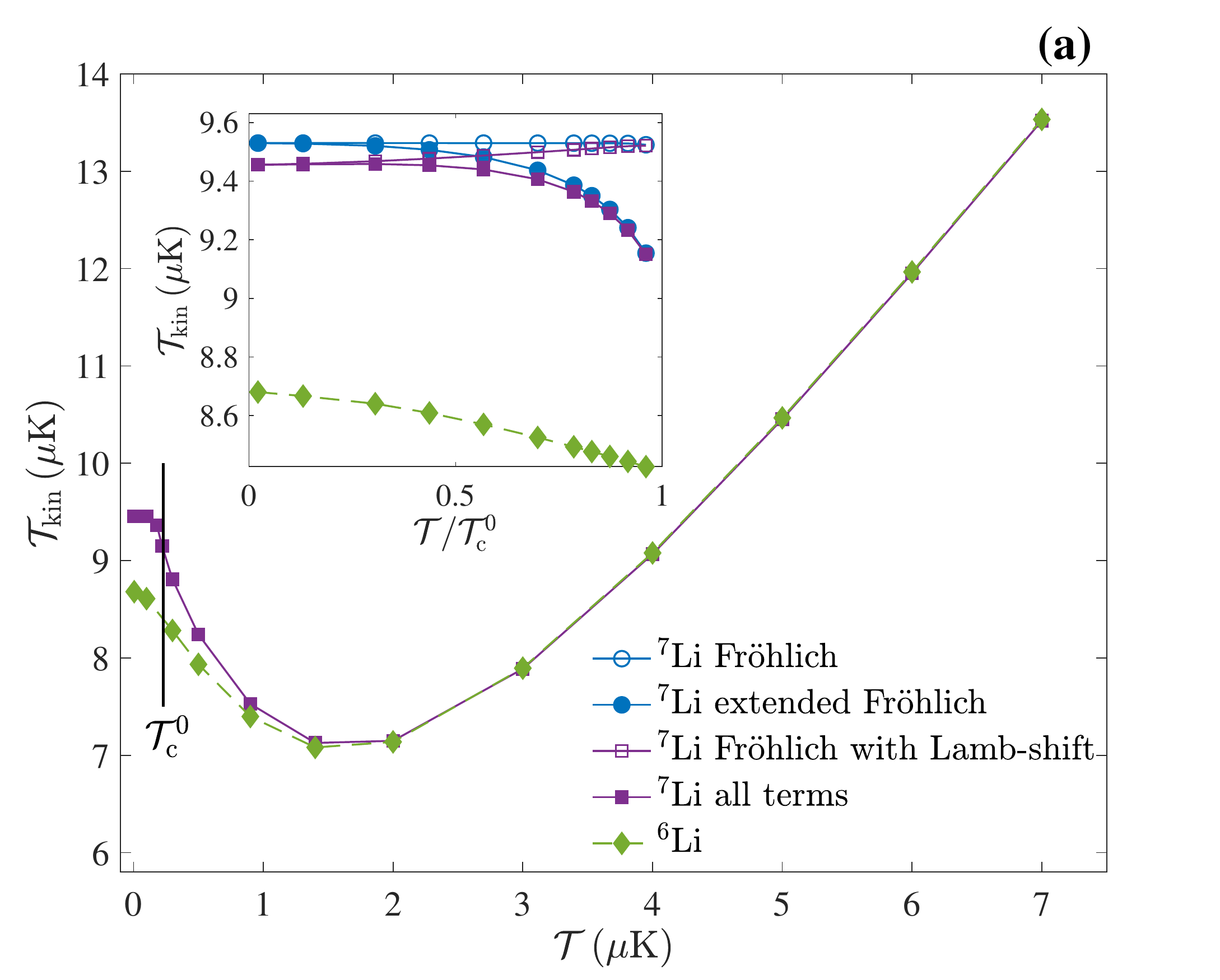}
\includegraphics[scale=0.4]{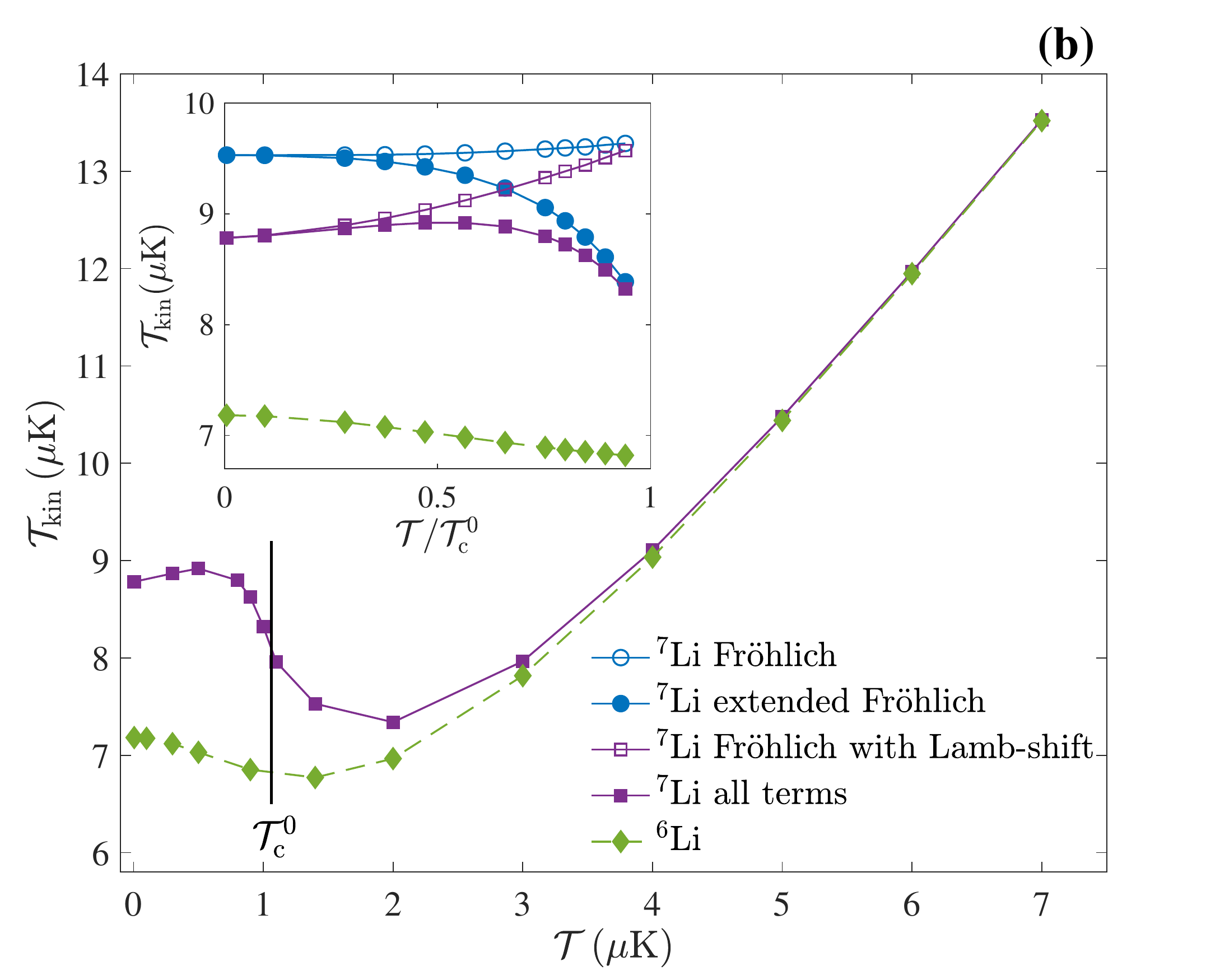}
\caption{\label{fig:Lithium_n18n19} (Color online). Ion temperature obtained from the averaged energy~(\ref{eq:meanH-Trf}) for $b \simeq 0.0780 \,R^\star$, $c\simeq 0.2239\,R^\star$, which correspond to $a_{\mathrm{ia}}^s \simeq R^\star$. The square and circle symbols refer to the bosons, while the diamonds to the fermions. The black vertical line indicates the position of the critical temperature of condensation $\mathcal{T}_{\mathrm{c}}^0$, while the Fermi temperature is not indicated. Panel (a): $n_{\mathrm{t}} = 10^{12}\,\text{cm}^{-3}$, $T=50\,\mathrm{ms}$, $\mathcal{T}_\mathrm{F}=0.61\,\mu\mathrm{K}$. Panel (b): $n_{\mathrm{t}} = 10^{13}\,\text{cm}^{-3}$, $T=6\,\mathrm{ms}$, $\mathcal{T}_\mathrm{F}=2.48\,\mu\mathrm{K}$.}
\end{figure*}
    
\begin{figure}
\centering
\includegraphics[width=0.5\textwidth]{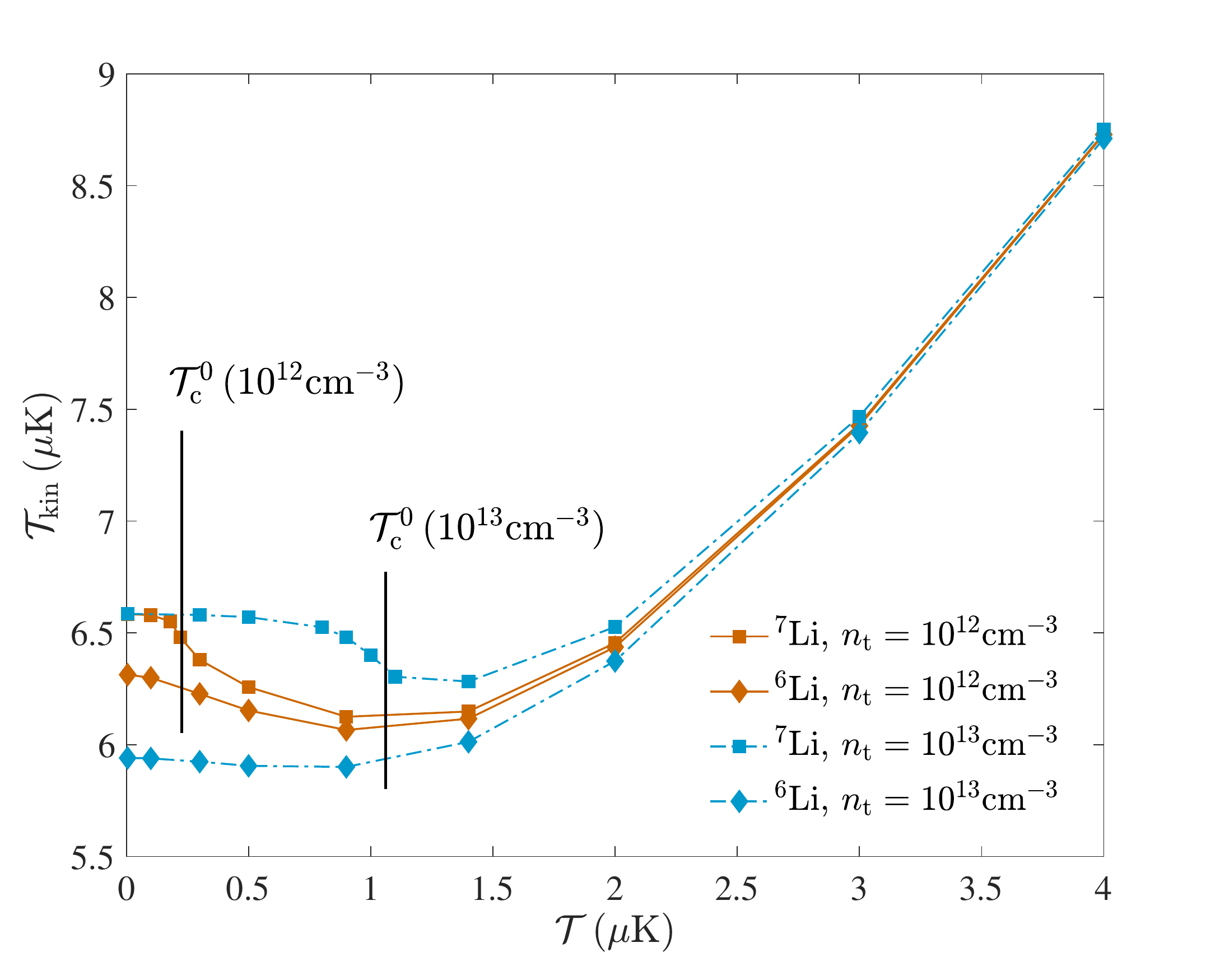}
\caption{\label{fig:pseudopotential} (Color online). Ion temperature obtained from the averaged energy~(\ref{eq:meanH-Trf}) with the pseudopotential and $a_{\mathrm{ia}}^s=R^\star$. The dot-dashed light blue lines correspond to $n_{\mathrm{t}} = 10^{13}\,\text{cm}^{-3}$, while the solid orange ones to $n_{\mathrm{t}} = 10^{12}\,\text{cm}^{-3}$. The simulation time is $T=10\,\mathrm{ms}$ and $T=100\,\mathrm{ms}$, respectively.}
\end{figure}

\begin{figure}
\centering
\includegraphics[width=0.5\textwidth]{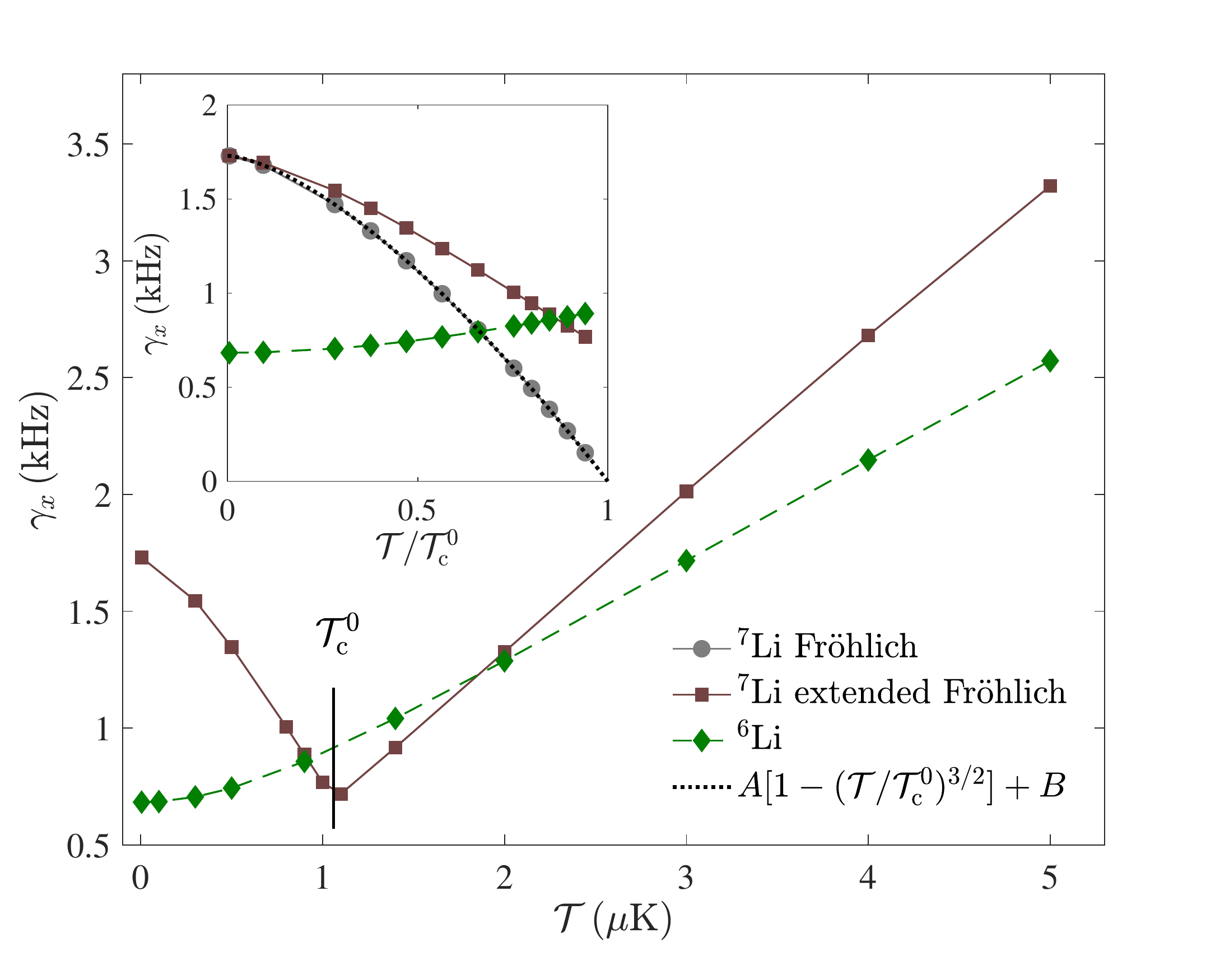}
\caption{\label{fig:damping_rates} (Color online). Damping rates of $\langle\hat{r}_\xi\rangle$ as a function of the gas temperature with a total density $n_{\mathrm{t}}=10^{13}\mathrm{cm}^{-3}$. The dotted line in the inset represents a fit with the condensate density dependence on the ratio $\mathcal{T}/\mathcal{T}_{\mathrm{c}}^0$.}
\end{figure}

\begin{figure}
\centering
\includegraphics[width=0.5\textwidth]{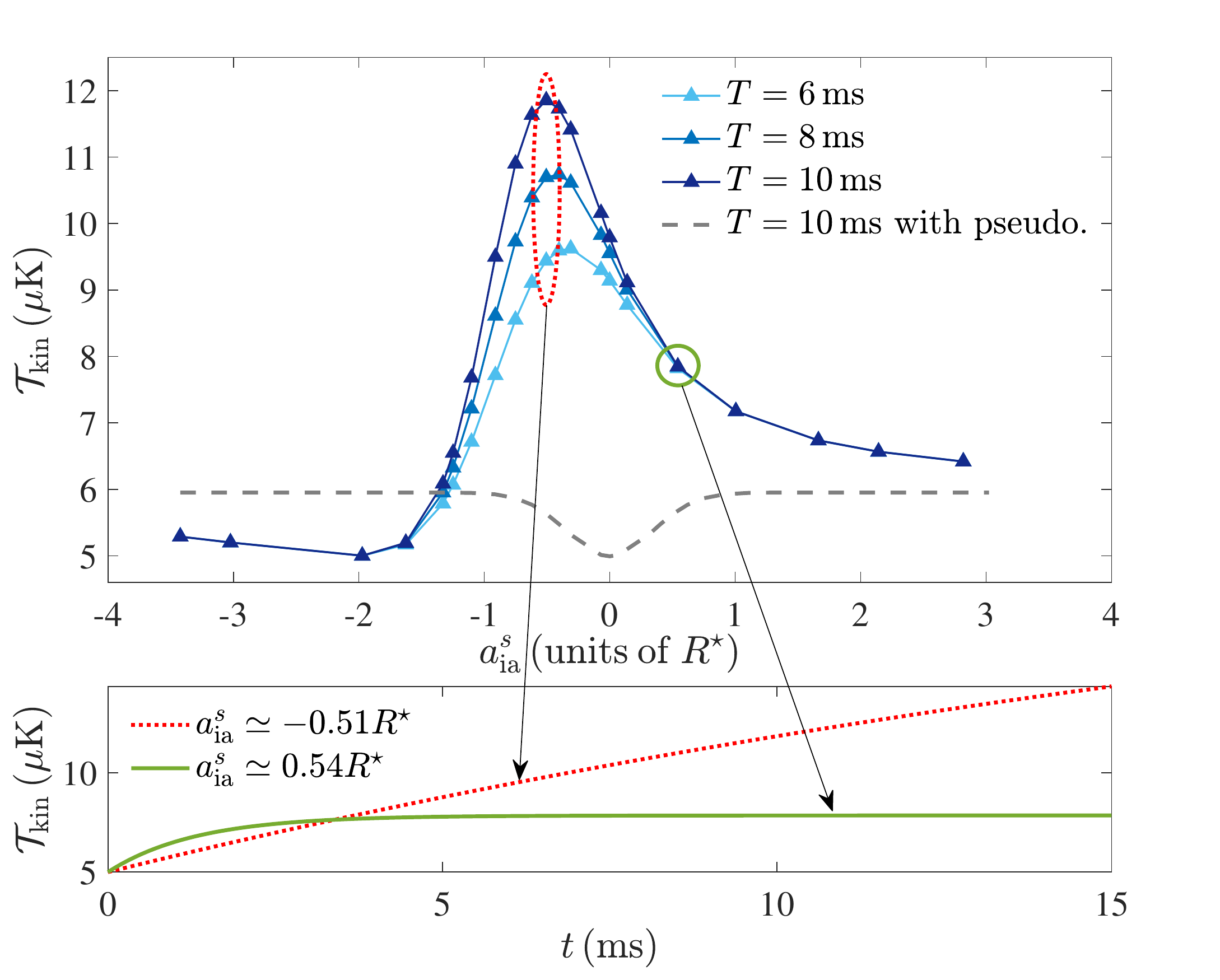}
\caption{\label{fig:Evsa} (Color online). Top panel: Ion temperature vs. the atom-ion scattering length for $^6$Li for $n_{\mathrm{t}} = 10^{13}\mathrm{cm}^{-3}$ and a gas temperature $\mathcal{T} = 0.1\mu\mathrm{K}$. Bottom panel: Ion temperature vs. time for two different scattering lengths.}
\end{figure}

In Fig.~\ref{fig:Lithium_n18n19} we show the averaged ion kinetic energy expressed as a temperature, $\mathcal{T}_\mathrm{kin} = 2\langle\langle\hat H_{I}^{\mathrm{kin}}(T)\rangle\rangle_{T_{rf}}/3k_{\mathrm{B}}$, at the final time $T$, namely when the ion has thermalised with the quantum gas, where the averaged energy is given by Eq.~(\ref{eq:meanH-Trf}). The definition of the kinetic temperature $\mathcal{T}_\mathrm{kin}$ includes the secular motion and the micromotion of the ion and the factor $2/3$ is due to the equipartition theorem \cite{Furst2018}. 
We note that the value of the thermalisation time $T$ relies on various system parameters, especially on the atomic density. For a fixed scattering length, decreasing the density by an order of magnitude implies an increase of the thermalisation time by roughly the same amount. As a consequence and for $\mathcal{T}<\mathcal{T}_{\mathrm{c}}^0$, $T$ strongly depends on the gas temperature, since the latter determines the density of the condensed fraction, i.e. $n_0$. A first estimation of the thermalisation time for each plot is found by looking at $\mathcal{T}_\mathrm{kin}$ as a function of time for a single value of the temperature (see e.g. the bottom panel of Fig.~\ref{fig:Evsa}). Then, the values of $\mathcal{T}_\mathrm{kin}$ at all temperatures are computed at the estimated time and at several larger times up to the one at which the difference between the curves becomes negligible. 
\\[1ex]
\textit{General remarks.} Let us first focus on the bosonic case below $\mathcal{T}_c^0$. As we can see from the insets of Fig.~\ref{fig:Lithium_n18n19}, the behavior of the ion kinetic energy is the result of the interplay among the different contributions. The empty blue circles correspond to the Fr\"ohlich model, i.e. they show the contribution of the condensed part of the gas only. The final temperature of the ion is basically independent of the gas temperature in this approximation. This result is consistent with the fact that the density barely affects the final energy of the ion. Indeed, in the Fr\"ohlich model the dependence of the equations on the gas temperature is almost entirely carried by the condensate density $n_0$, as the temperature-dependent factor $n_{q_{s,\xi}}$ in the definitions of $\Phi(t)$ and $\Lambda(t)$ is always much lower than unity. This behavior has to be traced back to the nature of the condensate, in which all the particles occupy the same single particle state. For this reason, the cooling effect of the condensate does not change when its temperature changes, as the latter only affects the fraction of condensed particles.
A similar reasoning can be applied to the Lamb-shift, whose contribution can be observed in the purple empty squares of Fig.~\ref{fig:Lithium_n18n19}. The additional cooling effect is stronger at temperatures much lower than $\mathcal{T}_\mathrm{c}^0$. This phenomenon is related to the condensate density $n_0$ increasing when the temperature is decreased, which implies a stronger coupling to the gas, and it is in agreement with what discussed in section \ref{sec:dadq} about the dependence of the $\delta a_\xi$ and $\delta q_\xi$ parameters on $n_0$. On the other hand, the contribution of the extended Fr\"ohlich model shows the opposite trend. As it can be seen from the full blue circles of the aforementioned insets, when $\mathcal{T}$ approaches the critical temperature the contribution of the interaction with the normal part of the gas bends the ion temperature downward. 
The extended Fr\"ohlich model, which is the only one contributing at $\mathcal{T}>\mathcal{T}_\mathrm{c}^0$ in the bosonic case and at every gas temperature $\mathcal{T}$ for the fermions, is responsible for a minimum in the final ion temperature, i.e. kinetic energy. In order to understand its nature, we studied the temperature dependence of some characteristic quantities involved in the equations such as $\eta_{s,\xi}^\pm$ [see Eq.~(\ref{eq:eta_sxi})]. For simplicity, let us now focus on the fermionic case, where no other contributions have to be considered. In this case, all the temperature dependence relies on $\eta_{s,\xi}^\pm$, which monotonically increases with $\mathcal{T}$ for all $s$ and $\xi = x, y, z$ (not shown). Moreover, by looking at the kinetic energy along the three directions at $t=T$  (also not shown), i.e. when thermalisation is achieved, we found that the temperature dependence along $\xi=x$ and $\xi=y$ presents the same minimum of $\mathcal{T}_\mathrm{kin}$ while along $\xi=z$ it is monotonic. We can therefore attribute the emergence of the minimum to the presence of the trap and, in particular, to the interference between terms with different $s$ due to the radiofrequency-induced micromotion. Finally, in Fig.~\ref{fig:pseudopotential} we observe that when the regularized polarization potential is substituted with the pseudopotential, the depth of the minimum is strongly decreased for $n_{\mathrm{t}}=10^{12}\mathrm{cm}^{-3}$ (orange data), and even barely visible at $n_{\mathrm{t}}=10^{13}\mathrm{cm}^{-3}$ (light blue data). We thus conclude that the long-range character of the atom-ion potential renders the occurrence of the minimum in the kinetic energy more pronounced.\\[1ex]
\textit{Density dependence.} 
While the overall behavior is not substantially affected by the value of the density, there are some differences that it is worth to remark. First, for large densities, the ion temperature in the case of $^6$Li (green diamonds in Fig.~\ref{fig:Lithium_n18n19}) is slightly lower at very low $\mathcal{T}$. This difference, though, is not substantial and is definitely negligible compared to the scale of $s$-wave energy threshold. Another difference concerns the contribution of the Lamb-shift (purple squares), which is enhanced at large densities, thus confirming what we discussed in section \ref{sec:dadq} and in the previous paragraph. Both of these differences, though, are only visible at very low temperatures. At high temperatures, nor the density and neither the statistics of the gas influence the result in a sensible way, a part from the time required for thermalization that, as anticipated, increases linearly with the decrease of the density.\\[1ex]
\textit{Damping rates.} We investigated the temperature dependence of the damping rates, $\gamma_x$, of $\langle \hat r_x \rangle$. In Fig.~\ref{fig:damping_rates} they are shown for $^6$Li and $^7$Li with a density $n_{\mathrm{t}}=10^{13}\mathrm{cm}^{-3}$. Before commenting on the result, let us briefly explain the procedure we followed in order to calculate the values of $\gamma_x$. We started from a non zero initial condition for $\langle \hat{r}_x\rangle$. Its value is not particularly important, because the damping rates do not rely on it anyway. By solving the system in Eq.~(\ref{eq:EOM1stMom}), we obtain $\langle \hat{r}_x(t)\rangle$, whose behavior is a damped oscillation with zero average value. 
We calculated the curve enveloping the oscillation and we fitted it with the exponential function $\alpha_x\,e^{-\gamma_x t}$ (the same procedure was applied to all other directions with similar findings). Interestingly, the temperature dependence of the coefficients $\gamma_\xi$ does not reproduce that of $\mathcal{T}_\mathrm{kin}$. In the case of $^6$Li (fermion), the curve is monotonic and so it is for $^7$Li (boson) above $\mathcal{T}_\mathrm{c}^0$. In the bosonic case, when the gas temperature is reduced below $\mathcal{T}_\mathrm{c}^0$, the damping rates increase with the density of the condensate. The grey circles in the inset of Fig.~\ref{fig:damping_rates} show that in the Fr\"ohlich model there is a one-to-one correspondence between the damping rates and the condensate density. The extended Fr\"ohlich model (brown squares) enhances the damping rates when the contribution of the normal part of the gas becomes stronger. This relation between the condensate density and the values of $\gamma_\xi$ strongly underlines the difference between bosonic and fermionic baths at low temperatures. Moreover, it could be exploited in experiments, where the condensate fraction may be extracted from the measurement of the damping rates.\\[1ex]
\textit{Scattering length dependence.} Figure~\ref{fig:Evsa} shows the dependence of the ion energy on the atom-ion scattering length at $\mathcal{T}=100\,\mathrm{nK}$ in the case of $^6$Li. As it can be seen in the top panel, for some values of the scattering length convergence is already achieved at $T=6\,\mathrm{ms}$. When the value of $a_\mathrm{ia}^s$ approaches roughly $-0.5R^*$, the thermalization time strongly increases, as shown in the bottom panel (red dotted line). Such (numerical) observation suggests an instability that could be related to the occurrence of a resonance, as experimentally observed recently~\cite{weckesser2021observation}. 
Indeed, although the master equation does not entail information about two-particle bound states, the behavior of the final energy could still give some hint about the microscopic dynamics, the latter emerging through the parameters of the atom-ion potential. As a reference, the grey dashed line shows the values of the converged ion temperature in the case of the pseudopotential approximation, where the only parameter involved is the scattering length. The curve is symmetric because $f(k)\propto-a_\mathrm{ia}$ and the latter enters only via $|f(k)|^2$. Hence, the long-ranged character of the polarisation potential together with the fact that there is no separation of length scales in the impurity system crucially affects the ion dynamical properties. 
\\
Let us conclude by mentioning that recently the polaronic properties of a free ion in a condensate have been investigated~\cite{Gregory2021}. There, quite different polaronic states have been identified on the basis that the atom-ion polarisation potential supports either one bound state or none. Without a two-body bound state, a polaron resembling that of a neutral impurity, albeit with a larger effective mass, has been found. Here, we have also investigated the scenario for which the potential does not support any bound state. In this case, however, we found for a broad range of negative atom-ion scattering lengths that the ion does not thermalise, that is, its dynamics is very unstable in the Paul trap. Thereby, even though in the framework of the master equation we cannot make quantitative predictions on the dynamical formation of many-body bound states such as its size, the presence of two-body bound states and the inclusion of the extended Fr\"ohlich model in the master equation description is of paramount importance for stabilising the ion dynamics in the rf-trap.


\subsection{Ion in a sodium gas}
\label{sec:Na}

\begin{figure}
\centering
\includegraphics[width=0.5\textwidth]{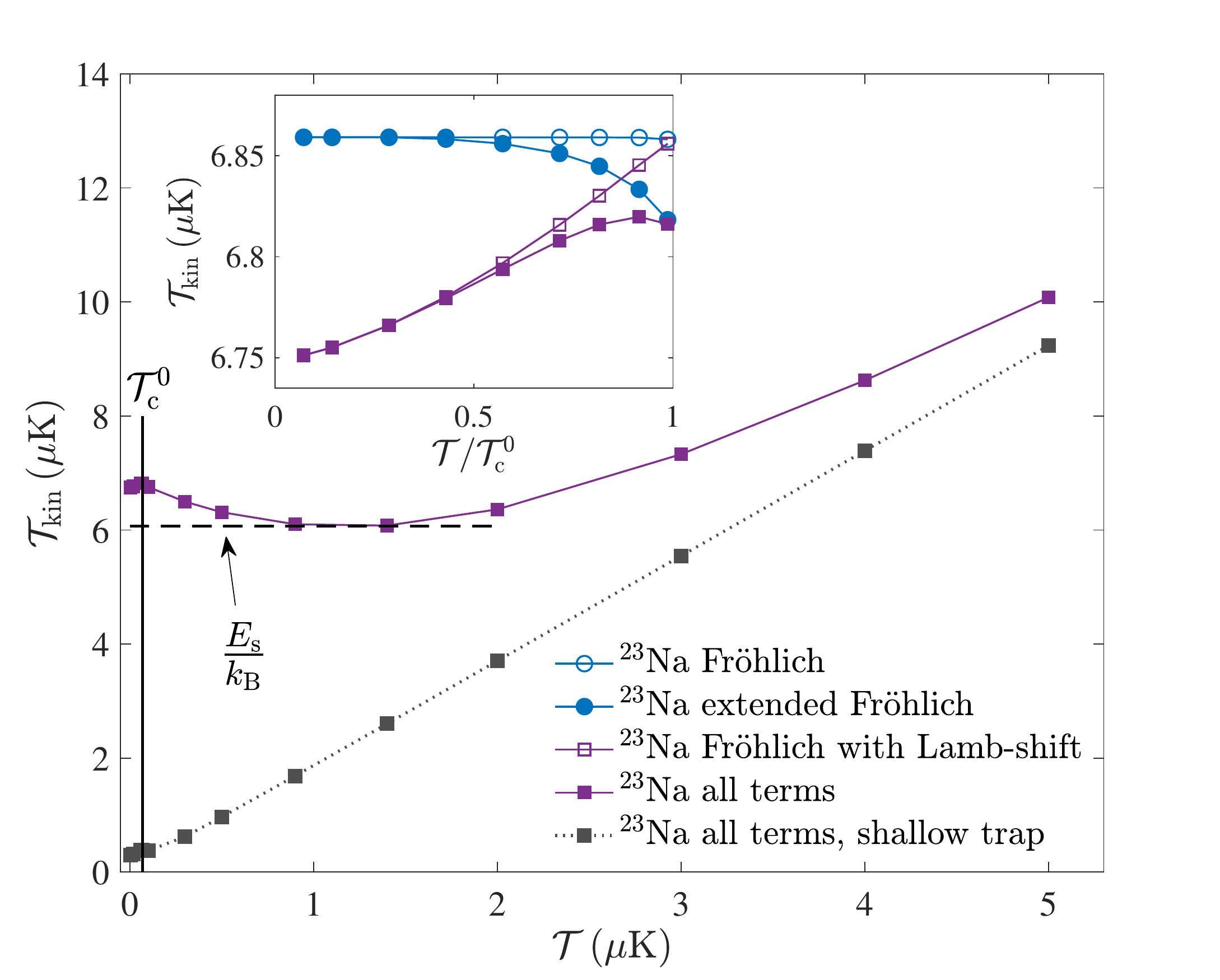}
\caption{\label{fig:Na23-lowtemp-018-b} (Color online). Ion temperature obtained from the averaged energy~(\ref{eq:meanH-Trf}) for $b \simeq 0.0780 \,R^\star$, $c\simeq 0.2239\,R^\star$, which correspond to $a_{\mathrm{ia}}^s \simeq R^\star$, and a total density $n_{\mathrm{t}} = 10^{12}\,\text{cm}^{-3}$. The grey squares with dotted line correspond to the shallow trap (see text for parameters). The black vertical lines indicate the position of the critical temperature of condensation $\mathcal{T}_{\mathrm{c}}^0$, while the dashed horizontal one to the $s$-wave threshold.}
\end{figure}

We have also investigated the ion dynamics in a heavier bosonic ensemble. In Fig.~\ref{fig:Na23-lowtemp-018-b} we illustrate the result of this analysis. With the linear Paul trap that we have assumed so far, the ion energy as a function of the gas temperature is shown by the purple squares. As it can be seen, the ion energy is always above the s-wave threshold and therefore no ultracold atom-ion collisions can be expected. 
Nonetheless, by reducing the rf-frequency by an order of magnitude, that is, $\Omega_{rf} = 2\pi\times$ 200 kHz, and by reducing the $q$-parameter by one half, i.e. $q_x = -q_y = 0.1$, which result in the trap frequencies $\nu_x \simeq 2 \pi\times$ 7 kHz, $\nu_y \simeq 2 \pi\times$ 6 kHz, $\nu_z \simeq 2 \pi\times$  4 kHz  and $\beta_{x,y} \simeq 0.0634$, $\beta_z \simeq 0.0447$, we find that at low gas temperatures (i.e. $\lesssim 400\,$nK) the $s$-wave limit can be beaten, as shown by the grey squares with dotted line in Fig.~\ref{fig:Na23-lowtemp-018-b}. The resulting trap frequencies indicate an almost isotropic ion trap. It has to be noticed, however, that with such a shallow Paul trap it will be experimentally challenging to suppress excess micromotion and to keep under control the ion heating due to the background electric noise. 

\section{Summary of the main results and conclusions}
\label{sec:disc}
We have investigated the quantum non-equilibrium dynamics of an ion in an rf-trap superimposed to a quantum gas of either bosons or fermions. To this end, in Sec.~\ref{sec:LDA-PL_approx}, we developed a quantum master equation by including the contribution of the Lamb-shift and the extended Fr\"ohlich model, that is, the non-condensate fraction. The final master equation can be found in Eq.~\eqref{eq:ME-LDA-SIunits}, where the definitions of the corresponding functions discern the case of the fermionic and bosonic bath. The equations of motion for the second and first moments are calculated in Sec.~\ref{sec:2ndMom} directly from the master equation and are given in Eq.~\eqref{eq:EOM2ndMom} and Eq.~\eqref{eq:EOM1stMom}, respectively. These systems of equations were numerically solved resulting in the findings exposed in Sec.~\ref{sec:results}. We also note that the master equation~\eqref{eq:ME5} is also an important result of our study, as it can be the starting point for other investigations, such as a free ion in a Bose-Einstein condensate.
We found significant differences in the ion dynamics between the bosonic and fermionic bath at low temperatures ($\mathcal{T}\lesssim\mathcal{T}_\mathrm{c}^0$), where the quantum nature of the gas emerges. As shown in Fig.~\ref{fig:Lithium_n18n19}, in this regime, a fermionic environment seems to ensure better cooling of the ion compared to the bosonic one, while at higher temperatures the difference gets less and less pronounced and it vanishes when $\mathcal{T}\gg\mathcal{T}_\mathrm{c}^0$. Similarly, the nature of the gas affects the damping of the ion below $\mathcal{T}_\mathrm{c}^0$ (see Fig.~\ref{fig:damping_rates}): the temperature dependence of the damping rates in the bosonic case is strongly characterised by the presence of the condensate and reflects the arising of its density. Moreover, we observed that the thermalisation time strongly increases for values of the scattering length around $-0.5R^\star$ (see Fig.~\ref{fig:Evsa}) and thermalisation may not even be achievable if the potential does not support any bound state. The former might be related to the occurrence of a resonance as recently found experimentally~\cite{weckesser2021observation} and it will be a subject of future investigations. Furthermore, as illustrated in Fig.~\ref{fig:Na23-lowtemp-018-b}, we found that a proper choice of the ion trap parameters enables to cool the ion motion in a sodium gas to the quantum regime, which affords prospects to quantum simulate impurity physics with large atom-ion mass ratios.
Finally, with the developed theoretical methods it will be interesting to investigate how the gas quantum statistics affects the Fock-state distribution of the ion motion, the coherence of ionic motional superpositions, and to develop interferometric protocols for measuring the gas temperature by reading out the thermal phonon distribution.



\section*{Acknowledgements}

This work is supported by the project ``NE 1711/3-1" of the Deutsche Forschungsgemeinschaft. We acknowledge H. F\"urst, Z. Idziaszek, and K. Jachymski for discussions. R.G. was supported by the Dutch Research Council (Vrije Programma 680.92.18.05). 


\appendix 
\section{Parameters of the regularised potential}
\label{sec:bc_reg}

To determine the parameters $b$ and $c$ of the regularised potential~(\ref{eq:Vaireg}), we follow the approach of Ref.~\cite{KrychPRA15}. Here, however, provide details that were not discussed in that reference. 

Since we have two free parameters, we need two physical conditions to determine them. To this aim, we impose that: 
\begin{itemize}
\item[(a)] The scattering length amplitude in first-order Born approximation~(\ref{eq:fq}) is exactly equal to minus the atom-ion scattering length at zero-energy; 
\item[(b)] The potential~(\ref{eq:Vaireg}) supports one bound state only. 
\end{itemize}
The condition (b) is motivated by that fact that the energy separation between bound states of the atom-ion polarisation potential~(\ref{eq:Vai}) is rather large (order of hundreds of $E^\star$), thus rendering very unlikely the population of deeper bound states at typical atomic gases densities.

In the zero-energy limit the three-dimensional s-wave ion-atom scattering length is indeed defined as 
\begin{align}
\label{eq:aisl}
a_{\mathrm{ia}}^s = -\lim_{q\rightarrow 0} f_s(q)
\end{align}
with $f_s(q)$ being the full s-wave scattering amplitude at energy $\hbar^2 q^2/(2 \mu)$, where $q = \vert\mathbf{k} - \mathbf{k}^\prime\vert$ is the magnitude of the momentum transfer in the relative frame of reference. Hence, the first aforementioned condition (a) reads 
\begin{align}
\label{eq:-f0}
a_{\mathrm{ia}}^s = -f(0).
\end{align}
In the zero-energy limit $q\rightarrow 0$, and therefore, by expanding the exponential functions in the last line of Eq.~(\ref{eq:fq}) to first order, we obtain 
\begin{align}
\label{eq:f0}
f(0) = \pi (R^\star)^2\frac{(b^2 + 2 b c - c^2)}{4 b (b + c)^2}.
\end{align}
Note that the scattering amplitude has the units of a length, which is consistent with the definition~(\ref{eq:aisl}). 

\begin{figure}
\centering
\includegraphics[scale=0.5]{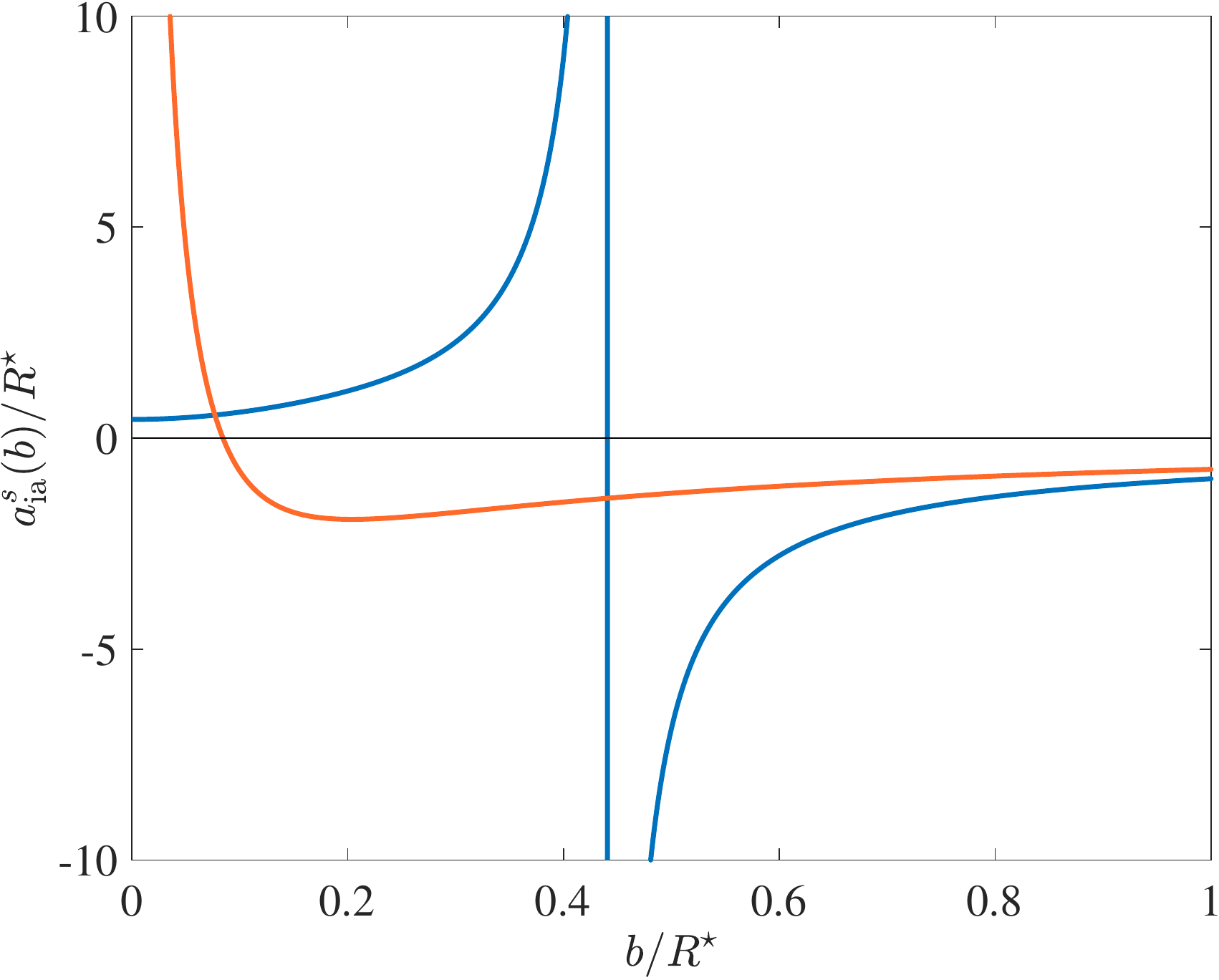}
\caption{\label{fig:scattering} (Color online). Atom-ion scattering length (blue line) computed via Eq.~(\ref{eq:aIA_MOD}) and minus the scattering amplitude (orange line) at zero-energy~(\ref{eq:f0}) as a function of the $b$ parameter of the regularised atom-ion interaction~(\ref{eq:Vaireg}). Here, we have chosen $c = 0.2039 R^\star$. The two lines intersect at $b \simeq 0.0770 R^\star$, which yields a scattering length $a_{\mathrm{ia}}^s \simeq 0.5385 R^\star$.}
\end{figure}

The fulfillment of the second condition (b) is attained by determining the s-wave scattering length as a function of either $b$ or $c$ by solving the scattering problem at zero-energy. 
To this aim, we solve numerically the radial time-independent Schr\"odinger equation
\begin{align}
\label{eq:SEscatt_SI}
\left[-\frac{\hbar^2}{2\mu}\frac{d^2}{d r^2}  + V_{\mathrm{ai}}^{r}(r)  \right]\psi(r)=0\qquad r\in [0,+\infty),
\end{align}
which in the $E^\star$ and $R^\star$ units reduces to 
\begin{align}
\label{eq:SEscatt}
\left[\frac{d^2}{d r^2}  +  \frac{r^2 - c^2}{r^2 + c^2} \frac{1}{(b^2 + r^2)^2} \right]\psi(r)=0. 
\end{align}
This differential equation is solved with boundary conditions $\psi(0) = 0$ and $\psi^\prime(0) = \epsilon$, where $\epsilon$ is a small number (e.g., 0.1). We note, however, that the result does not rely on the particular numerical value of $\epsilon$, as we have verified numerically. 
Thus, we fix the value of the parameter $c$ (in units of $R^\star$) and we solve iteratively Eq.~(\ref{eq:SEscatt}) for different values of the parameter $b$ by evaluating the corresponding scattering length, which becomes a function of $b$. We do the same for the scattering amplitude~(\ref{eq:f0}) and thus search for the value of $b$ where $a_{\mathrm{ia}}^s(b)$ and $- f(b)$ do cross, particularly where the first zero-energy resonance occurs, which indicates that we have one bound state only (see Fig.~\ref{fig:scattering}). 

Let us now briefly explain some details about the numerical calculation of the scattering length. First, we note that the solution to Eq.~(\ref{eq:SEscatt}) behaves like $r - a_{\mathrm{ia}}^s(b)$ at large distances, where the atom-ion interaction vanishes. Hence, we have 
\begin{align}
\label{eq:aIA_MOD}
a_{\mathrm{ai}}^s(b) = \lim_{r\rightarrow + \infty} \left[r - \frac{\psi(r)}{\psi^{\prime}(r)}\right].
\end{align}
Numerically, we have noticed that a large grid size has to be chosen (a few thousands of $R^\star$) such that the term on the right-hand side of the limit~(\ref{eq:aIA_MOD}) converges to a constant value, i.e. it is $r$-independent. An example of such a calculation is given in Fig.~\ref{fig:scattering}. \\For instance, by fixing $c=0.2239\,R^\star$, we find $b \simeq 0.078\,R^\star$, which yield $a_{\mathrm{ai}}^s \simeq 1.0054\,R^\star$. 
Finally, let us remark that such a strategy relies crucially on the first-Born approximation. Other strategies can be adopted in order to relax the latter (see, e.g., Refs.~\cite{IdziaszekPRA07,IdziaszekNJP11}). We chose, however, the method outlined above for consistency, since we make use of the first-Born scattering amplitude in the derivation of the master equation, as a consequence of the perturbative description of the open system.

\section{Thermal averages}
\label{sec:appendix:thermalave}

The thermal averages of the double commutator~(\ref{eq:ME3bis}) that yield non-zero contributions are:
\begin{align}
&\langle \tilde\Gamma^\dag_{\vek{q}}(t)\tilde\Gamma_{\vek{q}^\prime}(t)
\tilde\Gamma^\dag_{\vek{k}}(t^\prime)\tilde\Gamma_{\vek{k}^\prime}(t^\prime)\rangle_{B_0} = 
e^{\frac{i}{\hbar}[(\varepsilon({\vek{q}})-\varepsilon({\vek{q}^\prime}))t + 
(\varepsilon({\vek{k}})-\varepsilon({\vek{k}^\prime}))t^\prime]} \nonumber\\
&\times[
n_{\vek{q}}\delta_{\vek{q}^\prime,\vek{k}}\delta_{\vek{q},\vek{k}^\prime}
+ n_{\vek{q}} n_{\vek{q}^\prime}(
\delta_{\vek{q}^\prime,\vek{q}}\delta_{\vek{k},\vek{k}^\prime}
+\delta_{\vek{q}^\prime,\vek{k}}\delta_{\vek{q},\vek{k}^\prime}
)
]
\nonumber\\
&\langle \tilde\Gamma^\dag_{\vek{q}}(t)\tilde\Gamma_{\vek{q}^\prime}(t)
\tilde\Gamma_{\vek{k}}(t^\prime)\tilde\Gamma^\dag_{\vek{k}^\prime}(t^\prime)\rangle_{B_0} = 
e^{\frac{i}{\hbar}[(\varepsilon({\vek{q}})-\varepsilon({\vek{q}^\prime}))t + 
(\varepsilon({\vek{k}^\prime})-\varepsilon({\vek{k}}))t^\prime]} \nonumber\\
&\times
n_{\vek{q}} (1 + n_{\vek{k}^\prime})
(
\delta_{\vek{q}^\prime,\vek{q}}\delta_{\vek{k},\vek{k}^\prime}
+\delta_{\vek{q}^\prime,\vek{k}^\prime}\delta_{\vek{q},\vek{k}}
)
\nonumber\\
&\langle \tilde\Gamma^\dag_{\vek{q}}(t)\tilde\Gamma^\dag_{\vek{q}^\prime}(t)
\tilde\Gamma_{\vek{k}}(t^\prime)\tilde\Gamma_{\vek{k}^\prime}(t^\prime)\rangle_{B_0} = 
e^{\frac{i}{\hbar}[(\varepsilon({\vek{q}})+\varepsilon({\vek{q}^\prime}))t - 
(\varepsilon({\vek{k}^\prime})+\varepsilon({\vek{k}}))t^\prime]} \nonumber\\
&\times
n_{\vek{q}} n_{\vek{q}^\prime}
(
\delta_{\vek{q}^\prime,\vek{k}}\delta_{\vek{q},\vek{k}^\prime}
+\delta_{\vek{q}^\prime,\vek{k}^\prime}\delta_{\vek{q},\vek{k}}
)
\nonumber\\
&\langle \tilde\Gamma_{\vek{q}}(t)\tilde\Gamma_{\vek{q}^\prime}(t)
\tilde\Gamma^\dag_{\vek{k}}(t^\prime)\tilde\Gamma^\dag_{\vek{k}^\prime}(t^\prime)\rangle_{B_0} = 
e^{\frac{i}{\hbar}[(\varepsilon({\vek{k}})+\varepsilon({\vek{k}^\prime}))t^\prime - 
(\varepsilon({\vek{q}^\prime})+\varepsilon({\vek{q}}))t]} \nonumber\\
&\times
(1 + n_{\vek{q}} + n_{\vek{q}^\prime} + n_{\vek{k}} n_{\vek{k}^\prime})
(
\delta_{\vek{q}^\prime,\vek{k}}\delta_{\vek{q},\vek{k}^\prime}
+\delta_{\vek{q}^\prime,\vek{k}^\prime}\delta_{\vek{q},\vek{k}}
)
\nonumber\\
&\langle \tilde\Gamma_{\vek{q}}(t)\tilde\Gamma^\dag_{\vek{q}^\prime}(t)
\tilde\Gamma^\dag_{\vek{k}}(t^\prime)\tilde\Gamma_{\vek{k}^\prime}(t^\prime)\rangle_{B_0} = 
e^{\frac{i}{\hbar}[(\varepsilon({\vek{q}^\prime})-\varepsilon({\vek{q}}))t + 
(\varepsilon({\vek{k}})-\varepsilon({\vek{k}^\prime}))t^\prime]} \nonumber\\
&\times[
n_{\vek{q}^\prime}\delta_{\vek{q},\vek{k}}\delta_{\vek{q}^\prime,\vek{k}^\prime}
+n_{\vek{k}}\delta_{\vek{q},\vek{q}^\prime}\delta_{\vek{k},\vek{k}^\prime}\nonumber\\
&
+ n_{\vek{q}^\prime} n_{\vek{k}}(
\delta_{\vek{q}^\prime,\vek{q}}\delta_{\vek{k},\vek{k}^\prime}
+\delta_{\vek{q}^\prime,\vek{k}^\prime}\delta_{\vek{q},\vek{k}}
)
]
\nonumber\\
&\langle \tilde\Gamma_{\vek{q}}(t)\tilde\Gamma^\dag_{\vek{q}^\prime}(t)
\tilde\Gamma_{\vek{k}}(t^\prime)\tilde\Gamma^\dag_{\vek{k}^\prime}(t^\prime)\rangle_{B_0} = 
e^{\frac{i}{\hbar}[(\varepsilon({\vek{q}^\prime})-\varepsilon({\vek{q}}))t + 
(\varepsilon({\vek{k}^\prime})-\varepsilon({\vek{k}}))t^\prime]} \nonumber\\
&\times[
n_{\vek{k}}\delta_{\vek{q},\vek{k}^\prime}\delta_{\vek{q}^\prime,\vek{k}}
+(1 + n_{\vek{k}} + n_{\vek{q}})\delta_{\vek{q},\vek{q}^\prime}\delta_{\vek{k},\vek{k}^\prime}
\nonumber\\
&+ n_{\vek{q}^\prime} n_{\vek{k}^\prime}(
\delta_{\vek{q}^\prime,\vek{q}}\delta_{\vek{k},\vek{k}^\prime}
+\delta_{\vek{q}^\prime,\vek{k}}\delta_{\vek{q},\vek{k}^\prime}
)
].
\end{align}
For these identities we used the relation (4.7) of Ref.~\cite{EvansNPB96}. 


\section{Ion motion in a Paul trap}
\label{sec:appendix:ionmotion}

Here, we provide details on the analytical solution of the ion motion in a Paul trap using the notation of the review~\cite{Leibfried2003}. The goal is to provide the relevant steps of its derivation such that the interested reader can implemented them in numerics quickly and efficiently. 


\subsection{Classical solution of a charge in a Paul trap}
\label{sec:nuj}

Let us consider a particle of mass $M$ and charge $Z\vert e\vert$ in the quadruple field
\begin{align}
\label{eq:quadrupole}
\Phi(x,y,z,t) & = \frac{U}{2} (\alpha x^2 + \beta y^2 +\gamma z^2) \nonumber\\
\phantom{=} &+ \frac{\tilde{U}}{2}\cos(\Omega_{rf} t) (\alpha^\prime x^2 + \beta^\prime y^2 +\gamma^\prime z^2).
\end{align}
For a linear Paul trap we have: $0<\gamma = - (\alpha+\beta)$, $\alpha^\prime = - \beta^\prime$ and $\gamma^\prime = 0$ and Poisson equation, $\nabla^2\Phi = 0$, is fulfilled. The Newton equation of motion along the $x$-direction is given by (similarly for the other directions):
\begin{align*}
\ddot x(t) = -\frac{\vert e\vert Z}{m} \frac{\partial \Phi}{\partial x} =  -\frac{\vert e\vert Z}{m}\left[ U \alpha +\tilde{U}\alpha^\prime\cos(\Omega_{rf} t)\right] x.
\end{align*}
By introducing the dimensionless variable $\tau = \Omega_{rf} t /2$, the corresponding rescaled equation reads
\begin{align}
\label{eq:mathieu}
\ddot x(\tau) +\left[ a_x - 2 q_x \cos(2\tau)\right] x = 0,
\end{align}
where the newly introduced parameters are defined as: $a_x = 4\vert e\vert Z \alpha U/(M \Omega_{rf}^2)$ and $q_x = -2 \vert e\vert Z \alpha^\prime \tilde{U}/(M \Omega_{rf}^2)$. 
Note that for a linear Paul trap we have: $q_y=-q_x\equiv q$ and $q_z=0$, where $a_y = a_x = -a_z/2 \equiv a$. 
The above outlined equation is solved by using the following ansatz
\begin{align}
\label{eq:ansatz}
x(\tau) = A_x e^{i\beta_x \tau}\!\!\!\!  \sum_{n=-\infty}^\infty C_{2 n}^x e^{i2 n\tau} + B_x e^{-i\beta_x \tau} \!\!\!\!  \sum_{n=-\infty}^\infty C_{2 n}^x e^{-i2 n\tau}, 
\end{align}
where $A_x,\,B_x$ are constants that depend on the initial conditions, while the parameter $\beta_x$ and coefficients $C_{2n}^x$ have to be determined recursively. 
Here, we use the same notation of Ref.~\cite{Leibfried2003}, but we note that the coefficients $C_{2 n}^x$ could have been named $C_{n}^x$, as we actually do in Eq.~(\ref{eq:IonPosEvolution}). To this end, we insert the ansatz into Eq.~(\ref{eq:mathieu}) and we obtain 
\begin{align}
\label{eq:recursive}
C_{2n+2}^x - D_{2n}^x C_{2n}^x + C_{2n -2}^x = 0 \,\,\,\,\,\, D_{2n}^x = \frac{a_x -(\beta_x+2n)^2}{q_x}.
\end{align}
Iterative application of the above identities yield the continued fraction solutions
\begin{align}
\frac{C_{2n}^x}{C_{2n+2}^x} = \frac{1}{D_{2n}^x-\frac{1}{D_{2n-2}^x-\frac{1}{D_{2n-4}^x-\dots}}},\nonumber\\
\frac{C_{2n}^x}{C_{2n-2}^x} = \frac{1}{D_{2n}^x-\frac{1}{D_{2n+2}^x-\frac{1}{D_{2n+4}^x-\dots}}}.
\end{align}
With these expressions and Eq.~(\ref{eq:recursive}), we obtain
\begin{align}
D_{2n}^x = \frac{1}{D_{2n-2}^x-\frac{1}{D_{2n-4}^x\dots}} + \frac{1}{D_{2n+2}^x-\frac{1}{D_{2n+4}^x-\dots}}. 
\end{align}
Since $D_0^x = (a_x - \beta_x^2)/q_x$ and $D_{2n}^x$ above, we have

\begin{align}
\label{eq:betax}
\beta_x^2 = a_x - q_x \left[\frac{1}{D_{-2}^x-\frac{1}{D_{-4}^x\dots}} + \frac{1}{D_{2}^x-\frac{1}{D_{4}^x-\dots}}\right].
\end{align}
Note that the expression in the brackets $[\dots]$ of Eq.~(\ref{eq:betax}) still depends on $\beta_x$, $a_x$ and $q_x$. Nonetheless, given $a_x$ and $q_x$, Eq.~(\ref{eq:betax}) can be solved with respect to $\beta_x$ after a few iterations of the continued fraction. 

For the numerical assessment of the coefficients $C_{2n}^x$ we proceed as follows: First, we set the maximum of the $n$-index to some positive integer $N_F$ and $C_{0}^x=1$ such that $C_{\pm 2 n}^x=0$ $\forall n> N_F$. Second, using the previously obtained value of $\beta_x$ and the formula~(\ref{eq:recursive}), we solve an inhomogeneous linear system of equations given by
\begin{widetext}
\begin{align}
\left(
\begin{array}{cccccccccccc}
D_{2N_F}^x & -1 & 0 & 0 & 0 & \dots & \dots & \dots & \dots & \dots & \dots & \dots\\
-1 & D_{2(N_F-1)}^x & -1 & 0 & 0 & \dots & \dots & \dots & \dots & \dots & \dots & \dots\\
0 & -1 & D_{2(N_F-2)}^x & -1 & 0 & \dots & \dots & \dots & \dots & \dots & \dots & \dots\\
\vdots & \vdots & \vdots & \vdots & \vdots & \vdots & \vdots & \vdots & \vdots & \vdots & \vdots & \vdots\\
\vdots & \dots & \dots & \dots & 0 & -1 & D_2^x & 0 & 0 & \dots & \dots & \dots\\ 
\vdots & \dots & \dots & \dots & 0 & 0 & 0 & D_{-2}^x & -1 & 0 & \dots & \dots\\ 
\vdots & \vdots & \vdots & \vdots & \vdots & \vdots & \vdots & \vdots & \vdots & \vdots & \vdots & \vdots\\
\vdots & \dots & \dots & \dots & \dots & \dots & \dots & \dots & & \dots 0 & -1 & D_{-2N_F}^x\\ 
\end{array}
\right)
\left(
\begin{array}{c}
C_{2N_F}^x \\
C_{2(N_F-1)}^x \\
C_{2 (N_F-2)}^x \\
\vdots \\
C_2^x \\
C_{-2}^x \\
\vdots \\
C_{-2N_F}^x\\
\end{array}
\right)
=
\left(
\begin{array}{c}
0 \\
0 \\
0 \\
\vdots \\
1 \\
1 \\
\vdots \\
0 \\
\end{array}
\right).
\end{align}
\end{widetext}
The total number of numerically calculated coefficients is therefore $2 N_F$. Moreover, the normalisation condition 
\begin{align}
\sum_{n=-\infty}^{\infty}C_{2n}^x\simeq \sum_{n=-N_F}^{N_F}C_{2n}^x = 1
\end{align}
has to be satisfied, from which we obtain the final normalised coefficients $c_{2n}^x = C_{2n}^x/\sum_{j=-N_F}^{N_F}C_{2j}^x$. 
In this way we can then immediately evaluate the reference harmonic oscillator frequency 
\begin{align}
\label{eq:nux}
\nu_x  &= \Omega_{rf}\sum_{n=-N_F}^{N_F}c_{2n}^x \left(\frac{\beta_{x}}{2}+n\right).
\end{align}
Exactly the same procedure applies to the determination of the coefficients $C_{2n}^{y,z},\,c_{2n}^{y,z}$ and frequencies $\nu_{y,z}$. We note that the coefficients appearing in the ion solution~(\ref{eq:IonPosEvolution}) are precisely the coefficients $c_{2n}^{x,y,z}$. 

\begin{figure}
\centering
\includegraphics[scale=0.5]{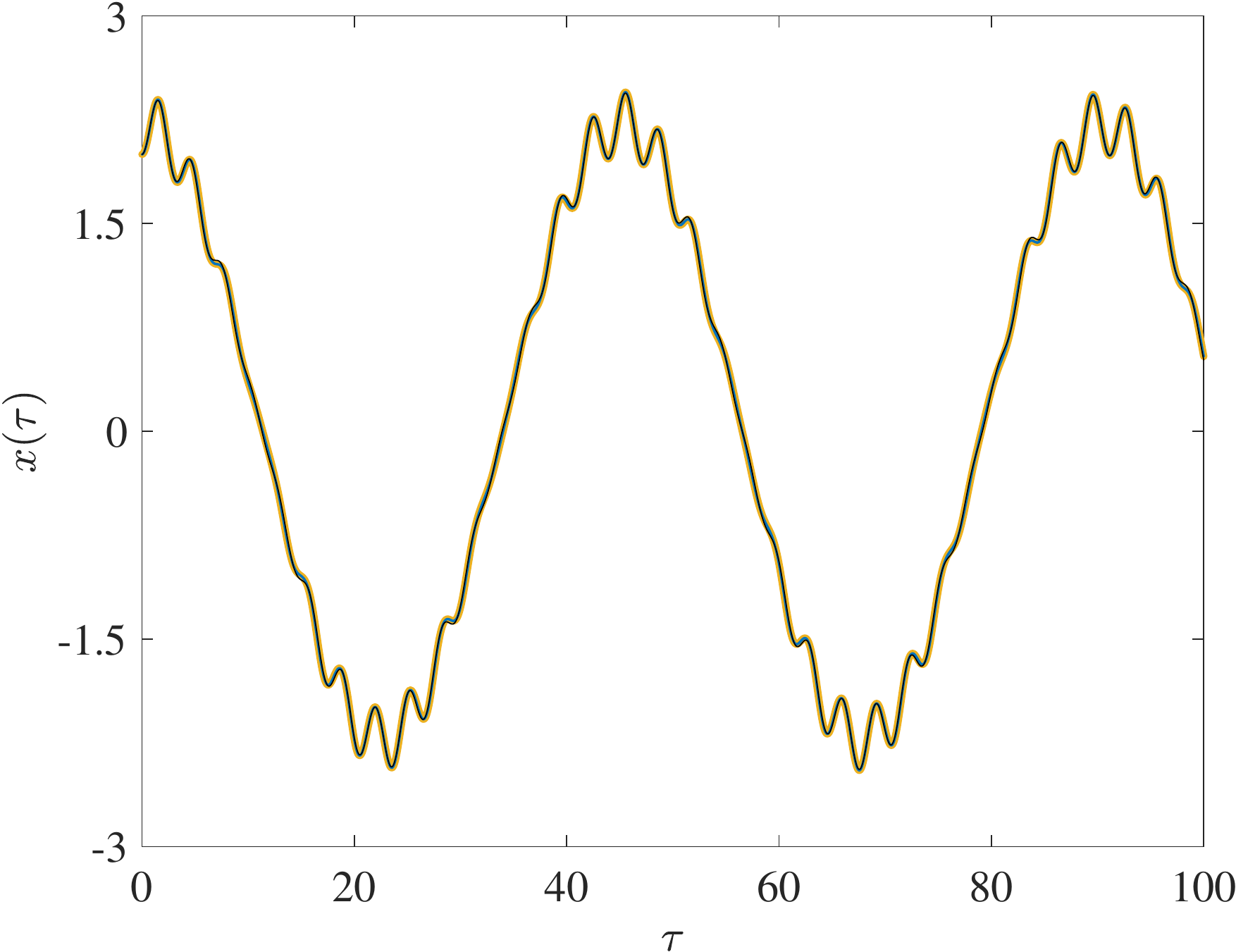}
\caption{\label{fig:comparison} (Color online). Comparison of the numerically exact solution obtained by solving the Newton equation of motion~(\ref{eq:mathieu}), the analytical solution~(\ref{eq:ansatz}), and the approximated one~(\ref{eq:solapx}). We have chosen the following parameters: $a_x = 0$, $q_x = 0.28$, and $N_F = 20$.}
\end{figure}

Finally, the classical trajectory is obtained as $x(\tau = 0) = A_x+B_x \equiv x_0$ and 
\begin{align*}
\dot x(\tau = 0) = i (A_x -B_x)\left[\beta_x + 2 \sum_{n=-N_F}^{N_F}n\,c_{2n}^x \right] \equiv \frac{p_0}{M}.
\end{align*}
From these equalities we obtain
\begin{align}
A_x =  \frac{x_0}{2} - i \frac{p_0}{2 M} \left[\beta_x + 2 \sum_{n=-N_F}^{N_F}n\,c_{2n}^x \right]^{-1},\\
B_x = \frac{x_0}{2} + i \frac{p_0}{2 M} \left[\beta_x + 2 \sum_{n=-N_F}^{N_F}nc\,_{2n}^x \right]^{-1}.
\end{align}
In the limit $a_x<\vert q_x\vert^2\ll 1$ and for $p_0=0$, an approximated solution is given by 
\begin{align}
\label{eq:solapx}
x_{\text{apx}}(\tau) = \frac{2 x_0}{2-q_x}\cos(\beta_x\tau)\left[
1-\frac{q_x}{2}\cos(2\tau)
\right].
\end{align}
This solution can be compared with the numerically obtained one from Eq.~(\ref{eq:ansatz}). In Fig.~\ref{fig:comparison} an example is shown, where we compare the solutions $x_{\text{apx}}(\tau)$ (thin black line), Eq.~(\ref{eq:ansatz}) (blue slighter ticker line than the black line), and the numerically solved Eq.~(\ref{eq:mathieu}) (tick yellow line).

\subsection{Quantum Hamiltonian}
\label{sec:ionHquant}

As it can be verified, Eq.~(\ref{eq:mathieu}) reproduces the motion of a parametric harmonic oscillator with squared frequency 
\begin{align}
\label{eq:PTfreq}
W_{\xi}(t) = \frac{\Omega_{rf}^2}{4}\left[a_{\xi} - 2 q_{\xi}\cos(\Omega_{rf}t)\right].
\end{align}
As it can be shown formally with the Heisenberg equations of motion for $\hat r_\xi$ and $\hat p_{\xi}$~\cite{Leibfried2003}, the following Hamiltonian in one spatial direction 
\begin{align}
\label{eq:Hion}
\hat H_I^{trap} = \frac{\hat p_{\xi}^2}{2M}+\frac{M}{2}W_{\xi}(t)\hat{r}_\xi^2\qquad \xi=x,y,z,
\end{align}
reproduces exactly the same equation of motion~(\ref{eq:mathieu}) for the operators $\hat r_\xi$. Thus, Eq.~(\ref{eq:Hion}) is the quantised version of the classical Hamiltonian for an ion in a Paul trap, where we have promoted the ion position and momentum variables to operators. Besides this, we note that the spatial directions are uncoupled because of the form of the quadrupole field~(\ref{eq:quadrupole}). 

The eigenfunctions of $\hat H_I^{trap}$ are given by:
\begin{align}
\psi_n(r_\xi,t) & = \left(\frac{M\nu_{\xi}}{\pi\hbar}\right)^{\frac{1}{4}}\frac{e^{-i n \nu_{\xi} t+i\frac{M}{2\hbar}\frac{\dot u_{\xi}(t)}{u_{\xi}(t)}r_\xi^2}}{\sqrt{2^n n! u_{\xi}(t)}} \nonumber\\
\phantom{=} &\times H_n\left(\sqrt{\frac{M\nu_{\xi}}{\hbar\vert u_{\xi}(t)\vert^2}} r_\xi\right)
\end{align}
where
\begin{align}
\label{eq:utnu}
u_{\xi}(t) &= e^{i\beta_{\xi}\Omega_{rf} t /2}\sum_{n=-\infty}^{\infty}C_{2n}^{\xi}e^{in\Omega_{rf} t}, 
\end{align}
with $u_{\xi}(0)=\sum_{n=-\infty}^{\infty}C_{2n}^{\xi}=1$, $\dot u_{\xi}(0) = i\nu_{\xi}$, and $\nu_{\xi}$ given by Eq.~(\ref{eq:nux}). 

Finally, as initial condition of the ion density matrix for the solution of the master equation we have chosen  
\begin{align}
\label{eq:rho_in}
\hat \rho_\xi = \ket{\psi_0(t=T_{rf})}\bra{\psi_0(t=T_{rf})}, \qquad\forall\xi = x,y,z,
\end{align}
since by starting from $\ket{\psi_0(t=0)}\bra{\psi_0(t=0)}$ and by computing the expectation value~(\ref{eq:meanHion}), one can numerically verify that the minimum of the ion energy occurs at the time $t = 2\pi/\Omega_{rf}$. This is the energy minimum we assumed in our analyses 
and Eq.~(\ref{eq:rho_in}) as initial condition. With that initial matrix we have computed the initial conditions for the moments of Sec.~\ref{sec:2ndMom}.  


\section{Ion master equation details}
\label{sec:appendix:ME}

In this section, we provide a few technical details on the calculation of the double integral as a consequence of the double summation in momentum space in Eq.~(\ref{eq:MElda}). 
The derivations are detailed for a bosonic bath only, while for a fermionic one we simply provide the final result, as they are very similar. Additionally, we provide details of the analytical calculation of the Cauchy principal value within the Fr\"ohlich model that yields the Lamb-shift. 


\subsection{Double integration in momentum space}
\label{sec:doubleintegral}

In Eq.~(\ref{eq:MElda}) we have to evaluate terms of the kind
\begin{align}
\label{eq:sumTBA1}
\sum_{\vek{q},\vek{q}^\prime}
n_{\vek{q}} (n_{\vek{q}^\prime} + 1) (q_\xi^\prime - q_\xi)^2 \vert c_{\vek{q}^\prime-\vek{q}} \vert^2
\delta(\omega_{\vek{q}^\prime} - \omega_{\vek{q}} \pm \omega_{s,\xi}),
\end{align}
where $\omega_{s,\xi} = \Omega_{rf}(\beta_\xi/2+s)$ and $ \omega_{\vek{q}} \equiv \varepsilon(\vek{q})/\hbar$. To this end, we first perform the centre-of-mass and relative coordinate transformation
\begin{align}
\vek{k} = \vek{q}^\prime-\vek{q},\qquad\qquad \vek{K} = \frac{\vek{q}^\prime+\vek{q}}{2},
\end{align}
with 
\begin{align}
\vek{q}^\prime & = \vek{K} + \frac{\vek{k}}{2}, \qquad \qquad \vek{q} = \vek{K} - \frac{\vek{k}}{2},
\end{align}
and therefore we have
\begin{align}
\omega_{\vek{q}^\prime} - \omega_{\vek{q}} = \frac{\hbar}{2 m} \left[(\vek{q}^\prime)^2 - \vek{q}^2\right] = \frac{\hbar}{m} (\vek{k} \cdot \vek{K}).
\end{align}
Given this, we can rewrite Eq.~(\ref{eq:sumTBA1}) as
\begin{align}
\label{eq:sumTBA2}
\sum_{\vek{k},\vek{K}}
n_{\vek{k},\vek{K}}^- (n_{\vek{k},\vek{K}}^+ + 1) k_\xi^2 \vert c_{\vek{k}} \vert^2
\delta\left(\frac{\hbar}{m} \vek{k} \cdot \vek{K} \pm \omega_{s,\xi}\right),
\end{align}
where
\begin{align}
n_{\vek{k},\vek{K}}^- & = 
\frac{1}{\exp\left\{
\beta_{\mathcal{T}}\left[
\frac{\hbar^2}{2m}
\left(
\vek{K}^2 + \frac{\vek{k}^2}{4} - \vek{k} \cdot \vek{K}
\right) 
-\mu_{\mathrm{G}}
\right]
\right\}
-1},
\nonumber\\
n_{\vek{k},\vek{K}}^+ & = 
\frac{1}{\exp\left\{
\beta_{\mathcal{T}}\left[
\frac{\hbar^2}{2m}
\left(
\vek{K}^2 + \frac{\vek{k}^2}{4} + \vek{k} \cdot \vek{K}
\right) 
-\mu_{\mathrm{G}}
\right]
\right\}
-1}.
\end{align}
We remind that $\mu_{\mathrm{G}}$ is the bosons' chemical potential and $\beta_{\mathcal{T}} = (k_{\mathrm{B}} \mathcal{T})^{-1}$.
Thus, we perform the continuum limit 
\begin{align}
\sum_{\vek{k},\vek{K}} \rightarrow \frac{L^6}{(2\pi)^6}\int_{\mathbb{R}^3}\mathrm{d}\vek{k} \int_{\mathbb{R}^3}\mathrm{d}\vek{K},
\end{align}
which transforms the double sum in Eq.~(\ref{eq:sumTBA2}) in the following two double integrals 
\begin{align}
\label{eq:intTBA1}
&\mathcal{I}_1 + \mathcal{I}_2 = \int_{\mathbb{R}^3}\mathrm{d}\vek{k} \int_{\mathbb{R}^3}\mathrm{d}\vek{K}\,
n_{\vek{k},\vek{K}}^- k_\xi^2 \vert c_{\vek{k}} \vert^2
\delta\left(\frac{\hbar}{m} \vek{k} \cdot \vek{K} \pm \omega_{s,\xi}\right)
\nonumber\\
&+\int_{\mathbb{R}^3}\mathrm{d}\vek{k} \int_{\mathbb{R}^3}\mathrm{d}\vek{K}\,
n_{\vek{k},\vek{K}}^- n_{\vek{k},\vek{K}}^+ k_\xi^2 \vert c_{\vek{k}} \vert^2
\delta\left(\frac{\hbar}{m} \vek{k} \cdot \vek{K} \pm \omega_{s,\xi}\right).
\end{align}
Here, we have neglected the common factor $\left(\frac{L}{2\pi}\right)^6$. In order to solve them, we first move to spherical coordinates 
\begin{align}
\label{eq:spherical}
k_x &= k \sin(\theta_k) \cos(\varphi_k),\nonumber\\
k_y &= k \sin(\theta_k) \sin(\varphi_k),\nonumber\\
k_z &= k \cos(\theta_k). 
\end{align}
Thus, the corresponding volume element is given by $\mathrm{d}\vek{k} = \mathrm{d}k \mathrm{d}\theta_k \mathrm{d}\varphi_k k^2 \sin(\theta_k)$ with $k\equiv\vert\vek{k}\vert$. While a similar change of variables applies to the centre-of-mass variable $\vek{K}$ as well, but with subscript $K$ for the angular variables, we note that we choose the ``z-axis" of the vector $\vek{K}$ along the relative variable $\vek{k}$. In such a way the scalar product appearing in the Dirac's delta can be written as
\begin{align}
\vek{k} \cdot \vek{K} = k K \cos(\theta_K).
\end{align}
Given this, the first integral becomes
\begin{align}
&\mathcal{I}_1 = 2\pi^2 (1+\delta_{z,\xi})
\int_0^\infty\mathrm{d}k k^4 \int_0^\pi \mathrm{d}\theta_k g_\xi^2(\theta_k)\sin(\theta_k) \vert c_{\vek{k}} \vert^2
\nonumber\\
&\times \int_0^\infty\mathrm{d}K K^2 \int_0^\pi \mathrm{d}\theta_K \sin(\theta_K)n_{\vek{k},\vek{K}}^- 
\nonumber\\
&\times\delta\left(\frac{\hbar}{m} k K \cos(\theta_K) \pm \omega_{s,\xi}\right),
\end{align}
where we have performed the integrations of the variables $\varphi_{k}$ and $\varphi_{K}$, since only $k_\xi$ relies on $\varphi_{k}$, while none of the functions in the integrand depend on $\varphi_{K}$. Besides this, because of $k_\xi$ we have introduced the angle function $g_\xi(\theta_k)=\delta_{z,\xi}\cos(\theta_k) + (1-\delta_{z,\xi})\sin(\theta_k)$ with $\delta_{z,\xi}$ being the Kronecker delta. Since neither $\vert c_{\vek{k}} \vert^2$ nor $n_{\vek{k},\vek{K}}^-$ rely on $\theta_k$, we can easily perform the integration
\begin{align}
\int_0^\pi \mathrm{d}\theta_k g_\xi^2(\theta_k)\sin(\theta_k) = 
\frac{2}{3}\left[
\delta_{z,\xi} + 2 (1 - \delta_{z,\xi})
\right].
\end{align}

Our next step is to integrate out the variable $K$. Towards this end, we first rewrite the Dirac's delta as
\begin{align}
&\delta\left(\frac{\hbar}{m} k K \cos(\theta_K) \pm \omega_{s,\xi}\right) = \frac{m}{\hbar k\vert\cos(\theta_K)\vert}
\nonumber\\
& \times
\delta\left(K \pm K_{s,\xi}(\theta_K,k)\right)
\delta_{0,1\pm\mathrm{sgn}(\omega_{s,\xi}\cos(\theta_K))},
\end{align}
where $\mathrm{sgn}(\cdot)$ is the sign function. The last Kronecker delta ensures that $K_{s,\xi}(\theta_K,k) = \frac{m\omega_{s,\xi}}{\hbar k \cos(\theta_K)}\ge 0$ in the minus case and $K_{s,\xi}(\theta_K,k)\leq 0$ in the plus case, since $K\in\mathbb{R}^+$ and the integral over $K$ would be zero otherwise, and so would be $\mathcal{I}_1$. Hence, we obtain
\begin{align}
\label{eq:I1}
&\mathcal{I}_1 = \frac{8}{3}\pi^2 \left(\frac{m}{\hbar}\right)^3 \omega_{s,\xi}^2
\int_0^\infty\mathrm{d}k k \vert c_{\vek{k}} \vert^2
\nonumber\\
&\times \int_0^\pi \mathrm{d}\theta_K \frac{\tan(\theta_K)n_{\vek{k},\mp\vek{K}_{s,\xi}(\theta_K,k)}^-}{\cos(\theta_K)\vert\cos(\theta_K)\vert}\delta_{0,1\pm\mathrm{sgn}(\omega_{s,\xi}\cos(\theta_K))}.
\end{align}
Finally, we perform the angular integral 
\begin{align}
\label{eq:ang_int}
&\int_0^\pi \mathrm{d}\theta_K \frac{\tan(\theta_K) n_{\vek{k},\mp\vek{K}_{s,\xi}(\theta_K,k)}^-}{\cos(\theta_K)\vert\cos(\theta_K)\vert} 
\delta_{0,1\pm\mathrm{sgn}(\omega_{s,\xi}\cos(\theta_K))} =
\nonumber\\
&\int_{-1}^1 \mathrm{d} u \frac{n_{\vek{k},\mp\vek{K}_{n,\xi}(u,k)}^-}{u^2\vert u\vert} \delta_{0,1\pm\mathrm{sgn}(\omega_{s,\xi} u)},
\end{align}
where we performed the change of variable $u = \cos(\theta_K)$. Hence,
\begin{align}
n_{\vek{k},\vek{K}_{s,\xi}(u,k)}^- & = 
\frac{1}{e^{
\frac{\beta\hbar\omega_{s,\xi}}{2 (k\ell_{s,\xi})^2}
\left[
\frac{1}{u^2} + \frac{(k\ell_{s,\xi})^4}{4}\pm(k\ell_{s,\xi})^2
\right] 
-\beta\mu_{\mathrm{G}}
}
-1}
\end{align}
with $\ell_{s,\xi}^2 = \frac{\hbar}{m\omega_{s,\xi}}$ and~(\ref{eq:ang_int}) can be rewritten as
\begin{align}
\label{eq:J}
\mathcal{J} = \int_{0}^1 \mathrm{d} u \frac{n_{\vek{k},\vek{K}_{s,\xi}(u,k)}^-}{u^3} = \int_{0}^1 \frac{\mathrm{d} u}{u^3\left(e^{\frac{\alpha_0}{u^2} + \alpha_\pm} - 1\right)},
\end{align}
where 
\begin{align}
\label{eq:alpha0pm}
\alpha_0 & = \frac{\beta_{\mathcal{T}}\hbar\omega_{s,\xi}}{2 (k\ell_{s,\xi})^2},\nonumber\\
\alpha_\pm & = \alpha_0 \left[
\frac{(k\ell_{s,\xi})^4}{4} \pm (k\ell_{s,\xi})^2
\right] 
-\beta_{\mathcal{T}}\mu_{\mathrm{G}}.
\end{align}
In order to solve $\mathcal{J}$, we perform the change of variable:
\begin{align}
\label{eq:var-change}
 z = \frac{\alpha_0}{u^2} + \alpha_\pm\,\, \Longrightarrow \,\, \mathrm{d} u = - \frac{\sqrt{\alpha_0}}{2}\frac{\mathrm{d} z}{(z-\alpha_\pm)^{3/2}}.
 \end{align}
 In such a way $\mathcal{J}$ is rewritten as 
 \begin{align*}
 \mathcal{J} = \frac{1}{2\alpha_0} \int_{\alpha_0+\alpha_\pm}^{+\infty} \frac{\mathrm{d} z}{e^z - 1} = 
 \frac{\alpha_0+\alpha_\pm-\ln(e^{\alpha_0+\alpha_\pm}-1)}{2\alpha_0},
 \end{align*}
 which holds as long as $\alpha_0>0$, as it is indeed the case. Hence, the integral~(\ref{eq:I1}) is given by
\begin{align}
\label{eq:I1final}
&\left(\frac{L}{2\pi}\right)^6 \mathcal{I}_1^\pm = \frac{1}{6} \left(\frac{m}{\pi\mu}\right)^2 \frac{\hbar}{\beta_{\mathcal{T}}}\, \mathcal{F}_{s,\xi}^{(1),\pm}
\end{align}
with 
\begin{align}
\label{eq:Fnj1}
&\mathcal{F}_{s,\xi}^{(1),\pm} = \int_0^\infty\mathrm{d}k\,k^3 \frac{\alpha_0(k)+\alpha_\pm(k)-\ln(e^{\alpha_0(k)+\alpha_\pm(k)}-1)}{\vert f(k) \vert^{-2}}.
\end{align}
The integral over $k$ in Eq.~(\ref{eq:Fnj1}) is computed numerically. We note that $\alpha_0(k)\sim k^{-2}$ and therefore the exponential diverges for $k\rightarrow 0$, which is not the case for $\alpha_\pm(k)\sim k^{2}$ that tends to zero. Because of the logarithm, however, the exponent of the exponential function compensates the -$\alpha_0(k)$ on the left-hand-side of the logarithm so that the overall behaviour of the integrand is zero when $k\rightarrow 0$. Instead, when $k\rightarrow\infty$, we have $\alpha_0(k)\rightarrow 0$, while $\alpha_\pm(k)$ diverges. For the same argument as before, the function of the integrand numerator tends to zero. Therefore, the integral converges, even if $\vert f(k) \vert^2 = 1$. For the fermionic bath, we get the expression 
\begin{align}
\label{eq:Fnj1_fermi}
&\mathcal{F}_{s,\xi}^{(1),\pm} = \int_0^\infty\mathrm{d}k\, k^3 \frac{\ln(1+e^{\alpha_0(k)+\alpha_\pm(k)})-\alpha_0(k)-\alpha_\pm(k)}{\vert f(k) \vert^{-2}}.
\end{align}
The result is very similar to the bosonic case, but one has to remember that the chemical potentials are different, especially for temperature below the Fermi temperature and the critical temperature for condensation.  

The integral $\mathcal{I}_2$ is almost the same, but $\mathcal{J}$ in Eq.~(\ref{eq:J}) is defined now as
\begin{align}
\mathcal{J} & = \int_{0}^1 \mathrm{d} u \frac{n_{\vek{k},\vek{K}_{s,\xi}(u,k)}^- n_{\vek{k},\vek{K}_{s,\xi}(u,k)}^+}{u^3} \nonumber\\
&= \int_{0}^1 \frac{\mathrm{d} u }{u^3\left(e^{\frac{\alpha_0}{u^2} + \alpha_-} - 1\right) \left(e^{\frac{\alpha_0}{u^2} + \alpha_+} - 1\right)}.
\end{align}
To solve it, we first perform the change of variable~(\ref{eq:var-change}), which yields
\begin{align}
\mathcal{J} & = \frac{1}{2\alpha_0} \int_{\alpha_0+\alpha_-}^{+\infty}  \frac{\mathrm{d} z}{(e^z - 1)(e^{z+\alpha_+ - \alpha_-}-1)}.
\end{align}
Thus, we perform the additional change of variable $y = e^z$ with d$z$=d$y/y$, and we obtain
\begin{align}
\label{eq:J_I2}
\mathcal{J} & = \frac{1}{2\alpha_0 e^{\alpha_+ - \alpha_-}} \int_{a}^{+\infty}  \frac{\mathrm{d} y}{y(y - 1)(y-b)},
\end{align}
with $a=e^{\alpha_0 + \alpha_-}$ and $b = e^{\alpha_- - \alpha_+}$. The integral in Eq.~(\ref{eq:J_I2}) can be solved analytically, which finally gives
\begin{align}
\label{eq:J_I2final}
\mathcal{J} & = \frac{ \ln(e^{\alpha_0 + \alpha_-} - e^{\alpha_- - \alpha_+}) -\alpha_0 - \alpha_-}{2\alpha_0 (1 - e^{\alpha_- - \alpha_+})} \nonumber\\
&-\frac{e^{\alpha_- - \alpha_+}\Re\{ \mathrm{arctanh}(1 - 2 e^{\alpha_0 + \alpha_-}) \}}{\alpha_0 (1 - e^{\alpha_- - \alpha_+})}.  
\end{align}
Hence, $\mathcal{I}_2$ preserves the structure of $\mathcal{I}_1$, that is,
\begin{align}
\label{eq:I2final}
&\left(\frac{L}{2\pi}\right)^6 \mathcal{I}_2 = \frac{1}{6} \left(\frac{m}{\pi\mu}\right)^2 \frac{\hbar}{\beta_{\mathcal{T}}}\, \mathcal{F}_{s,\xi}^{(2)},
\end{align}
where the radial integration in momentum space is given by
\begin{widetext}
\begin{align}
\label{eq:Fnj2}
&\mathcal{F}_{s,\xi}^{(2)} = \int_0^\infty\mathrm{d}k\, k^3 \vert f(k) \vert^{2}\,\left[
\frac{ \ln(e^{\alpha_0 + \alpha_-} - e^{\alpha_- - \alpha_+}) -\alpha_0 - \alpha_- -2 e^{\alpha_- - \alpha_+}\Re\{ \mathrm{arctanh}(1 - 2 e^{\alpha_0 + \alpha_-})\}}
{1 - e^{\alpha_- - \alpha_+}}
\right].
\end{align}
For the fermions we obtain a similar expression
\begin{align}
\label{eq:Fnj2_fermi}
&\mathcal{F}_{s,\xi}^{(2)} = \int_0^\infty\mathrm{d}k\, k^3 \vert f(k) \vert^{2}\,\left[
\frac{e^{\alpha_- - \alpha_+} \ln(1+e^{\alpha_0 + \alpha_-}) - \ln(e^{\alpha_- - \alpha_+}+e^{\alpha_0 + \alpha_-})}
{e^{\alpha_- - \alpha_+}-1}
-\alpha_0(k)-\alpha_-(k)
\right].
\end{align}
We note that in the numerical assessment of $\mathcal{F}_{s,\xi}^{(2)}$ we found that both for the bosons and the fermions the integral is essentially zero, as a consequence of the large numerical values taken by the exponents. For high densities and high temperatures, however, $\mathcal{F}_{s,\xi}^{(2)}$ is not negligible anymore. 
\end{widetext}


\subsection{Cauchy principal value calculation}
\label{sec:cpv}

We need to compute the Cauchy principal value
\begin{align}
\mathcal{I} = \mathcal{P} \int_{\mathbb{R}^3}d\vek{q} \frac{f^2(q) q_\xi q_{\xi^\prime}}{\omega_0 - \omega} \qquad \xi,\xi^\prime=1,2,3,
\end{align} 
where $\omega_0 \equiv \vert\Omega_{rf} (\beta_\xi/2 + s)\vert$, $\omega \equiv \varepsilon(\vek{q})/\hbar$, and $f(q)$ given by Eq.~(\ref{eq:fq}). Henceforth, we neglected the imaginary unit $i$ in Eq.~(\ref{eq:delta-CPV}). The above integral can be rewritten as [$q_{s,\xi}$ is defined in Eq.~(\ref{eq:qsj})]:
\begin{align}
&\mathcal{I}_-(q_{s,\xi}) = \frac{2 m}{\hbar} \mathcal{P} \int_{\mathbb{R}^3}d\vek{q}\, \frac{f^2(q) q_\xi q_{\xi^\prime}}{\vert q_{s,\xi}^2\vert - q^2} 
\nonumber\\
& = \frac{2 m}{\hbar} \mathcal{P} \int_{0}^\infty dq q^2\int_0^\pi d\theta \sin(\theta) \int_0^{2\pi}d\varphi \frac{f^2(q) q_\xi q_{\xi^\prime}}{\vert q_{s,\xi}^2\vert - q^2} 
\end{align} 
where we transformed the wave vector in spherical coordinates~(\ref{eq:spherical}).
The only angular dependence in the integrand comes from $q_\xi q_{\xi^\prime}$, as the other functions rely on $q$ only. Given this, one can verify that the angular integral of $q_\xi q_{\xi^\prime}$ always vanishes for $\xi\ne \xi^\prime$. Thus, we just need to compute the CPV for each direction separately. The angular part of the integration is the same for all directions, namely it yields $4\pi/3$. Thus, we need to assess the integral
\begin{align}
\label{eq:J-CPV}
\mathcal{I}_-(q_{s,\xi}) &= \frac{8 m\pi}{3\hbar} \mathcal{P} \int_{0}^\infty dq\, \frac{f^2(q) q^4}{\vert q_{s,\xi}^2\vert - q^2} 
= \frac{8 m\pi}{3\hbar} {\mathcal{J}}_-^\prime(q_{s,\xi}).
\end{align} 

In case the denominator of the integrand has a plus sign, we have an integration without any singularity:
\begin{align}
\label{eq:J-CPVplus}
\mathcal{I}_+(q_{s,\xi}) & = -\frac{2 m}{\hbar} \int_{\mathbb{R}^3}d\vek{q}\, \frac{f^2(q) q_\xi^2}{\vert q_{s,\xi}^2\vert + q^2} 
= - \frac{8 m\pi}{3\hbar} {\mathcal{J}}_+^\prime(q_{s,\xi}).
\end{align} 
This integration can be in principle carried out analytically, but we refrain to provide an expression, since it is quite involved and it includes Meijer G-functions. A similar argument holds for the integrals involving the bosonic occupation number, that is,
\begin{align}
\label{eq:J-CPVnq}
\mathcal{I}_\pm^{n_q}(q_{s,\xi}) & = \mp\frac{2 m}{\hbar} \int_{\mathbb{R}^3}d\vek{q}\, \frac{f^2(q) q_\xi^2}{\vert q_{s,\xi}^2\vert \pm q^2} 
\frac{1}{e^{\left[\beta_{\mathcal{T}}(\varepsilon(\mathbf{q})-\mu_{\text{G}})\right]}-1}
\nonumber\\
& = \mp \frac{8 m\pi}{3\hbar} {\mathcal{J}}_\pm^{n_q}(q_{s,\xi}).
\end{align} 
To compute them, we performed a numerical integration with Mathematica.


\section{Rescaled equations of motion}
\label{sec:rescaled-EoM}

In our numerical simulations we solve the differential equations~(\ref{eq:EOM2ndMom}) in rescaled units, namely we rescaled the time and space variables with respect to $\nu_{\xi}$ and $l_{\xi} = \sqrt{\hbar / (M\nu_\xi)}$, respectively, that is, with respect to the reference harmonic oscillator frequency and length scale. Hence we have: $\bar{r}_\xi^2 = (\langle r_\xi\rangle/l_\xi)^2$, $\bar{p}_{\xi}^2 = [\langle p_{\xi}\rangle / (l_{\xi}\nu_{\xi} M)]^2 = (l_\xi/\hbar)^2 \langle p_{\xi}^2\rangle$, and $\bar{c}_{\xi} = c_{\xi}/(l_{\xi}^2\nu_{\xi} M) = c_{\xi}/\hbar$, where we introduced a bar for indicating the dimensionless quantities. Given this, equations~(\ref{eq:EOM2ndMom}) in rescaled units read

\begin{align}
\label{eq:EOM2ndMom_scale}
&\frac{d}{d\tau} \bar{r}_\xi^2 = \bar{c}_{\xi} \nonumber\\[1ex]
&\frac{d}{d\tau} \bar{p}_{\xi}^2 = \left\{ \bar{\Gamma}_\xi \Im[\bar{\Phi}_\xi(\tau)] - \bar{{W}}_\xi^\prime(\tau)\right\} \bar{c}_{\xi} 
- 2\bar{\Gamma}_\xi\Im[\bar\Lambda_\xi(\tau)] \bar{p}_{\xi}^2 \nonumber\\
&+ \bar{\Gamma}_\xi \Re[\bar{\Phi}_\xi(\tau)] \nonumber\\[1ex]
&\frac{d}{d\tau} \bar{c}_{\xi} = 2 \left\{ \bar{\Gamma}_\xi \Im[\bar{\Phi}_\xi(\tau)] - \bar{{W}}_\xi^\prime(\tau)\right\} \bar{r}_\xi^2 + 2 \bar{p}_{\xi}^2 
\nonumber\\
& - \bar{\Gamma}_\xi\left\{
\Im[\bar\Lambda_\xi(\tau)]\bar{c}_{\xi} - \Re[\bar\Lambda_\xi(\tau)] 
\right\}.
\end{align}

with $\tau = \nu_\xi t$,
$\bar{\Omega}_{rf}^\xi = \Omega_{rf} / \nu_\xi$, and 
\begin{align}
\bar{\Gamma}_\xi & = \frac{2\pi}{3} \frac{m}{M} \left(\frac{M}{\mu}\right)^2 (n_0 l_\xi^3),\nonumber\\[1ex]
\bar{{W}}_\xi^\prime(\tau) & = \frac{\bar{{W}}_\xi(\tau)+\delta\bar{{W}}_\xi(\tau)}{\nu_\xi^2}
= \left(
\frac{\bar{\Omega}_{rf}^\xi}{2}
\right)^2
\nonumber\\
& \times 
\left[
a_\xi + \delta \bar a_\xi  - 2 (q_\xi + \delta \bar q_\xi)\cos(\bar{\Omega}_{rf}^\xi \tau) 
- 2 \delta \bar q_\xi^\prime \bar g_\xi(\tau)
\right]\nonumber\\[2ex]
\bar g_\xi(\tau) & = \!\!\!\sum_{s,s^\prime\notin S_{\mathrm{i}}} \!\!\! F_{s,s^\prime}^\xi \cos[(s-s^\prime)\bar{\Omega}_{rf}^\xi \tau]
\!\!\left[\bar{{\mathcal{J}}}_+^\prime(\bar q_{s,\xi}) - \bar{{\mathcal{J}}}_-^\prime(\bar q_{s,\xi})\right]
\nonumber\\[1ex]
\bar Q_\xi & = \frac{Q}{l_\xi} = \frac{32}{3}\frac{m}{M}\left(\frac{M}{\mu}\right)^2\frac{(n_0 l_\xi^3)}{(\bar{\Omega}_{rf}^\xi)^2}
\end{align}
and $\bar{{\mathcal{J}}}_\pm^\prime(\bar q_{s,\xi}) = {\mathcal{J}}_\pm^\prime(q_{s,\xi}) l_\xi$. Here, for the sake of simple notation, we just refer to the bosonic case. 
Moreover, the rescaled $\Phi$ and $\Lambda$ functions read:
\begin{equation}
    \begin{split}
        \bar{\Lambda}_\xi(\tau)=i&\sum_{s,s^\prime}C^\xi_s C^\xi_{s^\prime}\bigg\{|\bar{q}_{s,\xi}|^3\bar{f}(\bar{q}_{s,\xi})^2(1-\phi)\\[1ex]
        &\bigg[i\sin\big[(s-s')\bar{\Omega}_{rf}\tau\big](1+2n_{\bar{q}_{s,\xi}})\\[1ex]
        &+\cos\big[(s-s')\bar{\Omega}_{rf}\tau\big]\mathrm{sgn}\big(\beta_\xi/2+s\big)\bigg]\\[1ex]
        &-\bar{\eta}^+_{s,\xi}e^{-i(s-s')\bar{\Omega}_{rf}\tau}+\bar{\eta}^-_{s,\xi}e^{i(s-s')\bar{\Omega}_{rf}\tau}\bigg\}\\[1ex]
        -\frac{2}{\pi}\bar{\Omega}_{rf}&\sum_{s,s'}C^\xi_s C^\xi_{s^\prime}\bigg\{\cos\big[(s-s')\bar{\Omega}_{rf}\tau\big]\\[1ex]
       \Big[1-&2\Theta\big(\mathrm{sgn}(\beta_\xi/2+s)\big)\Big]\Big[\bar{\mathcal{J}}'_-(\bar{q}_{s,\xi})+\bar{\mathcal{J}}'_+(\bar{q}_{s,\xi})\\[1ex]
       &+2\big(\bar{\mathcal{J}}^{n_q}_-(\bar{q}_{s,\xi})+\bar{\mathcal{J}}^{n_q}_+(\bar{q}_{s,\xi})\Big]\\[1ex]
       -i\sin&\big[(s-s')\bar{\Omega}_{rf}\tau\big]\Big[\bar{\mathcal{J}}'_-(\bar{q}_{s,\xi})-\bar{\mathcal{J}}'_+(\bar{q}_{s,\xi})\Big]\bigg\}
    \end{split}
\end{equation}

\begin{equation}
    \begin{split}
       \bar{\Phi}_\xi(\tau)=\,\bar{\Omega}_{rf}&\sum_{s,s'}C^\xi_s C^\xi_{s^\prime}(\beta_\xi/2+s')\bigg\{|\bar{q}_{s,\xi}|^3\bar{f}(\bar{q}_{s,\xi})^2\\[1ex]
       &(1-\phi)\bigg[\cos\big[(s-s')\bar{\Omega}_{rf}\tau\big](1+2n_{\bar{q}_{s,\xi}})\\[1ex]
       &+i\sin\big[(s-s')\bar{\Omega}_{rf}\tau\big]\mathrm{sgn}\big(\beta_\xi/2+s\big)\bigg]\\[1ex]
       &+\bar{\eta}^-_{s,\xi}e^{-i(s-s')\bar{\Omega}_{rf}\tau}+\bar{\eta}^+_{s,\xi}e^{i(s-s')\bar{\Omega}_{rf}\tau}\bigg\}\\[2ex]
       -\frac{2}{\pi}\bar{\Omega}_{rf}&\sum_{s,s'}C^\xi_s C^\xi_{s^\prime}(\beta_\xi/2+s')\sin\big[(s-s')\bar{\Omega}_{rf}\tau\big]\\[1ex]
       \Big[2\Theta&\big(\mathrm{sgn}(\beta_\xi/2+s)\big)-1\Big]\Big[\bar{\mathcal{J}}'_-(\bar{q}_{s,\xi})+\bar{\mathcal{J}}'_+(\bar{q}_{s,\xi})\\[1ex]
       &+2\big(\bar{\mathcal{J}}^{n_q}_-(\bar{q}_{s,\xi})+\bar{\mathcal{J}}^{n_q}_+(\bar{q}_{s,\xi})\Big]
    \end{split}
\end{equation}

where $\bar{f}(\bar q_{s,\xi}) = f(q_{s,\xi})/ l_\xi$, $\bar{\beta}_{\mathcal{T}}^\xi = \beta_{\mathcal{T}}\hbar\nu_\xi$, 
$\bar{{\mathcal{J}}}_\pm^{n_q}(\bar q_{s,\xi}) = {\mathcal{J}}_\pm^{n_q}(q_{s,\xi}) l_\xi$, and 
\begin{align}
\label{eq:eta-qsj}
\bar{\eta}_{s,\xi}^{\pm} & = \frac{1}{16\pi^2} \frac{m}{M} \frac{\bar{\mathcal{F}}_{s,\xi}^{(1),\pm} +(-1)^\phi \bar{\mathcal{F}}_{s,\xi}^{(2)}}{(n_0 l_\xi^3) \bar{\beta}_{\mathcal{T}}^\xi},\nonumber\\[1ex]
\bar q_{s,\xi} & = l_\xi q_{s,\xi} = \sqrt{\frac{2 m \Omega_{rf}}{M\nu_\xi}\left\vert
\frac{\beta_\xi}{2} + s
\right\vert}.
\end{align}
For the numerical assessment of the integrals involved in the functions ${\mathcal{F}}_{s,\xi}^{(1,2)} = \bar{\mathcal{F}}_{s,\xi}^{(1,2)}/l_\xi^2$, we note that they have the units of a wave vector. Given this, we replace the integral variable by $k\mapsto \bar{k}_\xi = k l_\xi$ in Eqs.~(\ref{eq:Fnj1},\ref{eq:Fnj2}) and we define the rescaled parameters in Eq.~(\ref{eq:alpha0pm}) as:
\begin{align}
\alpha_0(\bar{k}_\xi) & = \frac{m}{M} \frac{\varpi_{s,\xi}^2}{2} \frac{\bar{\beta}_{\mathcal{T}}^\xi}{\bar{k}^2_\xi}\nonumber\\
\alpha_\pm(\bar{k}_\xi) & = \frac{\bar{\beta}_{\mathcal{T}}^\xi \varpi_{s,\xi}}{2}\left[
\frac{\bar{k}_\xi^2}{4} \frac{M}{m}\frac{1}{\varpi_{s,\xi}} \pm 1 
\right] - \bar{\beta}_{\mathcal{T}}^\xi \bar{\mu}_{\text{G}}^\xi\nonumber\\
\varpi_{s,\xi} & = \frac{\Omega_{rf}}{\nu_\xi}\left(\frac{\beta_\xi}{2} + s\right),
\qquad
\bar{\mu}_{\text{G}}^\xi = \frac{\mu_{\text{G}}}{\hbar\nu_\xi}.
\end{align}

Finally, let us comment on the rescaling of the scattering amplitude $f(q)$. The expression given in Eq.~(\ref{eq:fq}) assumes that the regularisation parameters $b,\,c$ are given in units of $R^\star$, as it is more convenient to work with that unit length when solving the Schr\"odinger equation~(\ref{eq:SEscatt_SI}). Thus, if everything is in that unit length, that is, also the $q$ wave vector, then the scattering amplitude is in $R^\star$ units as well. As a consequence, if we wish to have it in $l_\xi$ units, we have to multiply $f(q)$ by $R^\star / l_\xi$. Hence, when we have to assess the scattering amplitude in the $\bar{\Lambda}$ and $\bar{\Phi}$ functions, we need first to provide $q_{s,\xi}$ in $R^\star$ units as well as $b$ and $c$, and then multiply the obtained result by $R^\star / l_\xi$. The wave vector $q_{s,\xi}$ in $R^\star$ units is given by
\begin{align}
q_{s,\xi}^\star & = R^\star q_{s,\xi} = \frac{R^\star}{l_\xi} \bar q_{s,\xi},
\end{align}
where $\bar q_{s,\xi}$ is defined in Eq.~(\ref{eq:eta-qsj}). The situation is slightly different when the integrations involved in the functions ${\mathcal{F}}_{s,\xi}^{(1,2)}$, ${\mathcal{J}}_\pm^\prime(q_{s,\xi})$, and ${\mathcal{J}}_\pm^{n_q}(q_{s,\xi})$ are considered. We can rescale the integrands in $R^\star$ units, as the scattering amplitude~(\ref{eq:fq}), and then we rescale the result in $l_\xi$ units. Alternatively, we first rescale the scattering amplitude in units of $l_\xi$, thus we perform the integrations in $l_\xi$ units. We have chosen the second option, as the regularisation parameters $b,\,c$ have been obtained in $R^\star$ units. In this case the scattering amplitude in $l_\xi$ units is given by
\begin{align}
\bar f(\bar q)& = \frac{\bar c^2 \pi (R^\star)^2 l_\xi^{-2}}{(\bar b^2 - \bar c^2)^2\, \bar q}\left\{
e^{-\bar b \bar q}\left[ 1 + \frac{(\bar b^4 - \bar c^4) \bar q}{4 \bar b \bar c^2}\right] - e^{-\bar c \bar q}
\right\},
\end{align}
where we have introduced the factor $\left(R^\star/l_\xi\right)^2$. The regularisation parameters $\bar b,\,\bar c$ are in $l_\xi$ units, which can be obtained from the $b,\,c$ in $R^\star$ units using the relations
\begin{align}
\label{eq:bar_bc}
\bar b = b \frac{R^\star}{l_\xi}, \qquad \bar c = c \frac{R^\star}{l_\xi}.
\end{align}
A similar reasoning applies for the integral~(\ref{eq:J-CPV}). Indeed, using the definitions~(\ref{eq:bar_bc}) and Eq.~(\ref{eq:J-CPV}), one has to replace $R^\star$ in Eq.~(\ref{eq:J-CPV}) by $\left(R^\star/l_\xi\right)^4$. Furthermore, the free particle dispersion relation is rescaled as $\bar\varepsilon(\mathbf{q}) = \varepsilon(\mathbf{q}) / (\hbar\nu_\xi) = M \bar q^2/(2 m)$.


\section{Lindblad form of the master equation}
\label{sec:lindbland}

The ion master equation~(\ref{eq:ME-LDA-SIunits}) cannot be transformed in a Lindblad-type Markovian master equation, as a crucial assumption to obtain such a form is the rotating-wave approximation. Using the definitions for the position and momentum operators for each spatial direction $\xi=x,y,z$
\begin{align}
\hat r_\xi = \sqrt{\frac{\hbar}{2 M\nu_\xi}}(\hat a^\dag_\xi + \hat a_\xi)
\qquad
\hat p_\xi = i \sqrt{\frac{M\hbar\nu_\xi}{2}}(\hat a^\dag_\xi - \hat a_\xi)
\end{align}
we can rewrite Eq.~(\ref{eq:ME-LDA-SIunits}) as
\begin{align}
\label{eq:ME-LDA-Lindbland}
\dot{\hat\rho}_\xi &= -\frac{i}{\hbar} [\hat H_S^\xi +\delta\hat H_S^\xi + \tilde{H}_S^\xi,\hat\rho_\xi]
+\gamma_\xi^a\hat{\mathcal{D}}[a_\xi]\hat\rho_\xi 
+ \gamma_\xi^{a^\dag}\hat{\mathcal{D}}[a^\dag_\xi]\hat\rho_\xi\nonumber\\
&+\gamma_\xi^+\hat G_+[\hat a_\xi,\hat a_\xi^\dag]\hat\rho_\xi - \frac{i}{\hbar}\gamma_\xi^-\hat G_-[\hat a_\xi,\hat a_\xi^\dag]\hat\rho_\xi.
\end{align}
Here, we have introduced the operators
\begin{align}
\label{eq:Lindbland-operators}
&\tilde{H}_S^\xi = \hbar\Delta_\xi\left[
\hat a_\xi^\dag \hat a_\xi + \frac{(\hat a_\xi^\dag)^2 + \hat a_\xi^2}{2}
\right] + i\hbar\Delta^\prime_\xi [(\hat a_\xi^\dag)^2 - \hat a_\xi^2],\nonumber\\
&\hat{\mathcal{D}}[a_\xi]\hat\rho_\xi = \hat a_\xi \hat\rho_\xi \hat a_\xi^\dag - \frac{\hat a_\xi^\dag \hat a_\xi \hat\rho_\xi + \hat\rho_\xi \hat a_\xi^\dag \hat a_\xi}{2},\nonumber\\
&\hat G_\pm[\hat a_\xi,\hat a_\xi^\dag]\hat\rho_\xi = 
\hat a_\xi^\dag \hat\rho_\xi \hat a_\xi^\dag \pm \hat a_\xi \hat\rho_\xi \hat a_\xi 
\nonumber\\
&- \frac{[(\hat a_\xi^\dag)^2 \pm \hat a_\xi^2] \hat\rho_\xi + \hat\rho_\xi [(\hat a_\xi^\dag)^2 \pm \hat a_\xi^2]}{2},
\end{align}
and the damping rates as well as energy shifts
\begin{align}
\label{eq:rates-lindbland}
&\gamma_\xi^a = \hbar\Gamma\left[
\frac{\Re(\Phi)}{M\nu_\xi} + \Im(\Lambda)
\right],\qquad 
\gamma_\xi^{a^\dag} = \hbar\Gamma\left[
\frac{\Re(\Phi)}{M\nu_\xi} - \Im(\Lambda)
\right],\nonumber\\
&\gamma_\xi^+ = \frac{\hbar\Gamma\Re(\Phi)}{M\nu_\xi},\qquad
\gamma_\xi^- = \hbar^2\Gamma\Re(\Lambda),\nonumber\\
&\Delta = -\frac{\hbar\Gamma\Im(\Phi)}{M\nu_\xi},\qquad
\Delta^\prime = \frac{\hbar\Gamma\Im(\Lambda)}{2}.
\end{align}
The first line of Eq.~(\ref{eq:ME-LDA-Lindbland}) has the structure of the usual Lindbland master equation with damping rates $\gamma_\xi^a$, $\gamma_\xi^{a^\dag}$ and damping operators $\hat a_\xi$ and $\hat a_\xi^\dag$, respectively. The second line of Eq.~(\ref{eq:ME-LDA-Lindbland}), however, cannot be recasted in either a unitary term like the commutator in the first line or in a dissipative term as the second and third terms of the first line of Eq.~(\ref{eq:ME-LDA-Lindbland}). Those two last terms originate from the fact that we did not apply the rotating-wave approximation. The additional Hamiltonian term $\hat H_S^\prime$ is also a consequence of this fact. Now, looking at the structure of the $\Phi$~(\ref{eq:Phi}) and $\Lambda$ functions~(\ref{eq:Lambda},\ref{eq:PhiLambda-lamb}) and at the fact that we consider a linear Paul trap for which the most relevant coefficients $C_n^\xi$ are those for $n=0,\pm 1$, we see that while $\Re(\Phi^{\mathcal{P}})=\Phi^{\mathcal{P}}$ has a negligible effect, $\Re(\Phi^{\delta})$ provides a non-negligible effect such that it renders $\gamma_\xi^+$ non-zero. A similar argument holds for $\gamma_\xi^-$ for which $\Re(\Lambda^{\mathcal{P}})$ yields a significant contribution, but not $\Re(\Lambda^{\delta})$. On the other hand, $\Im(\Phi^{\delta})$ is almost negligible so that $\Delta\simeq 0$, but $\Im(\Lambda^{\delta})$ produces a non negligible contribution, while $\Im(\Lambda^{\mathcal{P}})$ is negligible, so that $\Delta^\prime$ provides an important contribution to the ion dynamics.

In conclusion, the ion master equation like the ones for a neutral impurity in a condensate~\cite{DaleyPRA04,NielsenNJP19,LenaPRA20} cannot be recasted in Lindblad form, unless the counter rotating terms are neglected. In the future, however, it would be interesting to explore another approach that has been recently proposed~\cite{nathan2020universal}. Here, it is shown that one does not need to apply the rotating-wave approximation and, by using another strategy to apply the Markov approximation, it is possible to derive a different Markovian quantum master equation in Lindblad form, but with time-dependent decay operators. The advantage of this approach is that the master equation can be equivalently simulated by a stochastic Schr\"odinger equation, similarly to the well-known Monte Carlo wavefunction approach~\cite{MCWF1992,Molmer:93}. The reduction from a density matrix to a ket state description, albeit averaging over many quantum trajectories, could be especially useful for simulating the impurity dynamics fully in three-dimensions.


\section{Self-consistency of the master equation}
\label{sec:consistency}

\begin{figure}
\centering
\includegraphics[scale=0.45]{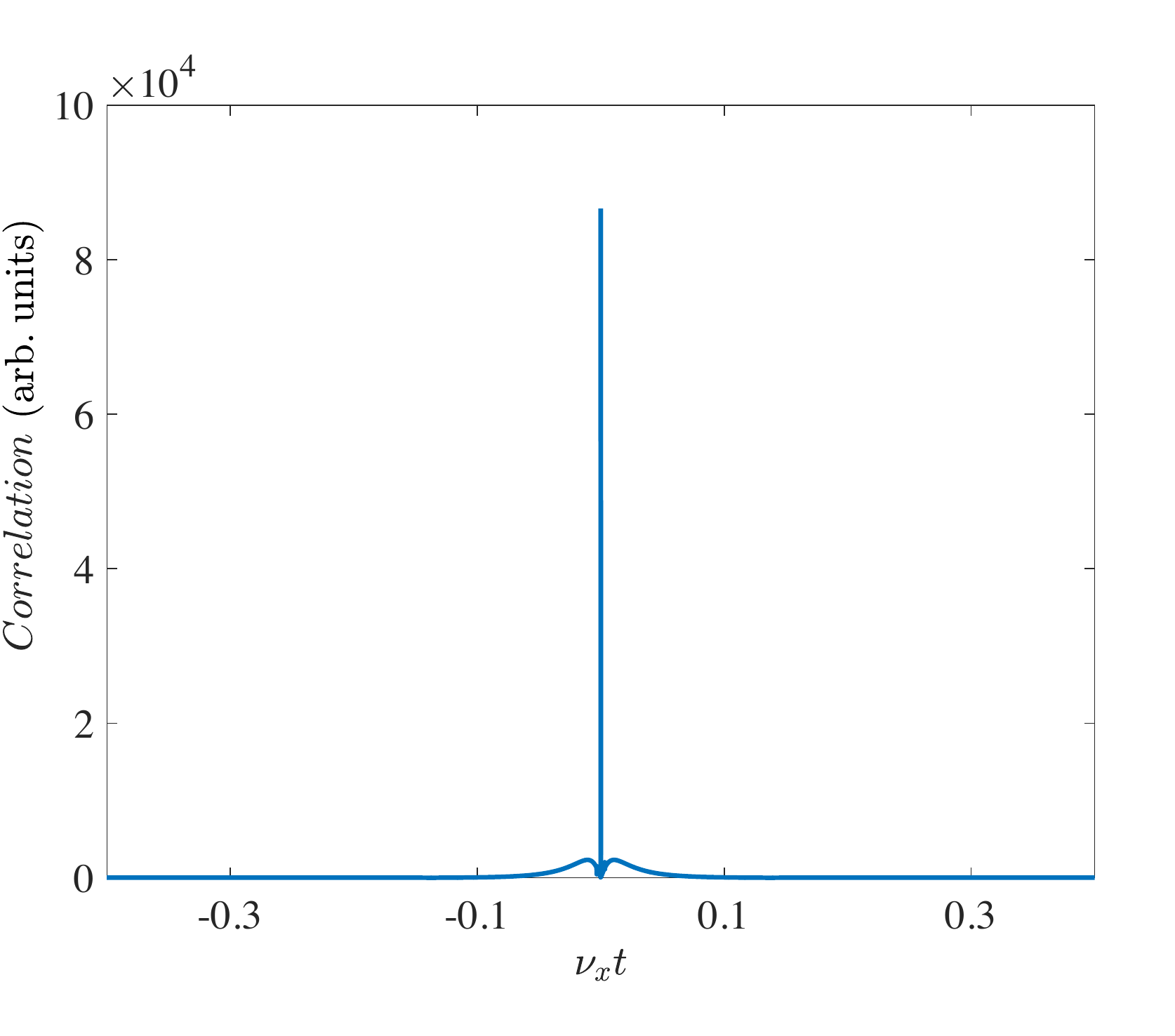}
\caption{\label{fig:bathcorr} (Color online). Example of bath correlation function as evaluated by Eq.~(\ref{eq:bathcorr}) for the $x$-direction.}
\end{figure}

As we already pointed out, the dissipative damping rate must be smaller than the thermal energy and the typical system's transition frequencies. In this case, the dissipative rate in the $\xi$-th direction is proportional to [see also Eq.~(\ref{eq:rates-lindbland})]
\begin{align}
\label{eq:gamma}
\gamma_\xi \sim \Gamma \sum_s\vert {q}_{s,\xi}\vert^3 f({q}_{s,\xi})^2 n_{{q}_{s,\xi}} \vert F_{s,s}^\xi\vert,
\end{align}
where we have neglected the contribution of the terms for which $s\ne s^\prime$, since these, on average, vanish due to the fast rf-oscillations. 
The dissipative rate has to satisfy the two conditions: $\hbar\gamma_\xi/(k_B \mathcal{T}) \ll 1$ and $\gamma_\xi / \nu_\xi \ll 1$. For instance, for the $^{23}$Na/$^{174}$Yb$^+$ pair with a gas temperature of $ \mathcal{T} = 200$ nK and trap parameters $a = -0.001$, $q = 0.2$, and $\Omega_{rf} = 2\pi\,$2 MHz we obtain the trap frequencies $\nu_x = 2 \pi\,$ 112 kHz, $\nu_y = 2 \pi\,$ 169 kHz, $\nu_z = 2 \pi\,$45 kHz, $k_B  \mathcal{T}/\hbar = 2\pi\,$4 kHz, for which the dissipative rate fulfils the above outlined requirements rather well, i.e. the ratios are smaller than 3$\times$10$^{-4}$ for an atomic peak density $10^{14} $cm$^{-3}$. These conclusions can be further corroborated by an evaluation of the bath correlation functions. For example, starting from Eq.~(\ref{eq:MElda}) and by performing the replacement~(\ref{eq:replacement}), the first correlation function due to the single sum over $\mathbf{q}$ in the curl brackets is given by 
\begin{align}
\label{eq:bathcorr}
& \sum_{\vek{q}} \Omega^2_{\vek{q}} \sin(\varepsilon(\vek{q})\tau/\hbar) q_j q_s \nonumber\\
\phantom{=}&\propto \int_0^\infty \mathrm{d}\bar{q}\,\bar{q}^2 \left\{ 
e^{-\bar{b} \bar{q}}\left[ 1 + \frac{(\bar{b}^4 - \bar{c}^4)q}{4 \bar{b} \bar{c}^2}\right] - e^{- \bar{c} \bar{q}}
\right\}
\sin(\Xi \bar{q}^2\bar{\tau})
\end{align}
where $\Xi = \mu E^\star/(m \hbar\nu_\xi)$, $\bar{\tau} = \nu_\xi \tau$, and the regularisation parameters $b,\,c$ as well as the wave vector $q$ have been rescaled with respect to $R^\star$ and $1/R^\star$, respectively. An example of such a correlation function is given 
in Fig.~\ref{fig:bathcorr} for the spatial direction $x$. As it can be seen, the function decays rapidly to zero, i.e. for times larger than, approximatively, 0.15/$\nu_x$ it vanishes. Hence, the Markov approximation in our setting is satisfied reasonably well. 
 
\bibliography{ion-atom,liter}

\end{document}